%% file: aa201629512.tex
\documentclass[longauth]{aa_gaia} 

\usepackage{graphicx}
\usepackage[varg]{txfonts}
\usepackage{natbib,twoopt}
\usepackage{xcolor}
\usepackage[breaklinks=true]{hyperref} 
\bibpunct{(}{)}{;}{a}{}{,}             
\makeatletter
  \newcommandtwoopt{\citeads}[3][][]{\href{http://adsabs.harvard.edu/abs/#3}%
    {\def\hyper@linkstart##1##2{}%
     \let\hyper@linkend\@empty\citealp[#1][#2]{#3}}}
  \newcommandtwoopt{\citepads}[3][][]{\href{http://adsabs.harvard.edu/abs/#3}%
    {\def\hyper@linkstart##1##2{}%
     \let\hyper@linkend\@empty\citep[#1][#2]{#3}}}
  \newcommandtwoopt{\citetads}[3][][]{\href{http://adsabs.harvard.edu/abs/#3}%
    {\def\hyper@linkstart##1##2{}%
     \let\hyper@linkend\@empty\citet[#1][#2]{#3}}}
  \newcommandtwoopt{\citeyearads}[3][][]%
    {\href{http://adsabs.harvard.edu/abs/#3}
    {\def\hyper@linkstart##1##2{}%
     \let\hyper@linkend\@empty\citeyear[#1][#2]{#3}}}
\makeatother

\newcommand\gaia{\textit{Gaia}}
\newcommand\hip{\textsc{Hipparcos}}
\newcommand\tyc{\textit{Tycho}}
\newcommand\tyctwo{\textit{Tycho}-2}
\newcommand\gdr{\gaia~DR1}
\newcommand\secref[1]{Sect.~\ref{#1}}

\newcommand\figref[1]{Fig.~\ref{#1}}
\newcommand\figsref[1]{Figs.~\ref{#1}}
\newcommand\figrefalt[1]{Figure~\ref{#1}}

\newcommand\equref[1]{Eq.~\eqref{#1}}

\newcommand\tabref[1]{Table~\ref{#1}}

\newcommand\gdrtotnum{\ensuremath{1\,142\,679\,769}}
\newcommand\gdrsecnum{\ensuremath{1\,140\,622\,719}}
\newcommand\tgasnum{\ensuremath{2\,057\,050}}
\newcommand\tycnum{\ensuremath{1\,963\,415}}
\newcommand\hipnum{\ensuremath{93\,635}}
\newcommand\varnum{\ensuremath{3194}}
\newcommand\cepnum{\ensuremath{599}}
\newcommand\cepnumnew{\ensuremath{43}}
\newcommand\rrlnum{\ensuremath{2595}}
\newcommand\rrlnumnew{\ensuremath{343}}
\newcommand\muas{\ensuremath{\mu\text{as}}}

\newcommand\kms{\ensuremath{\text{km~s}^{-1}}}

\begin{document} 

\title{{\gaia} Data Release 1}

\subtitle{Summary of the astrometric, photometric, and survey properties}

\input{authors_full}

\date{Received ; accepted }

\abstract{At about 1000 days after the launch of {\gaia} we present the first {\gaia} data release,
{\gdr}, consisting of astrometry and photometry for over 1 billion sources brighter than magnitude
$20.7$.}
{A summary of {\gdr} is presented along with illustrations of the scientific quality of the data,
followed by a discussion of the limitations due to the preliminary nature of this release.}
{The raw data collected by {\gaia} during the first 14 months of the mission have been processed by
the {\gaia} Data Processing and Analysis Consortium (DPAC) and turned into an astrometric and
photometric catalogue.}
{{\gdr} consists of three components: a primary astrometric data set which contains the positions,
parallaxes, and mean proper motions for about 2 million of the brightest stars in common with the
{\hip} and {\tyctwo} catalogues -- a realisation of the {\tyc}-{\gaia} Astrometric Solution (TGAS)
-- and a secondary astrometric data set containing the positions for an additional $1.1$ billion
sources. The second component is the photometric data set, consisting of mean $G$-band magnitudes
for all sources. The $G$-band light curves and the characteristics of $\sim3000$ Cepheid and
RR Lyrae stars, observed at high cadence around the south ecliptic pole, form the third component.

For the primary astrometric data set the typical uncertainty is about $0.3$~mas for the positions
and parallaxes, and about $1$~mas~yr$^{-1}$ for the proper motions. A systematic component of
$\sim0.3$~mas should be added to the parallax uncertainties. For the subset of $\sim94\,000$ {\hip}
stars in the primary data set, the proper motions are much more precise at about
$0.06$~mas~yr$^{-1}$. For the secondary astrometric data set, the typical uncertainty of the
positions is $\sim 10$~mas. The median uncertainties on the mean $G$-band magnitudes range from the
mmag level to $\sim0.03$ mag over the magnitude range $5$ to $20.7$.}
{{\gdr} is an important milestone ahead of the next {\gaia} data release, which will feature
five-parameter astrometry for all sources. Extensive validation shows that {\gdr} represents a major
advance in the mapping of the heavens and the availability of basic stellar data that underpin
observational astrophysics. Nevertheless, the very preliminary nature of this first {\gaia} data
release does lead to a number of important limitations to the data quality which should be carefully
considered before drawing conclusions from the data.}


\maketitle

\titlerunning{{\gaia} Data Release 1: Summary} 
\authorrunning{{\gaia} Collaboration}

%
%

\section{Introduction}
\label{sec:intro}

The {\gaia} satellite was launched at the end of 2013 to collect data that will allow the
determination of highly accurate positions, parallaxes, and proper motions for $>1$ billion sources
brighter than magnitude $20.7$ in the white-light photometric band $G$ of {\gaia} (thus going deeper
than the originally planned limit of $G=20$). The astrometry is complemented by multi-colour
photometry, measured for all sources observed by {\gaia}, and radial velocities which are collected
for stars brighter than $G\approx17$. The scientific goals of the mission are summarised in
\cite{DPACP-1}, while a more extensive scientific motivation for the mission is presented in
\cite{2001A&A...369..339P}.

The spacecraft, its scientific instruments, and the observing strategy have been designed to meet
the performance requirement of $24$ {\muas} accuracy on the parallax of a $15$th magnitude
solar-type star at the end of the nominal 5 year mission lifetime. The entity entrusted with the
data processing for the {\gaia} mission, the {\gaia} Data Processing and Analysis Consortium
\citep[DPAC, described in][]{DPACP-1}, is expected to deliver the final data products (at their
ultimately achievable accuracy) only at the end of post-operational phase of the mission, currently
foreseen for 2022--2023. It was therefore agreed at the time of the creation of DPAC that the
astronomical community should have access to the {\gaia} data at an earlier stage through
intermediate data releases. It was understood that these intermediate releases are based on
preliminary calibrations and only on a subset of the measurements available at the end of the
mission, and therefore will not be representative of the end-of-mission {\gaia} performance.

In this paper we present the first such intermediate {\gaia} data release ({\gaia} Data Release 1,
\gdr), which is based on the data collected during the first 14 months of the nominal mission
lifetime (60 months). In \secref{sec:instruments} we provide a short summary of the {\gaia}
instruments and the way the data are collected. We summarise the astrometric, photometric and
variable star contents of {\gdr} in \secref{sec:gdrsummary}. A summary of the validation of the
results is provided in \secref{sec:gdrvalidation} and a few illustrations of the contents of {\gdr}
are provided in \secref{sec:sciencedemos}. The known limitations of this first release are presented
in \secref{sec:gdrlimitations}. In \secref{sec:access} we provide pointers to the {\gdr} data access
facilities and documentation available to the astronomical community. We conclude in
\secref{sec:conclusions}. Although {\gdr} is the first major catalogue release with results from the
{\gaia} mission, {\gaia} data has already been made publicly available as `Science Alerts' on
transient sources, which for example led to the discovery of only the third known eclipsing AM
CVn-system \citep{2015MNRAS.452.1060C}.

We stress at the outset that {\gdr} represents a preliminary release of {\gaia} results with many
shortcomings. We therefore strongly encourage a detailed reading of \secref{sec:gdrlimitations} and
the documentation associated with the release as well as carefully taking into account the listed
limitations when drawing conclusions based on the data contained in {\gdr}.

%
%

\section{{\gaia} instruments and measurements}
\label{sec:instruments}

We provide a brief overview of the {\gaia} instruments and the way measurements are collected in
order to introduce some of the technical terms used in the rest of the paper. A full description of
the {\gaia} spacecraft, instruments, and measurement principles can be found in \cite{DPACP-1}.

{\gaia} continuously scans the sky with two telescopes pointing in directions separated by the basic
angle of $106.5^\circ$. The images produced by the telescopes are projected onto the same focal
plane composed of 106 CCDs which function as the detectors of the various instruments in the {\gaia}
payload. The scanning is achieved through the continuous revolution of {\gaia} about its spin axis
with a period of 6 hours. The spin axis direction precesses around the direction to the Sun (as seen
from {\gaia}), which allows complete coverage of the sky. Statistics of the sky coverage achieved
for {\gdr} are presented in \cite{DPACP-14} and \cite{DPACP-12}, while the properties of the {\gaia}
scanning law with respect to variable star studies are described in \cite{DPACP-15}.

The spinning motion of the spacecraft results in the source images moving across the focal plane.
This necessitates the operation of the {\gaia} CCDs in time-delayed integration (TDI) mode so as to
allow the accumulation of charge as the images move across the CCDs. The CCDs are not fully read
out, only the pixels in a `window' around each source are read out and stored for transmission to
the ground. These windows come in various sizes and sampling schemes.

The astrometric instrument takes up most of the focal plane and collects source images in the
{\gaia} white-light pass band $G$ \citep[covering the range
$330$--$1050$~nm,][]{DPACP-9,2010A&A...523A..48J}. The fundamental inputs to the astrometric data
processing consist of the precise times when the image centroids pass a fiducial line on the CCD
\citep{2012A&A...538A..78L}. The image centroid and the flux contained in the image are determined
as part of the pre-processing \citep{DPACP-7}. The sensitivity of the astrometric instrument is such
that sources brighter than about $G=12$ will lead to saturated images. This effect is mitigated
through the use of TDI gates, which are special structures on the CCDs that can be activated to
inhibit charge transfer and hence to effectively reduce the integration time for bright sources.

The photometric instrument is realised through two prisms dispersing the light entering the field of
view of two dedicated sets of CCDs. The Blue Photometer (BP) operates over the wavelength range
$330$--$680$ nm, while the Red Photometer (RP) covers the wavelength range $640$--$1050$ nm
\citep{DPACP-9,2010A&A...523A..48J}. The data collected by the photometric instrument consists of
low resolution spectrophotometric measurements of the source spectral energy distributions. This
colour information is intended for use in the astrometric processing (to correct for chromatic
effects) and to provide the astrophysical characterisation of all sources observed by {\gaia}. The
$G$-band photometry is derived from the fluxes measured in the astrometric instrument. Results from
the photometric instrument are not presented as part of {\gdr}. The photometry in this first release
only concerns the fluxes measured in the $G$ band.

The spectroscopic instrument, also called the radial-velocity spectrometer (RVS) collects medium
resolution ($R\sim11\,700$) spectra over the wavelength range $845$--$872$~nm, centred on the
Calcium triplet region \citep{2011EAS....45..181C}. The spectra are collected for all sources to
$G\approx17$ (16th magnitude in the RVS filter band) and serve primarily to determine the radial
velocity of the sources, although at the bright end \citep[$G<12.5$,][]{2016A&A...585A..93R}
astrophysical information can be derived directly from the spectra. Results from this instrument are
not contained in {\gdr}.

Observations of sources by {\gaia} can be referred to in several ways. `Focal plane transits' refer
to a crossing of the entire focal plane by a given source, which corresponds to a `visit' by {\gaia}
of a specific coordinate on the sky. `CCD transits' refer to the crossing by a source of a
particular CCD in the focal plane. Thus the focal plane transit of the astrometric field typically
consists of 10 transits across individual CCDs, while a photometric instrument transit (BP or RP)
consists of only one CCD transit, and a transit across the RVS instrument consists of three CCD
transits \citep[see][for more details on the focal plane layout and functionalities, and the
in-flight performance of the {\gaia} CCDs]{DPACP-1,DPACP-20}. This distinction is important when it
comes to the difference between the number of measurements (CCD transits) collected for a source and
the number of times it was observed (focal plane transits) by {\gaia}. In the rest of the paper we
will refer to an `observation' or a `focal plane transit' to indicate that a source was observed by
{\gaia} and we refer to `CCD transit' whenever individual CCD measurements are discussed.

Events onboard {\gaia} are labelled by the so-called onboard mission time line (OBMT), which is a
time scale defined by the onboard clock. This time scale is eventually transformed into the
physical barycentric coordinate time (TCB) \citep{DPACP-1,DPACP-14}. By convention OBMT is expressed
in units of 6 hour ($21\,600$~sec) spacecraft revolutions since launch and this unit is often used
in figures of some quantity versus time, including in the papers accompanying {\gdr} and in the data
release documentation (see \secref{sec:access}). For the practical interpretation of time lines
expressed in OBMT the following approximate relation between the OBMT (in revolutions) and TCB at
{\gaia} (in Julian years) can be used:
\begin{equation}
  \text{TCB} \simeq \text{J}2015.0 + (\text{OBMT} - 1717.6256~\text{rev})/(1461~\text{rev}) \, .
\end{equation}
This relation is precise to $\pm2$~sec and is valid only for the time span corresponding to {\gdr}.
The time interval covered by the observations used for {\gdr} starts at OBMT 1078.3795~rev =
J2014.5624599~TCB (approximately 2014~July~25, 10:30:00~UTC), and ends at OBMT 2751.3518~rev =
J2015.7075471~TCB (approximately 2015~September~16, 16:20:00~UTC), thus spanning 418~days. This time
interval contains a significant number of gaps which are caused by: events or operations onboard
{\gaia} that prevent the collection of data or make the raw data unusable for a while (such as the
decontamination of the payload); problems in the pre-processing leading to effective gaps in the
available raw {\gaia} data \citep[which has to be reconstructed from the raw telemetry,][]{DPACP-7};
gaps in the spacecraft attitude solution deliberately introduced around the times when
micro-meteoroid hits occurred \citep{DPACP-14}. Telemetry losses along the spacecraft to ground link
are only a very minor contribution to the data gaps. As a result of these gaps the amount of data
processed for {\gdr} comprises slightly less than 12 (out of the above mentioned 14) months. The
data gaps inevitably affect the quality of the {\gdr} results. In future releases the gaps related
to the on-ground processing will disappear.

%
%

\section{Overview of the contents of {\gdr}}
\label{sec:gdrsummary}

\begin{table}[t]
  \caption{Basic statistics on the contents of {\gdr}}
  \label{tab:gdr1stats}
  \centering
  \begin{tabular}{lr}
    \hline\hline
    \noalign{\smallskip}
    \multicolumn{2}{c}{\textbf{Source numbers}} \\
    \noalign{\smallskip}
    \hline
    \noalign{\smallskip}
    Total number of sources & \gdrtotnum \\
    No.\ of primary (TGAS) sources & \tgasnum \\
    \quad {\hip} & \hipnum \\
    \quad {\tyctwo} (excluding Hipparcos stars) & \tycnum \\
    No.\ of secondary sources & \gdrsecnum \\
    No.\ of sources with light curves & \varnum \\
    \quad Cepheids & \cepnum \\
    \quad RR Lyrae & \rrlnum \\
    \noalign{\smallskip}
    \hline
    \noalign{\smallskip}
    \multicolumn{2}{c}{\textbf{Magnitude distribution percentiles ($G$)}} \\
    \noalign{\smallskip}
    \hline
    \noalign{\smallskip}
    $0.135$\% & $11.2$ \\
    $2.275$\% & $14.5$ \\
    $15.866$\% & $17.1$ \\
    $50$\% & $19.0$ \\
    $84.134$\% & $20.1$ \\
    $97.725$\% & $20.7$ \\
    $99.865$\% & $21.0$ \\[3pt]
    \noalign{\smallskip}
    \hline
  \end{tabular}
\end{table}

{\gdr} contains astrometry, $G$-band photometry, and a modest number of variable star light curves,
for a total of {\gdrtotnum} sources. Basic statistics for {\gdr} are listed in
\tabref{tab:gdr1stats}. The three main components of {\gdr} are:
\begin{enumerate}
  \item The {\em astrometric data set} which consists of two subsets:

    The {\em primary astrometric data set} contains the positions, parallaxes, and mean proper
    motions for {\tgasnum} stars in common between the \gdr, {\hip} and {\tyctwo} catalogues. This
    data set represents the realisation of the {\tyc}-{\gaia} astrometric solution (TGAS), of which
    the principles were outlined and demonstrated in \cite{2015A&A...574A.115M}. The typical
    uncertainty is about $0.3$~mas for the positions, and about $1$~mas~yr$^{-1}$ for the proper
    motions. For the subset of {\hipnum} {\hip} stars in the primary astrometric data set the proper
    motions are much more precise, at about $0.06$~mas~yr$^{-1}$. The typical uncertainty for the
    parallaxes is $0.3$~mas, where it should be noted that a systematic component of $\sim0.3$~mas
    should be added (see \secref{sec:gdrlimitations}). 

    The {\em secondary astrometric data set} contains the positions for an additional {\gdrsecnum}
    sources. For the secondary data set the typical uncertainty on the positions is $\sim 10$~mas.

    The positions and proper motions are given in a reference frame that is aligned with the
    International Celestial Reference Frame (ICRF) to better than $0.1$~mas at epoch J2015.0, and
    non-rotating with respect to ICRF to within $0.03$~mas~yr$^{-1}$. The detailed description of
    the production of the astrometric solution, as well as a more detailed statistical summary of
    the astrometry contained in {\gdr} can be found in \cite{DPACP-14}. An in-depth discussion of
    the {\gdr} reference frame and the optical properties of ICRF sources is presented in
    \cite{DPACP-26}.
    
  \item The {\em photometric data set} contains the mean {\gaia} $G$-band magnitudes for all the
    sources contained in {\gdr}. The brightest source in {\gdr} has a magnitude $G=3.2$, while the
    majority of the sources ($99.7$\%) are in the range $11.2\leq G\leq21$. The small fraction of
    sources at $G>21$ \citep[where the nominal survey limit is $G=20.7$,][]{DPACP-1} most likely have
    erroneously determined $G$-band fluxes, but nevertheless passed the data quality filters
    described in \secref{sec:gdrvalidation}. The typical uncertainties quoted on the mean value of $G$
    range from a milli-magnitude or better at the bright end ($G\lesssim 13$), to about $0.03$ mag
    at the survey limit. The details of the photometric data set, including the data processing and
    validation of the results is described in \cite{DPACP-12,DPACP-9,DPACP-10,DPACP-11}.

  \item The {\em Cepheids and RR Lyrae data set} contains the $G$-band light curves and
    characteristics of a modest sample of {\cepnum} Cepheid ({\cepnumnew} newly discovered) and
    {\rrlnum} RR Lyrae ({\rrlnumnew} new) variables located around the south ecliptic pole and
    observed at high cadence during a special scanning period in the first four weeks of the
    operational phase of {\gaia}. The variable star contents of {\gdr} are described in detail in
    \cite{DPACP-15} and \cite{DPACP-13}.
\end{enumerate}

\begin{figure}[t]
  \resizebox{\hsize}{!}{\includegraphics{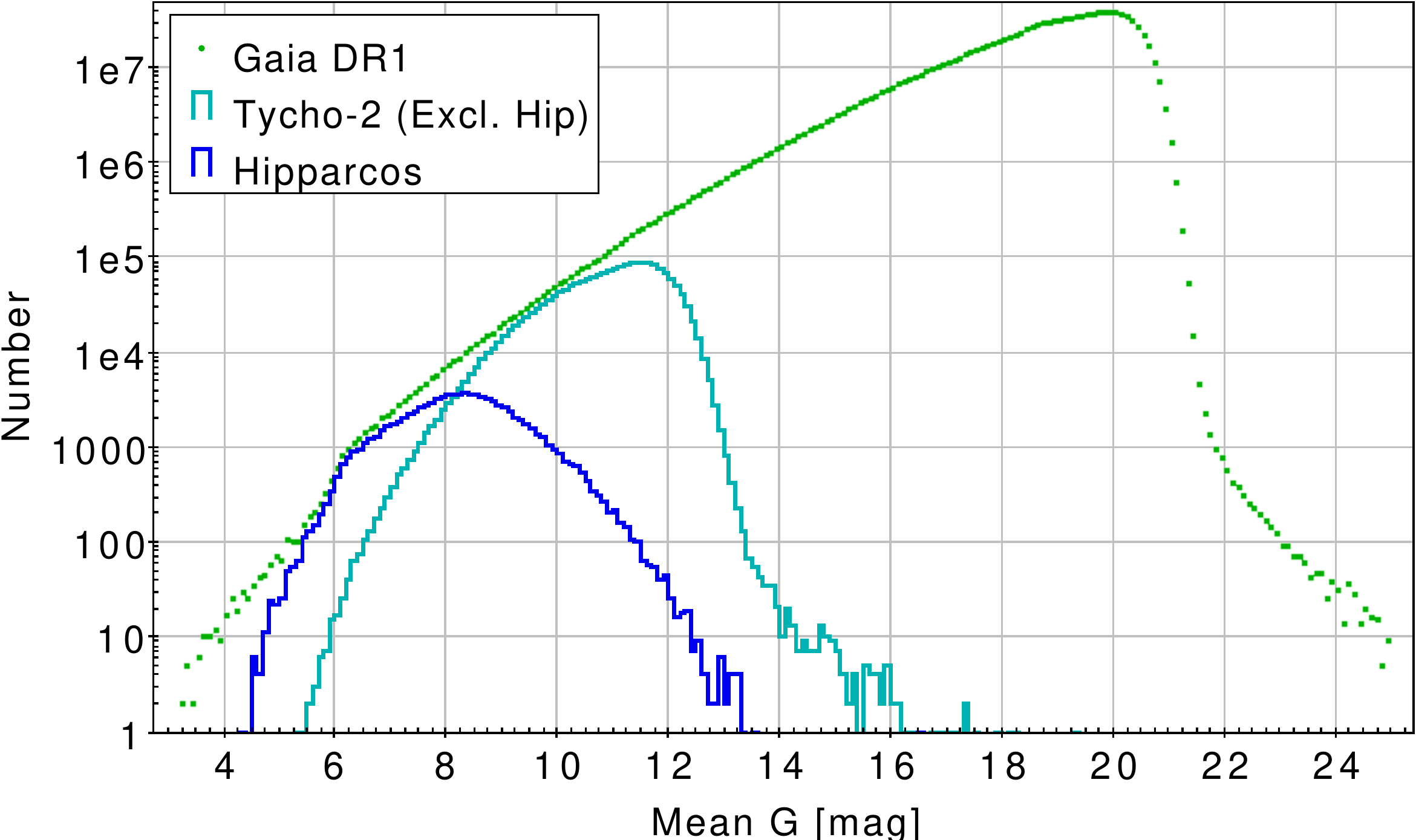}}
  \caption{Distribution of the mean values of $G$ for all {\gdr} sources shown as histograms with
  $0.1$~mag wide bins. The distributions for the {\hip} and {\tyctwo} (excluding the {\hip} stars)
  subsets are also shown. Note the lack of bright sources at $G\lesssim7$.
  \label{fig:magdistro}}
\end{figure}

The distribution of the sources in magnitude is shown in \figref{fig:magdistro}. The magnitude
distribution of the sources reveals a drop-off at $G\lesssim7$. Neither all {\hip} nor all {\tyctwo}
sources are included in {\gdr} and at the faint end the magnitude limit is sky position dependent
and ill-defined. At magnitudes below $G\sim5$ the total number of sources in {\gdr} is larger than
the number of {\hip} sources in {\gdr}. This is however only apparent as most of these sources are
in fact in common with the {\hip} catalogue but have been treated as secondary astrometric sources,
because a good 5-parameter astrometric solution could not be derived. The limitations of {\gdr},
including its completeness, are discussed in \secref{sec:gdrlimitations}.

Of the $1141$ million sources in the secondary astrometric data set 685 million are in common with
the Initial {\gaia} Source List \citep[IGSL,][]{2014AA...570A..87S} and 456 million are new sources
\citep{DPACP-14}. The IGSL formed the starting point for the process of assigning {\gaia}
observations to sources \citep{DPACP-7}. Hence the term `new' should strictly speaking be
interpreted as referring to sources that could not be matched to known IGSL sources. No attempt was
made to establish how many sources are truly new discoveries by {\gaia} but this is likely to be a
substantial fraction (over 400 million) of the new sources mentioned above. The IGSL has been
publicly available for some time and we caution that when looking up a source in {\gdr} through its
already known IGSL source identifier, it should be kept in mind that a large fraction of the $1.2$
billion sources in the IGSL does not appear in {\gdr}.

%
%

\section{{\gdr} validation and source filtering}
\label{sec:gdrvalidation}

A substantial effort was dedicated to the validation of the results contained in {\gdr}. This is a
complex task which takes place at various levels within the DPAC. The outputs produced by the DPAC
subsystems \citep[described in][]{DPACP-1} are validated first through an `internal' quality control
process. For the astrometric data set in {\gdr} this internal validation is described in
\cite{DPACP-14}, while that for the photometric and variable star data sets is described in
\citep{DPACP-11} and \cite{DPACP-15}, respectively. A second validation stage is carried out by the
DPAC unit responsible for the data publication \citep[cf.][]{DPACP-1}, which examines all the data
contained in {\gdr} together and thus provides an independent quality check. This global validation
process is described in \cite{DPACP-16}. Here we summarise only the most important findings
from the validation and provide complementary illustrations of the quality of {\gdr} in
\secref{sec:sciencedemos}.

Numerous tests were done during the validation stage of the {\gdr} production, ranging from basic
consistency checks on the data values to the verification that the data is scientifically correct.
No problems were revealed that would prevent the timely publication of {\gdr}. However, a number of
minor problems were found that have been addressed either by a filtering of the available DPAC
outputs before their incorporation into the data release, or by documenting the issues found as
known limitations to {\gdr} (see \secref{sec:gdrlimitations}). The filtering applied to the
astrometric and photometric processing outputs before the global validation stage was as follows:
\begin{itemize}
  \item For the primary astrometric data set only sources for which the standard uncertainties on the
    parallaxes and positions are less than 1~mas and 20~mas, respectively, were kept. In addition it
    was required that the sources have valid photometric data. For the secondary astrometric data
    set the sources were filtered by requiring that they were observed by {\gaia} at least 5 times
    (i.e.\ at least 5 focal plane transits), and that their astrometric excess noise (which
    indicates the astrometric modelling errors for a specific source) and position standard
    uncertainty are less than 20~mas and 100~mas, respectively. More details can be found in
    \cite{DPACP-14}. We stress that no filtering was done on the actual value of the source
    astrometric parameters.
  \item Although the photometric results were not explicitly filtered before their incorporation
    into {\gdr}, a number of filters internal to the photometric data processing effectively leads
    to filtering at the source level. In particular sources with extremely blue or red colours will
    not appear in {\gdr}.
  \item The only filtering done on the outputs of the variable star processing was to remove a
    handful of sources that were very likely a duplicate of some other source (see below for more
    discussion on duplicate sources).
\end{itemize}

\begin{figure*}[t]
  \includegraphics[width=\textwidth]{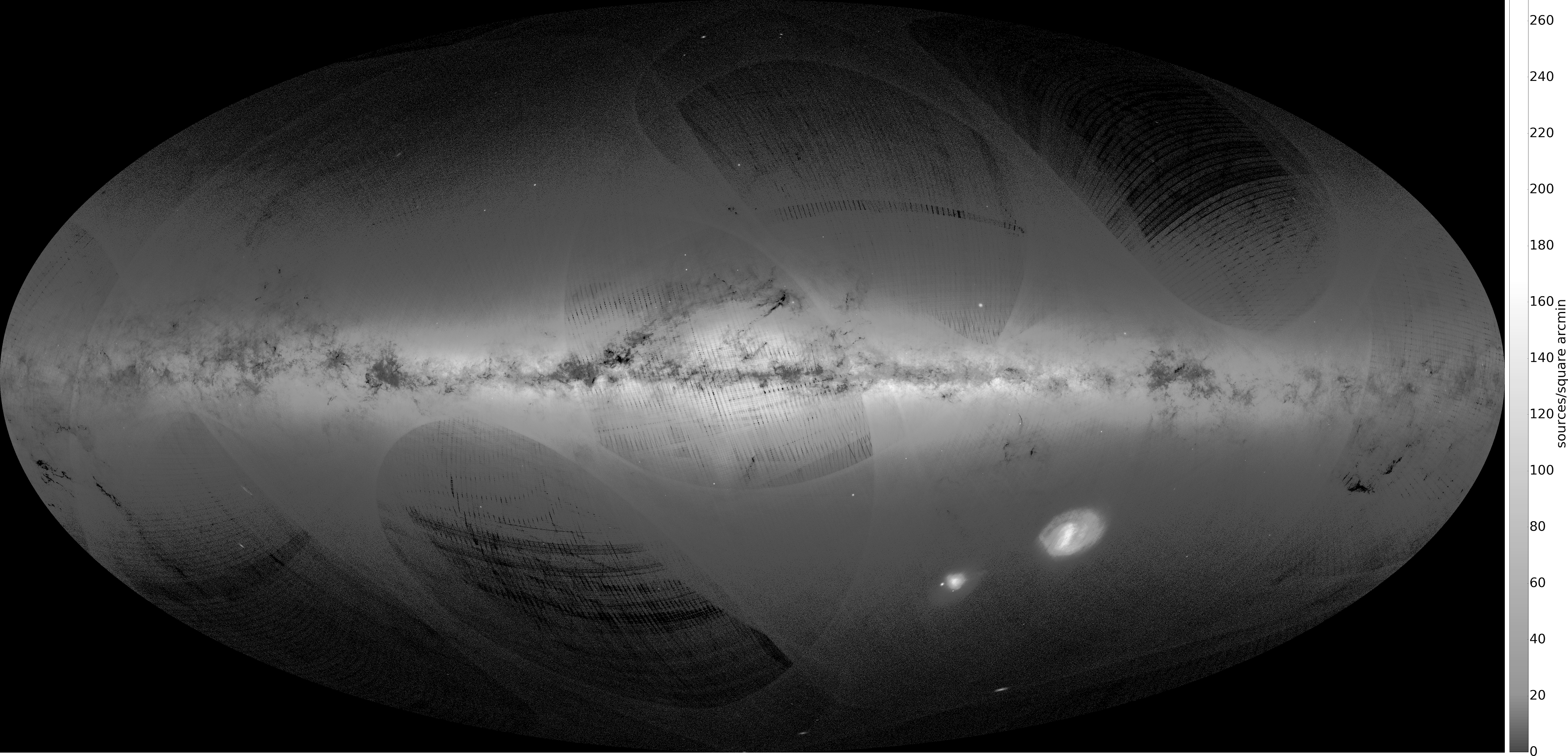}
  \caption{Sky distribution of all {\gdr} sources in Galactic coordinates. The source density is
  shown with a grey scale chosen to highlight both the impressive amount of detail in the outlines
  of the well-known dust features along the Galactic plane, and the non-astronomical artefacts in
  the source distribution (see text). Image credits: CENTRA - University of Lisbon (part of the
  DPAC-CU9 visualisation team).
  \label{fig:skymap}}
\end{figure*}

The second validation stage \citep{DPACP-16} revealed the following problems that were addressed
through a further filtering of the astrometric and photometric processing outputs before their final
incorporation into {\gdr}. The filters described below were thus applied after the filters above.
\begin{itemize}
  \item Some 37 million source pairs were found which are separated by less than 1 {\gaia} focal
    plane pixel size on the sky (i.e.\ 59~mas), or are separated by less than 5 times their combined
    positional standard uncertainty (where the factor 5 accounts for a possible underestimation of
    the standard uncertainties). The vast majority of these pairs are created during the cross-match
    stage, when observations (focal plane transits) get grouped together and assigned to sources
    \citep[see][]{DPACP-7}. The main underlying cause is sources appearing twice in the IGSL, which
    was evident from the many close pairs occurring along photographic survey plate boundaries
    \citep[the IGSL is based to a large extent on photographic surveys,][]{2014AA...570A..87S}. A
    large fraction of these pairs are likely to be two instances of the same physical source (i.e.\
    the source appears twice in the {\gaia} source list with two different identifiers). One member
    of each of these close pairs was filtered out of the {\gdr} source list and the remaining
    sources were flagged as having a duplicate associated to them in the {\gaia} source list. This
    flag thus indicates that the source in question has fewer observations contributing to its
    astrometry and photometry because part of the observations were assigned to another (fictitious)
    source. This filtering will in a fraction of the cases inevitably have removed one component
    from a real double source (be it a binary or an optical pair). This problem of duplicate sources
    will disappear in future {\gaia} data releases due to improvements in the cross-match algorithm
    and the moving away from the Initial {\gaia} Source List as the basis for assigning observations
    to sources.
  \item For some 1 million sources the mean $G$ values were grossly inconsistent with either
    existing photometry (for example some TGAS stars were assigned $G$-band magnitudes much fainter
    than the {\tyctwo} survey limit) or with the broad-band $G_\mathrm{BP}$ and $G_\mathrm{RP}$
    magnitudes determined from the {\gaia} Blue and Red Photometers. In either case data processing
    problems are indicated and sources were removed from {\gdr} when there were fewer than 11
    measurements in the $G$ band (i.e.\ CCD transits in the astrometric part of the focal plane),
    or if both $(G-G_\mathrm{BP})$ and $(G-G_\mathrm{RP})$ were larger than $+3$.
\end{itemize}

Although the filtering described above will have removed the vast majority of problem cases from the
DPAC outputs before the publication of {\gdr}, it will nevertheless not be perfect. Genuine sources
will have been removed and the filtering criteria do not guarantee the absence of a small fraction
of problematic sources in {\gdr}.

The decision to filter out the problematic cases rather than publish them with, e.g.\ indicator
flags, was driven by data quality considerations and by the need to remove the large number of
spurious sources created in the process of matching observations to sources \citep[see][]{DPACP-7,
DPACP-14}. The filtering thus reflects the preliminary nature of the first {\gaia} data release. In
future intermediate releases the shortcomings in the data processing will be addressed and more
measurements will be added, which means that reliable results can be derived for more sources. The
level of filtering is thus expected to go down and more sources will enter the published catalogue.

%
%

\begin{figure*}[t]
  \includegraphics[width=\textwidth]{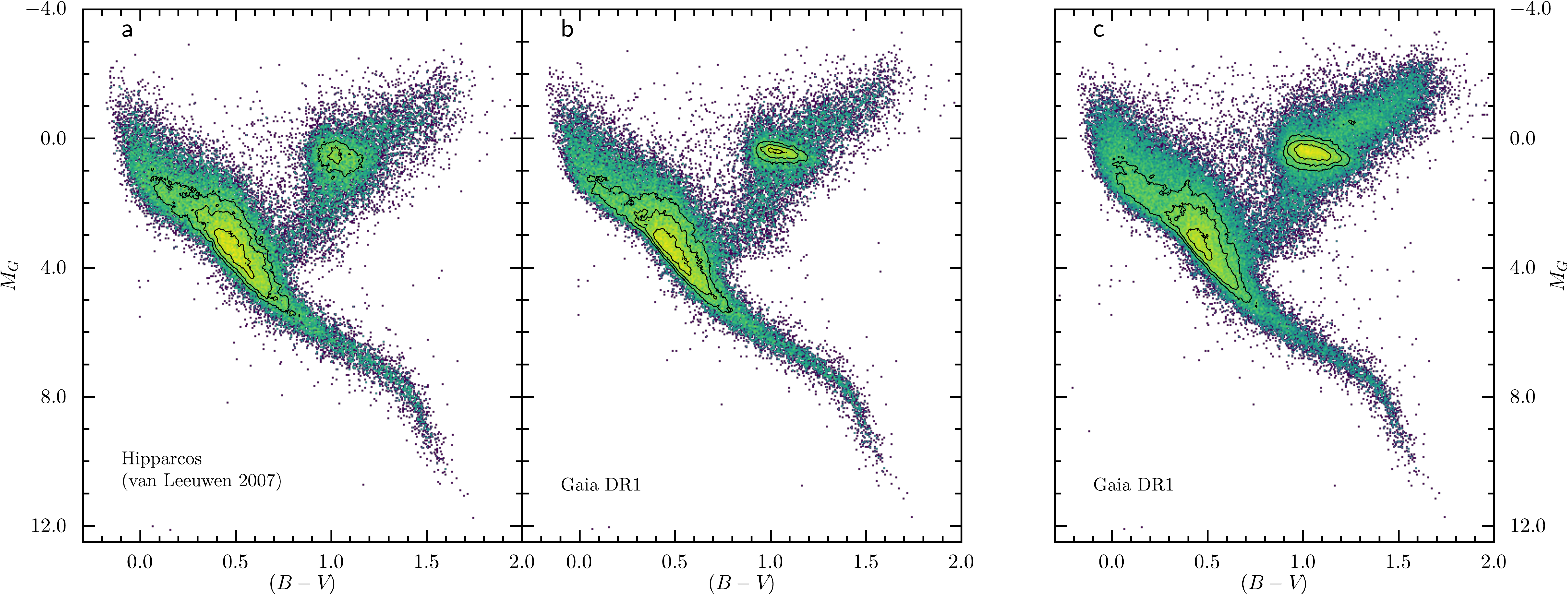}
  \caption{Comparison of the observational HR-diagram in the $M_G$ vs.~$(B-V)$ plane for the {\hip}
  stars in {\gdr}, using their {\hip} \citep{book:newhip} parallaxes (\textbf{a}) and their
  parallaxes as listed in {\gdr} (\textbf{b}, \textbf{c}). The relative standard uncertainties on
  the parallax are less than 20\% for both the {\hip} and {\gdr} parallaxes in panels \textbf{a} and
  \textbf{b}, while in panel \textbf{c} all stars with relative parallax uncertainties better than
  20\% in {\gdr} are shown. The stars were otherwise selected as described in the text. All panels
  show the stars as individual symbols where possible and where the symbols overlap the relative
  source density is shown, with colours varying from purple (dark) to yellow (light) indicating
  increasing density on a logarithmic scale. The contours enclose 10, 30, and 50 per cent of the
  data.}
  \label{fig:hrdcomp}
\end{figure*}

\section{Illustrations of the {\gdr} contents}
\label{sec:sciencedemos}

Here we provide a few illustrations of the contents of {\gdr}. The purpose is not to provide a
scientific analysis but to demonstrate through astronomically relevant examples the overall quality
of the {\gaia} data. A more detailed examination of the scientific quality of {\gdr} is provided in
two studies on open clusters \citep{DPACP-23} and the Cepheid period-luminosity relation
\citep{DPACP-24}. We end this section with a comment on the Pleiades cluster distance.

\subsection{The {\gaia} sky}
\label{sec:gaiasky}

The distribution of all {\gdr} sources on the sky is illustrated in \figref{fig:skymap}. The source
density shown in \figref{fig:skymap} is based on the accurate positions of the $1.1$ billion sources
in {\gdr} and represents the most detailed all-sky map in the optical to date. This can be
appreciated in particular in the very fine outlining of the dust features along the Galactic plane.
Also noteworthy are the Magellanic clouds, where in the Large Magellanic Cloud the individual
features in the star forming regions north of the bar are outlined in the source distribution; the
M31 and M33 galaxies which are both outlined in individual detections made by \gaia; and the Orion A
and B clouds which can be seen against the backdrop of the sources detected by \gaia. Also
recognisable are globular clusters, such as $\omega$ Centauri with over two hundred thousand sources
individually appearing in {\gdr}, and the Fornax dwarf spheroidal galaxy ($\sim30\,000$ sources in
{\gdr}) near $(\ell,b)\approx(237^\circ,-66^\circ)$. The full detail of this sky map is impossible
to convey in print. An interactive and zoomable version will be available, through the Aladin sky
atlas application \citep{2000A&AS..143...33B,2014ASPC..485..277B} and a dedicated visualisation
service, both as part of the {\gaia} data access facilities (see \secref{sec:access}). The sky map
also reveals a number of prominent non-astronomical artefacts which reflect the preliminary nature
of the first {\gaia} data release. They are further discussed in \secref{sec:gdrlimitations}.

The depth of the {\gaia} survey, its all-sky reach, the high angular resolution, combined with the
highly accurate source positions, promises a revival of classical star count studies, in particular
with future {\gaia} data releases where the shortcomings in the completeness and angular resolution
of {\gdr} (see \secref{sec:gdrlimitations}) will have been addressed. The {\gaia} sky map is also of
immediate interest to studies of minor solar system bodies through stellar occultations, the
predictions of occultation tracks on the Earth benefiting from the dense distribution of sources
with accurately known positions.

Finally, the {\gaia} sky map will be the standard reference in the optical for some time to come, in
particular when in future releases the {\gaia} catalogue will be more complete in sky, magnitude,
and colour coverage, and the source positions are further refined, with parallaxes and proper
motions becoming available for all {\gaia} sources. This is to the benefit of all (optical)
telescope guidance applications, especially large-mirror telescopes with small fields of view.
Space missions will also benefit from the {\gaia} sky map. As an example, it is planned to improve
the recently released Hubble Source Catalog \citep{2016AJ....151..134W} through a re-reduction of
the astrometry with respect to the {\gaia} source positions.

\subsection{Hertzsprung-Russell diagrams based on {\gdr} parallaxes}
\label{sec:hrd}

With the advent of {\gdr} we now for the first time have access to two large samples of parallaxes
accurate at the (sub-)milliarcsecond level. As explained in \cite{DPACP-14} the {\gaia} and {\hip}
parallaxes are independent and can thus sensibly be compared to each other. The comparison described
in the appendix of \cite{DPACP-14} shows that overall the {\gdr} and {\hip} parallaxes are the same
to within the combined uncertainties. A closer look at the parallaxes near zero reveals that for the
{\hip} stars in {\gdr} the number of negative parallaxes is much smaller, which is expected for a
data set that is more precise. This comparison is furthermore exploited in \cite{DPACP-14} to derive
the relation between the formal and actual (published) uncertainties for the astrometric source
parameters in the primary astrometric data set of {\gdr}.

We illustrate the better overall precision of the {\gaia} parallaxes by constructing observational
Hertzsprung-Russell diagrams in $M_G$ vs.~$(B-V)$ using the {\hip} parallaxes from
\cite{book:newhip} and the parallaxes from {\gdr}. The result is shown in \figref{fig:hrdcomp}. The
$43\,546$ {\hip} stars included in the left two panels (\textbf{a} and \textbf{b})
were selected according to:
\begin{equation}
  \begin{split}
    (\varpi/\sigma_\varpi)_{\text{\gaia}} \geq 5 \quad \wedge \quad
    (\varpi/\sigma_\varpi)_{\text{\hip}} \geq 5 \quad \wedge \quad \\
    \sigma_G \leq 0.05 \quad \wedge \quad
    \sigma_{(B-V)} \leq 0.05\,,
  \end{split}
  \label{eq:hrdselect}
\end{equation}
where $\varpi$ is the parallax and $\sigma_\varpi$ the corresponding standard uncertainty. The
values of $(B-V)$ and their standard uncertainties were taken from the {\hip} Catalogue
\citep[][]{book:newhip}. The $74\,771$ stars in the rightmost panel (\textbf{c}) were selected only
on the value of the relative uncertainty in the {\gdr} parallax but with the same criteria on the
uncertainty in $G$ and $(B-V)$. The median {\gdr} parallax for the smaller sample is $7.5$ and for the
larger sample it is $5.0$~mas, while 90 per cent of the stars have a parallax larger than $3.6$
(smaller sample) and $2.2$~mas (larger sample). A comparison of the left two panels shows that with
the {\gdr} parallaxes the main sequence is better defined, being somewhat narrower and with a
sharper boundary along the faint end. The distribution of red clump giants is much narrower in
luminosity, with the effect of extinction and reddening clearly seen as an elongation in the
direction of fainter magnitudes and redder colours. 

\begin{figure}
  \resizebox{\hsize}{!}{\includegraphics{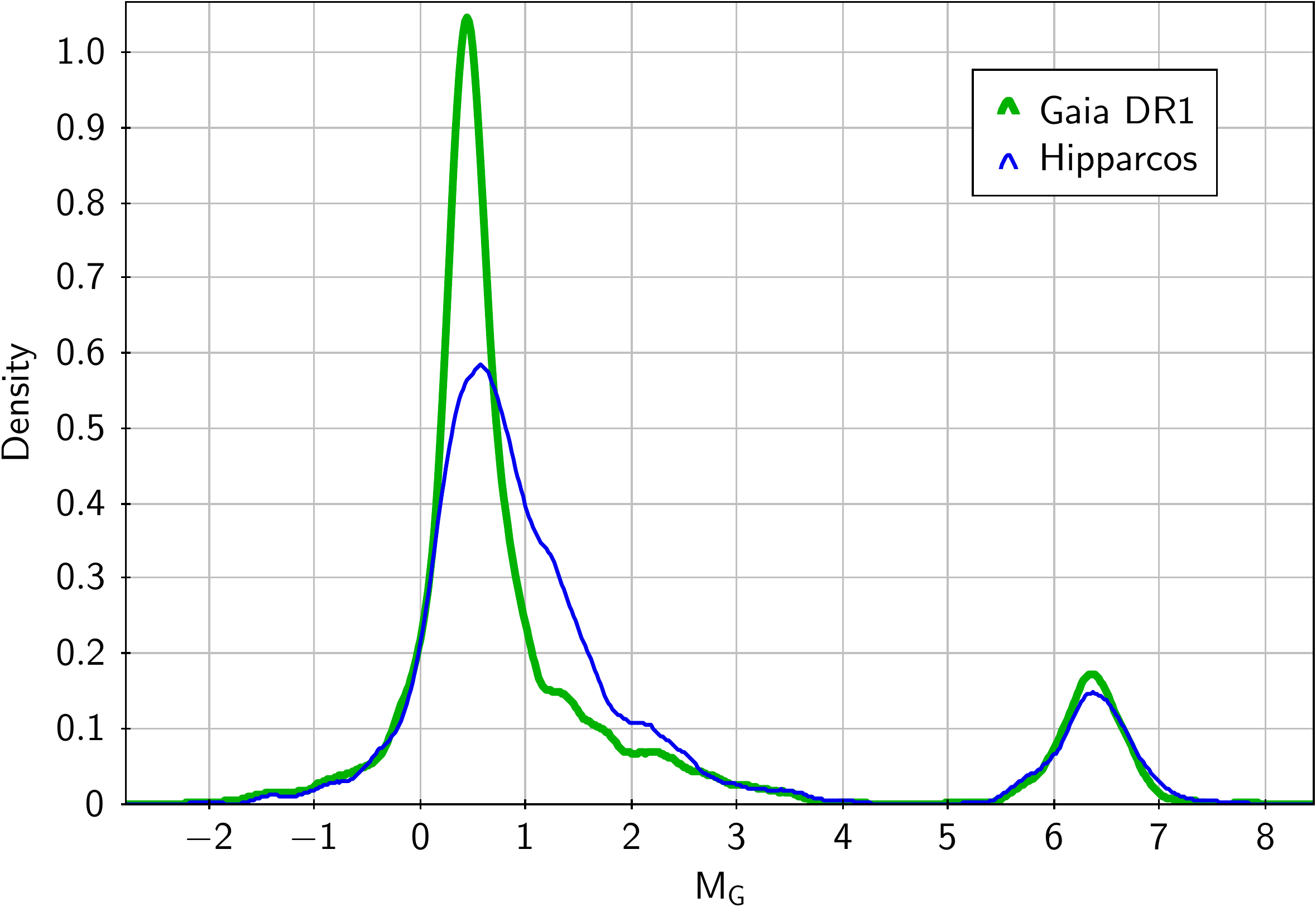}}
  \caption{Distribution of absolute magnitudes $M_G$ for the stars from \figref{fig:hrdcomp} (panels
  a and b) within the colour range $1.0\leq(B-V)\leq1.1$. The thick green solid line shows the
  distribution of $M_G$ derived from {\gdr} parallaxes, while the thin blue line shows the same for
  the {\hip} parallaxes. The distributions are represented as kernel density estimates, using an
  Epanechnikov kernel \citep{doi:10.1137/1114019} with a band-width of $0.2$~mag.}
  \label{fig:mgdistcolourslice}
\end{figure}

The narrower luminosity distribution of the red clump giants and main sequence dwarfs in {\gdr} is
further illustrated in \figref{fig:mgdistcolourslice}. The luminosity distribution is shown for the
stars in the left two panels of \figref{fig:hrdcomp} that have colours in the range
$1.0\leq(B-V)\leq1.1$ ($3174$ stars), including both the clump stars around $M_G\sim0.5$ and the
main sequence dwarfs around $M_G\sim6$, as well as a fraction of sub-giants (at $1\lesssim
M_G\lesssim3$). The luminosity distributions for both the dwarfs and the clump giants are narrower
for {\gdr} than for {\hip}. For the dwarfs (defined as stars with $M_G>4.5$) the robust scatter
estimates for the width of the distribution of $M_G$ \citep[see][for the definition of this
quantity]{DPACP-14} are $0.32$ for {\gdr} and $0.38$ for {\hip}. For the clump giants the numbers
are sensitive to the range of $M_G$ used to isolate the clump and whether that range is defined
using the {\gdr} or {\hip} luminosities. Using the broad selection $-0.5\leq M_G(\hip)\leq1.5$ the
robust scatter estimates are $0.37$ for {\gdr} and $0.46$ for {\hip}. When isolating the clump
giants using {\gdr} luminosities ($-0.2\leq M_G(\text{\gdr})\leq1.2$) the robust scatter estimates
are $0.30$ for {\gdr} and $0.49$ for {\hip}. The detailed interpretation of the scatter in $M_G$ for
the red clump giants and how this relates to the parallax quality of {\gdr} and {\hip} is
complicated by the {\hip} survey selection function, the filtering applied for {\gdr}, the parallax
systematics present in {\gdr} (see \secref{sec:gdrlimitations}), and the biases introduced by the
{\hip} magnitude limit and the selection on relative parallax error. These effects lead to an
incomplete and non-representative sample of red clump giants. A proper interpretation of the scatter
in the luminosities (and of the mean observed luminosity) requires the modelling of the population
of red clump giants and of the {\gaia} and {\hip} survey properties, which we consider beyond the
scope of the illustrations provided in this section.

The rightmost panel in \figref{fig:hrdcomp} shows how in {\gdr} a larger volume is covered by
relatively precise parallaxes; the overall width in colour of the upper main sequence and red clump
is larger due to the larger extinction values probed, and the upper main sequence and giant branch
are better populated. In numbers the median relative uncertainty on the {\hip} parallax for the
stars selected according to \equref{eq:hrdselect} is $0.1$, while for the {\gdr} parallaxes it is
$0.04$.

\begin{figure}
  \resizebox{\hsize}{!}{\includegraphics{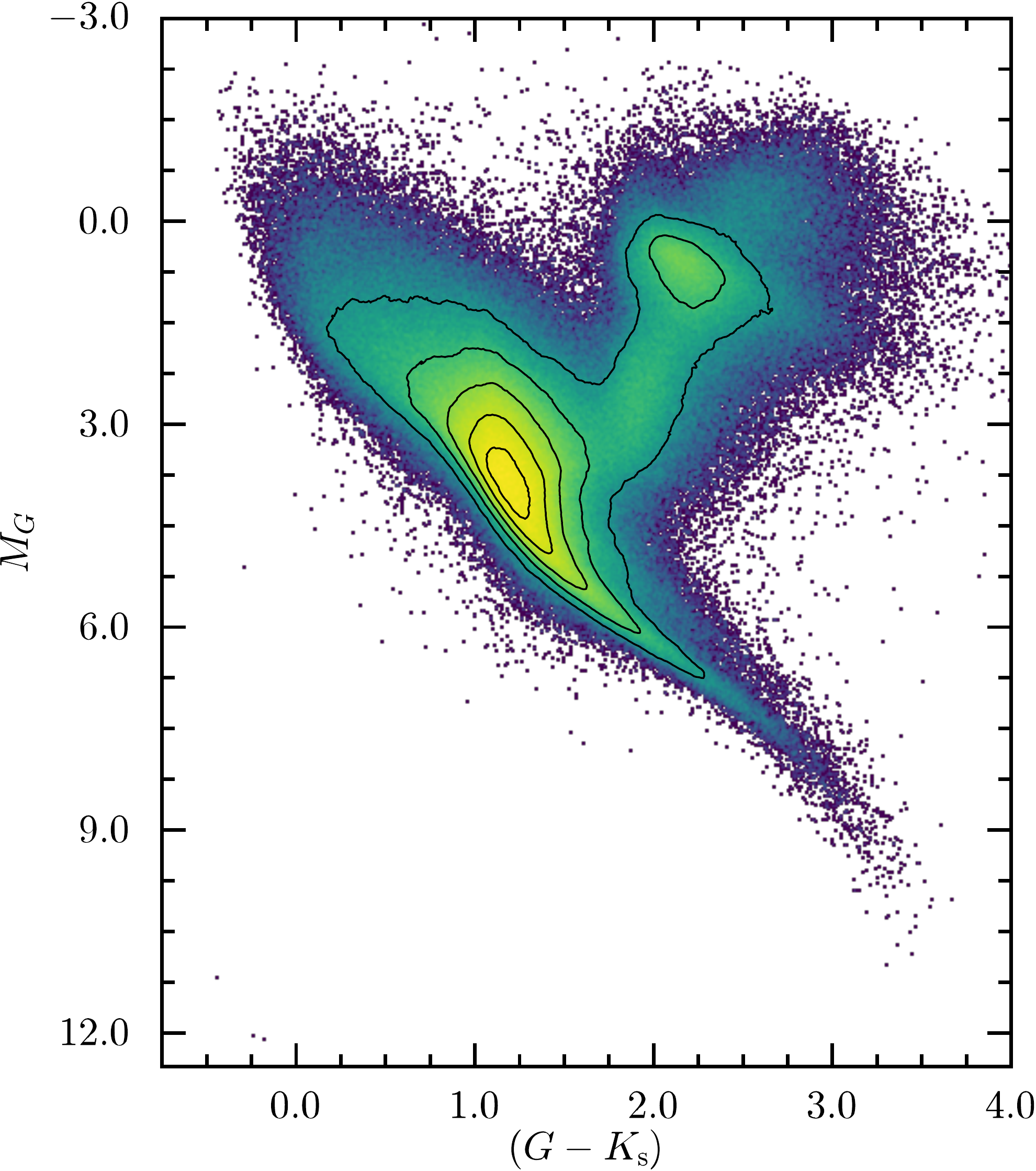}}
  \caption{Observational HR-diagram for all stars in {\gdr} selected as explained in the text for
  which the $(G-K_\mathrm{s})$ colour index can be calculated from {\gdr} and the data in the 2MASS
  Point Source Catalogue. The visualisation is the same as in \figref{fig:hrdcomp} with the contours
  enclosing 10\%, 30\%, 50\%, 70\%, and 90\% of the data.}
  \label{fig:hrdlarge}
\end{figure}

In \figref{fig:hrdlarge} we show the observational HR-diagram for a much larger sample of stars from
{\gdr} for which the $(G-K_\mathrm{s})$ colour index can be calculated from {\gdr} and the data in
the 2MASS Point Source Catalogue \citep{2006AJ....131.1163S}. The selection of the sources in this
diagram is according to \equref{eq:hrdselect}, where the limit on the {\hip} relative parallax error
does not apply and the limit on the standard uncertainty in the colour index now applies to
$(G-K_\mathrm{s})$. In addition the 2MASS photometric quality flag was required to be equal to `A'
for each of the $J$, $H$, and $K_\mathrm{s}$ magnitudes. The resulting sample contains $1\,004\,204$
stars (there are $1\,037\,080$ stars with $\varpi/\sigma_\varpi\geq5$ in total in {\gdr}). The
sample covers a substantially larger volume, the median parallax being $2.9$~mas, while 90 per cent
of the stars have a parallax larger than $1.7$~mas. The larger volume covered is evident from the
large number of luminous stars in the HR-diagram: $42\,333$ stars at $M_G<2$ in the rightmost panel
of \figref{fig:hrdcomp}, compared to $190\,764$ in \figref{fig:hrdlarge}. In addition the effect of
extinction is now more prominently visible as a broadened upper main sequence and turn-off region,
as well as in the elongation of the red clump. A hint of the binary sequence in parallel to the main
sequence can be seen around $(G-K_\mathrm{s})\sim2.2$ and $M_G\sim6$. Note the three white dwarfs at
$(G-K_\mathrm{s})<0$ and $M_G>11$; from the brightest to the faintest their 2MASS designations and
{\gaia} source identifiers are 2MASS J21185627+5412413, 2MASS J16482562+5903228, 2MASS
J19203492-0739597, and 2176116580055936512, 1431783457574556672, 4201781727042370176, respectively.
This diagram is also an illustration of the use of pre-computed cross-match tables, linking {\gdr}
and other large surveys, which are provided along with the data release \citep[see
\secref{sec:access} and][]{DPACP-17}.

\begin{figure}
  \resizebox{\hsize}{!}{\includegraphics{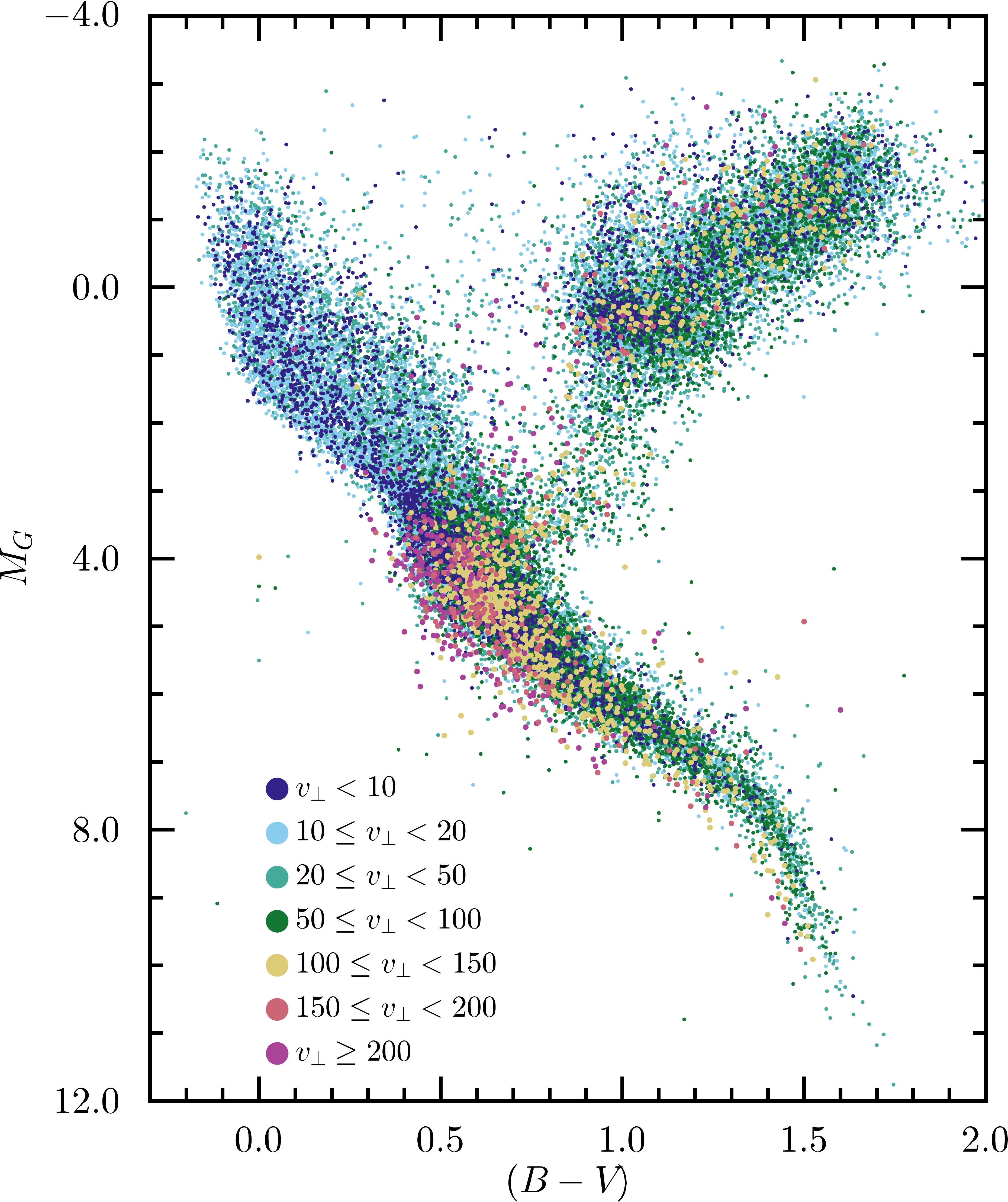}}
  \caption{Observational HR-diagram showing where stars with specific values of the transverse
  velocity $v_\perp$ tend to occur. The colour coding of the points is according to tangential
  velocity interval, as indicated in the legend (in km~s$^{-1}$).}
  \label{fig:hrdvperp}
\end{figure}

An HR-diagram can also be produced with the $(B-V)$ colour index. However, this requires the use of
different sources for the colour index values. When we combined {\hip}
\citep{1997ESASP1200.....E,book:newhip}, {\tyctwo} \citep{2000A&A...355L..27H}, and APASS
\citep{2014CoSka..43..518H} a diagram containing only a third as many stars resulted. This reflects
the lack of high quality all-sky optical photometry over the brightness range in between that
covered by {\hip} and modern digital sky surveys, such as the Sloan Digital Sky Survey
\citep{2000AJ....120.1579Y}, which usually only cover apparent magnitudes fainter than $\sim15$.
This situation will be remedied with the second {\gaia} data release through the publication of the
$G_\mathrm{BP}$ and $G_\mathrm{RP}$ magnitudes obtained from the integrated fluxes measured with the
Blue and Red Photometers.

Finally, following \cite{2004astro.ph..3506G}, in \figref{fig:hrdvperp} we show a version of the
HR-diagram which is colour coded according to the transverse velocity of the stars $v_\perp =
\mu/\varpi\times4.74\ldots$ (in km~s$^{-1}$), where $\mu$ is the length of the proper motion vector
of the star. The stars in this diagram were selected according to the criteria in
\equref{eq:hrdselect} (without selecting on the {\hip} relative parallax error), using the $(B-V)$
colour index as listed in the {\hip}, {\tyctwo}, or APASS catalogues (in that order of preference),
where the {\tyctwo} colours were transformed to approximate Johnson colours according to:
$(B-V)_\mathrm{J} \approx 0.85(B-V)_\mathrm{T}$ \citep[][Sect.~1.3, Vol.~1]{1997ESASP1200.....E}. It
was further demanded that $G\leq 7.5$, or $\mu\geq200$~mas~yr$^{-1}$, or $\varpi\geq10$~mas. The
$41\,136$ stars in this diagram are represented by symbols which are colour coded by tangential
velocity interval as indicated in the figure legend. This nicely illustrates the well-known mix of
stellar populations in a local sample (the median parallax for this sample being $10.7$~mas, while
90 per cent of the stars have a parallax larger than $2.8$~mas). At low velocities the young disk
stars along the main sequence are outlined ($v_\perp<50\kms$). The turn-off region for the old disk
is visible at $50\kms\leq v_\perp<100\kms$, while at higher velocities halo stars are visible, which
along the main sequence are clearly shifted to the lower metallicity region.

\begin{figure}[t]
  \resizebox{\hsize}{!}{\includegraphics{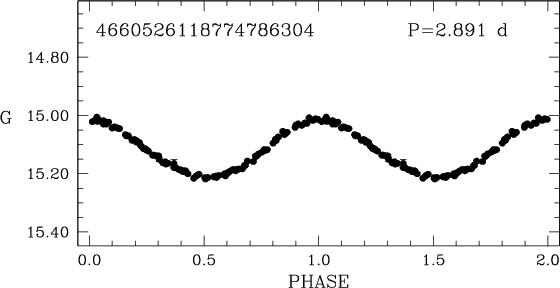}}\\[5pt]
  \resizebox{\hsize}{!}{\includegraphics{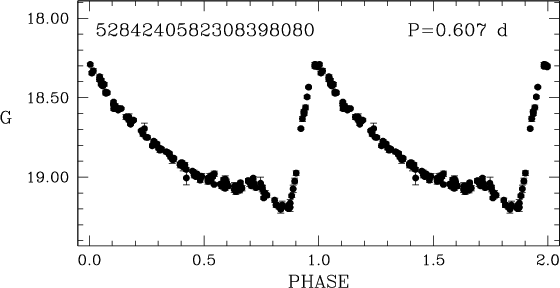}}
  \caption{Example light curves from the Cepheids and RR Lyrae data set in {\gdr}. The top panel
  shows the light curve for a fundamental mode classical Cepheid in the Large Magellanic Cloud
  (period $2.891$~days), while the bottom panel shows the light curve for a fundamental mode RR
  Lyrae star (RRab, period $0.607$~days), also in the Large Magellanic Cloud.}
  \label{fig:variables}
\end{figure}

\subsection{{\gdr} proper motions}

Given the different time spans that underlie the determinations of proper motions listed in the
{\hip} ($\Delta\text{epoch}\sim3.5$~yr), {\gdr} ($\Delta\text{epoch}\sim24$~yr), and {\tyctwo}
($\Delta\text{epoch}\sim90$~yr) catalogues, it is interesting to look for sources with discrepancies
between the proper motions listed in the three catalogues. The proper motion differences may point
to the presence of non-modelled astrometric components (such as orbital motion in a binary), and
thus to sources worthy of further investigation.

If this is attempted, very large discrepancies between {\gdr} and {\tyctwo} proper motions may occur
(of order $100$--$250$~mas~yr$^{-1}$), which seems surprising at first sight. We performed a close
inspection of 39 such cases and examined proper motion solutions for these sources for which the
{\tyctwo} position was not used (these solutions are not published in {\gdr}). In all cases there is
close agreement (to within a few mas~yr$^{-1}$ in both coordinates) between the {\gaia}-only proper
motion and the proper motion listed in {\gdr}. The fact that {\gaia} measures the same proper
motions over a 14 month time span as over the 24~yr time span used for the primary astrometric data
set implies that the large discrepancies mentioned above are due to errors in the {\tyctwo} proper
motions. These errors are most likely caused by mismatches of the {\tyc} sources to old photographic
catalogues, as was confirmed by inspecting the surroundings of a few sources among the 39 mentioned
above.

The above example points to the high quality of the {\gdr} proper motions and serves as a warning
not to over-interpret discrepancies between {\gdr} proper motions and those in existing proper motion
catalogues.

\subsection{Photometry of variable stars}

\figrefalt{fig:variables} shows two examples of phase-folded light curves from the Cepheids and RR
Lyrae data set in {\gdr}, one of a Cepheid and one of an RR Lyrae variable. Both curves highlight
the quality of the $G$-band photometry in {\gdr}. In the case of the Cepheid variable the error bars
are comparable to or smaller than the symbol size, while for the RR Lyrae variable the uncertainties
on the individual measurements are $\sim0.02$~mag. More light curves and an extensive description of
the Cepheids and RR Lyrae variables in {\gdr} are presented in \cite{DPACP-13}. The high cadence at
which these stars were observed is not representative for the nominal {\gaia} mission, but reflects
the special Ecliptic Pole Scanning Law used during the first weeks of the mission \citep{DPACP-1}.

\subsection{Comment on the Pleiades cluster mean parallax}
\label{sec:pleaides}

\begin{figure}
  \resizebox{\hsize}{!}{\includegraphics{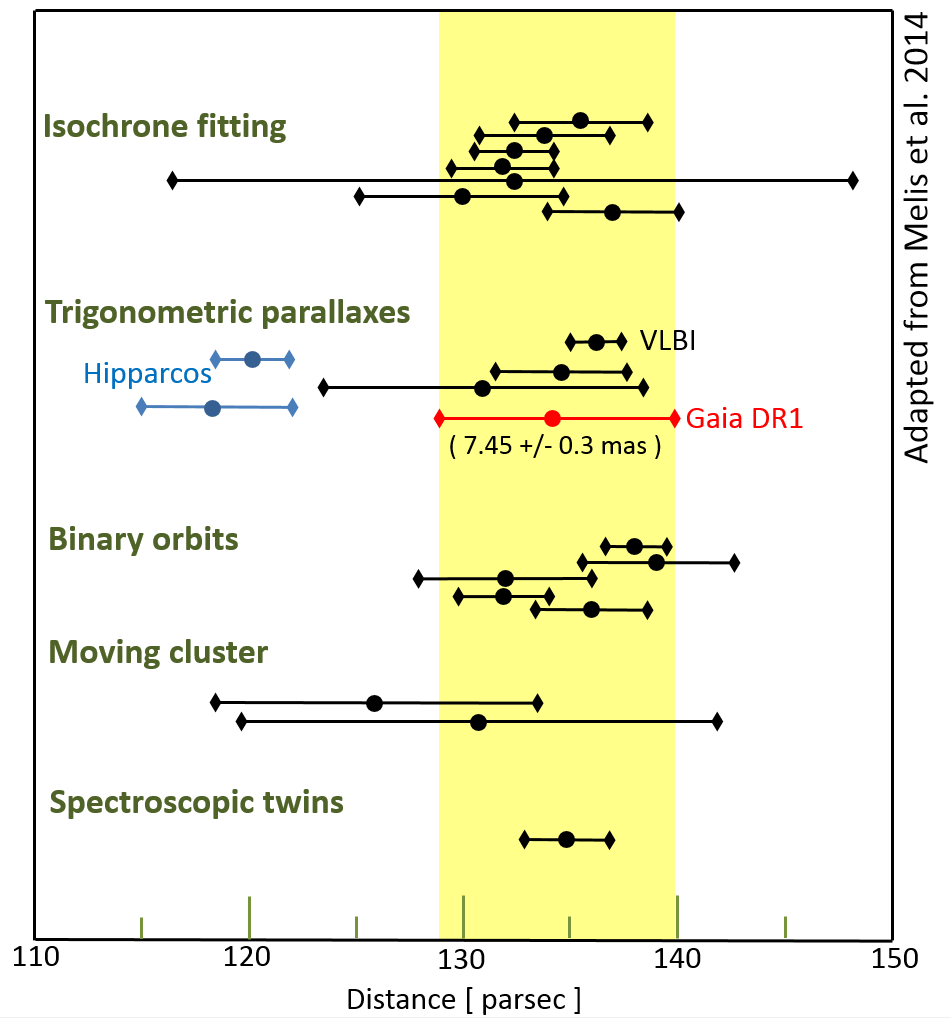}}
  \caption{Existing measurements of the parallax or distance modulus for the Pleiades cluster or
  individual cluster members, all expressed in parsecs. Figure adapted from
  \cite{2014Sci...345.1029M}. The point indicated with `VLBI' is the distance corresponding to the
  parallax determined by \cite{2014Sci...345.1029M}, while the point indicated with `Spectroscopic
  twins' is the distance corresponding to the parallax determined by \cite{2016arXiv160603015M}. The
  references for the rest of the points can be found in \cite{2014Sci...345.1029M}.}
  \label{fig:pleiades1}
\end{figure}

Since the publication of Hipparcos-derived trigonometric cluster parallaxes for the Pleiades
\citep{1999A&A...341L..71V, 2009A&A...497..209V} there has been a discrepancy between the Hipparcos
values and a number of other distance determinations derived with various methods.
\figrefalt{fig:pleiades1} displays the set of existing measurements of either the parallax or the
distance modulus of the cluster or of individual cluster members, all expressed as distances in
parsecs. The {\gdr} adds another item to this set. It is indicated in \figref{fig:pleiades1} by
the yellow shaded area.

\begin{figure}
  \resizebox{\hsize}{!}{\includegraphics{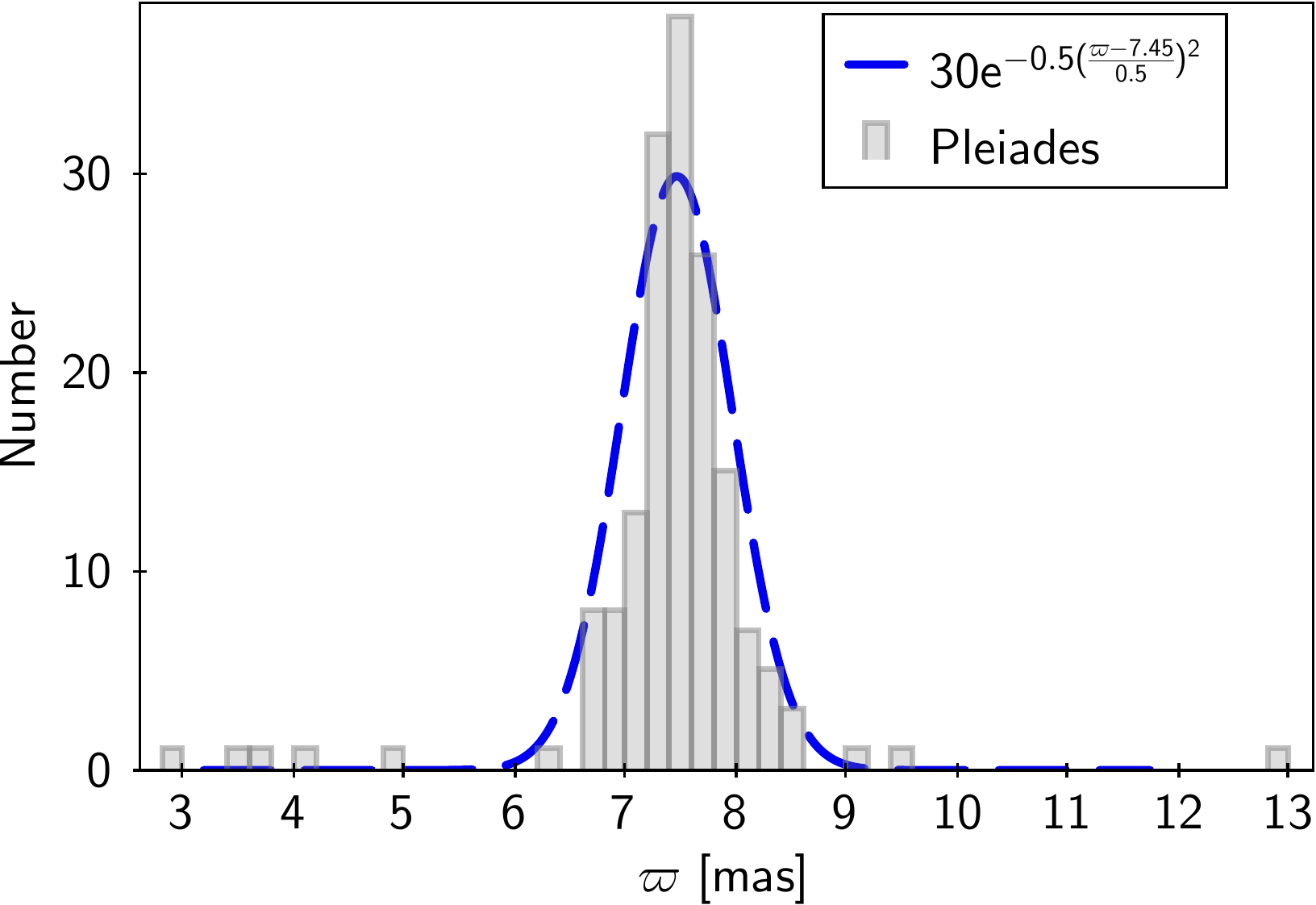}}
  \caption{Histogram of all {\gdr} parallaxes of proper motion selected Pleiades cluster members
  (using the proper motions of {\gdr} as the sole selection criterion). The over-plotted Gaussian
  distribution has a mean of $7.45$~mas, a standard deviation of $0.5$~mas and is normalised to a
  maximum value of 30 for comparison purposes.}
  \label{fig:pleiades2}
\end{figure}

A simplistic selection of Pleiades members can be done solely on the basis of the {\gdr} positions
and proper motions by demanding that the selected stars lie within 5 degrees from the
position $(\alpha,\delta)=(56.75^\circ,24.12^\circ)$ and that the proper motions obey:
\begin{equation}
  \left[ (\mu_{\alpha*}-20.5)^2+(\mu_\delta+45.5)^2 \right]^{1/2} \leq 6\text{ mas yr}^{-1}\,.
  \label{eq:pmselectionpleiades}
\end{equation}
This leads to the selection of 164 stars from the {\gdr} primary astrometric data set.
\figrefalt{fig:pleiades2} shows the histogram of the parallaxes of these 164 stars, which apart from
a few outliers (field stars not belonging to the Pleiades) are well clustered in a peaked
distribution. The median of this distribution is at $\varpi=7.45$~mas, and the standard
deviation (robustly estimated) of the distribution is $0.49$~mas. If the observations were
independent, this would lead to a standard uncertainty in the mean of $0.49/\sqrt{N}=0.04$~mas.
However, as described in the paper on the astrometric solution for {\gdr} \citep{DPACP-14} and in
the paper on the validation of {\gdr} \citep{DPACP-16}, a not precisely known systematic uncertainty
of the order of $0.3$~mas must be added to the parallax uncertainties (see also
\secref{sec:gdrlimitations}). These systematic terms are correlated over small spatial scales, which
means that the parallax uncertainties are not independent for the Pleiades members considered here,
leading to no reduction of the uncertainties by averaging. Therefore the best estimate we can make
at this time for the mean Pleiades parallax is $7.45\pm0.3$~mas, corresponding to a distance of
about $134\pm6$~pc. This is indicated by the half-width of the yellow shaded area in
\figref{fig:pleiades1}.

We want to emphasise that, taking this systematic uncertainty into account, {\gdr} cannot be
considered as giving a final and definite answer on the so-called Pleiades distance discrepancy. In
particular an explanation for the discrepancy between {\gdr} and {\hip} cannot be provided at this
stage. A proper and more extensive analysis of the {\gdr} astrometry for nearby open clusters
(including the Pleiades) is presented in \cite{DPACP-23}, with the results providing further
arguments as to why the Pleiades distance estimated from {\gdr} parallaxes cannot be considered
definitive. A conclusive answer to the question on the Pleiades distance -- in the form of a
sufficiently precise and systematically reliable trigonometric parallax for the cluster -- can,
however, be expected from future {\gaia} data releases (probably already {\gaia}~DR2). What the present
release definitely does is to make another significant addition to the accumulating information on
the Pleiades distance which is summarised in \figref{fig:pleiades1}.

%
%

\section{Known limitations of {\gdr}}
\label{sec:gdrlimitations}

{\gdr} represents a major advance in terms of the availability of high-accuracy parallaxes and
proper motions for the 2 million stars in the primary astrometric data set and in terms of accurate
positions and homogeneous all-sky photometry for all sources out to the {\gaia} survey limit.
Nevertheless the data release is based on an incomplete reduction of a limited amount of raw {\gaia}
data and is thus of a very preliminary nature. We summarise the major shortcomings of {\gdr} in this
section both to warn the users of the data and to enable a careful scientific exploitation of the
{\gdr} data set. We stress however, that all the shortcomings listed below will be addressed in
future {\gaia} data releases, with major improvements already expected for the second data release.

\subsection{Data processing simplifications for \gdr}
\label{sec:shortcuts}

\begin{figure}
  \resizebox{\hsize}{!}{\includegraphics{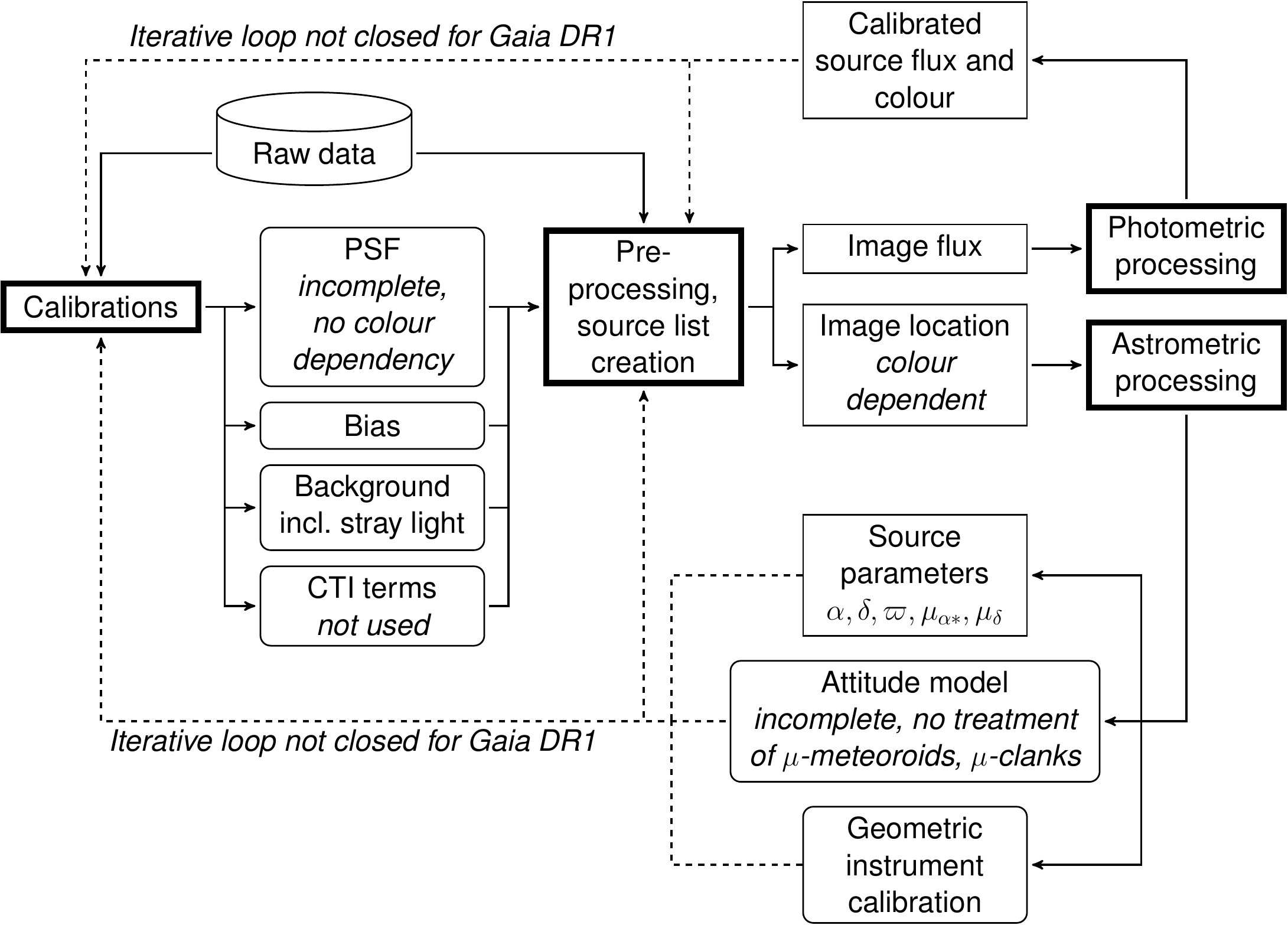}}
  \caption{The DPAC data processing flow as used for {\gdr} in schematic and simplified form. Thick
  lined boxes show processing steps, rounded boxes represent calibrations derived during the
  processing, while thin-lined boxes show processing outputs. The solid lines indicate the
  processing flow as realised for {\gdr}, while the dashed lines indicate processing flows that were
  not implemented for {\gdr}. The remarks in italics highlight important shortcomings in the {\gdr}
  processing.}
  \label{fig:procflow}
\end{figure}

We show in \figref{fig:procflow} in highly simplified form the DPAC data processing flow for the
astrometric and photometric data reduction. The purpose of the diagram is to highlight the
shortcomings in the data processing for {\gdr} compared to the intended data processing for future
data releases (for simplicity many processing steps are left out, including the processing of the
RVS data and the derivation of higher level results, such as source astrophysical parameters). The
steps that should be taken during the processing are:
\begin{enumerate}
  \item From the raw data derive (initial) calibrations of the {\gaia} PSF, the CCD bias, the
    astrophysical and stray light induced background flux in the image, and the parameters
    describing the charge transfer inefficiency (CTI) effects in the CCDs.
  \item Use the calibrations to determine from the raw CCD-level measurements both the source flux
    and the source location within the observation window.
  \item Use the spacecraft attitude to create the source list, by assigning observations (focal plane
    transits) to existing sources or by creating new sources if needed.
  \item Process the image fluxes to derive calibrated $G$-band photometry and process the BP/RP data
    to derive the source colours. Process the image locations in order to derive the astrometric
    source parameters, the attitude model for the {\gaia} spacecraft, and the geometric instrument
    calibrations.
  \item Introduce the known source locations on the sky, the geometric instrument calibrations, the
    attitude model, and the source colours into step 1 above and improve the accuracy of the
    calibrations.
  \item Repeat steps 2 and 3 using the improved astrometry and calibrations from step 5.
    Subsequently repeat step 4 using the improved image locations and fluxes.
  \item Iterate the above steps, including progressively more data, until convergence on the final
    astrometric and photometric results at their ultimately attainable accuracy.
\end{enumerate}

As illustrated in \figref{fig:procflow} steps 5--7 above were not carried out during the processing
for {\gdr}, which means that the inputs for the astrometric and photometric processing are limited
in quality due to the use of immature calibrations, in particular an incomplete PSF model which does
not account for source colour effects on the detailed image shape, or for PSF variations across the
focal plane and in time. The source locations within the images and the astrometry derived from
those will be strongly affected by systematics related to source colour \citep[see][appendix
C]{DPACP-14}. Systematic effects related to the PSF model can also be expected in the $G$-band
photometry derived from the image fluxes. A further limitation to the quality of {\gdr} astrometry,
indicated in \figref{fig:procflow}, is that the attitude modelling within the astrometric solution
is incomplete. No treatment of micro-meteoroid hits or micro-clanks was included (except for the
exclusion of the data from short time intervals affected by large hits) leading to attitude
modelling errors which in turn will limit the astrometric accuracy that can be attained
\citep[see][in particular appendix D]{DPACP-14}. The treatment of CTI effects was not included in
{\gdr}, which is justified given the present low levels of radiation damage to the {\gaia} CCDs
\citep{2016arXiv160801476C}.

We stress that the above description of the data processing for {\gdr} is mainly illustrative and
not intended as a complete description of all the simplifications that were introduced to enable a
timely first {\gaia} data release. For details on the actual processing for {\gdr} refer to
\cite{DPACP-7} (pre-processing and source list creation), \cite{DPACP-12,DPACP-9,DPACP-10}
(photometric processing), \cite{DPACP-15} (variable star processing), and \cite{DPACP-14}
(astrometric processing). In particular the latter paper contains an extensive description of the
known problems introduced by the preliminary astrometric processing.

In the following subsections we summarise the most prominent issues with {\gdr} which should be
taken into consideration when using the data for scientific analyses. These concern catalogue
completeness, and systematics in the astrometric and photometric results which were revealed during
the validation of the DPAC outputs produced for {\gdr}. Much more detail on the validation of {\gdr}
can be found in \cite{DPACP-14, DPACP-11, DPACP-15, DPACP-16}.

\begin{figure}
  \resizebox{\hsize}{!}{\includegraphics{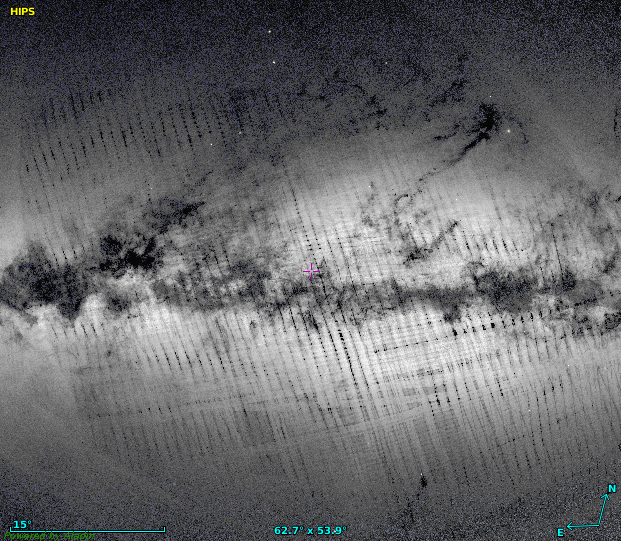}}
  \caption{{\gdr} source density distribution on the sky in the direction of the Milky Way bulge region.
  Note the prominent `striping' and the gaps in the source distribution.}
  \label{fig:bulge}
\end{figure}

\subsection{{\gdr} source list and completeness}

The {\gdr} celestial source density distribution shown in \figref{fig:skymap} contains a number of
clearly non-astronomical artefacts, which is illustrated in more detail in \figref{fig:bulge} for
the Milky Way bulge region. In particular away from the Milky Way plane, but also across the Bulge
region, \figref{fig:skymap} shows obvious source under-densities as well as {\em apparent}
over-densities, where the latter surround the regions (along the ecliptic) dominated by the former.

The patterns in \figref{fig:skymap} are related to the {\gaia} scanning law \citep[cf.][]{DPACP-1}
and are caused by the source filtering applied for {\gdr}. The areas around the ecliptic are
inherently observed less often due to the characteristics of the scanning law, and in particular
have been rather poorly observed over the first 14 months of the mission (covering {\gdr}), both in
terms of the number of visits and the coverage in scanning direction. This results in the sources in
the less well covered areas having a larger probability of being filtered out, which gives the
regions in between (with far fewer sources filtered out) the appearance of containing more sources.
Hence \figref{fig:skymap} shows primarily a deficit of sources in the less well observed regions of
the sky.

\begin{figure}
  \resizebox{\hsize}{!}{\includegraphics{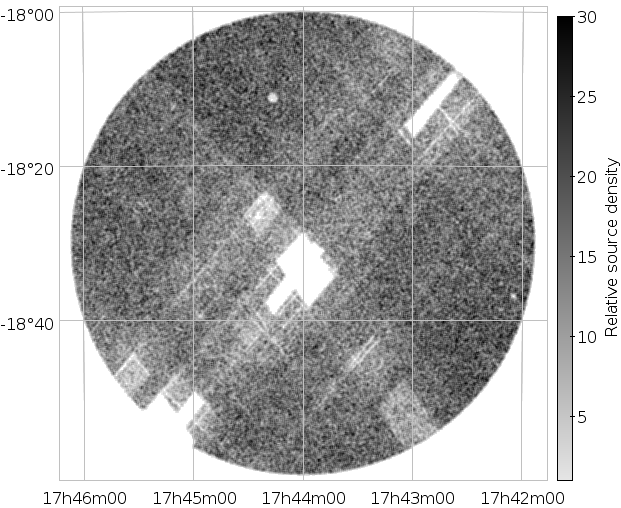}}\\
  \resizebox{\hsize}{!}{\includegraphics{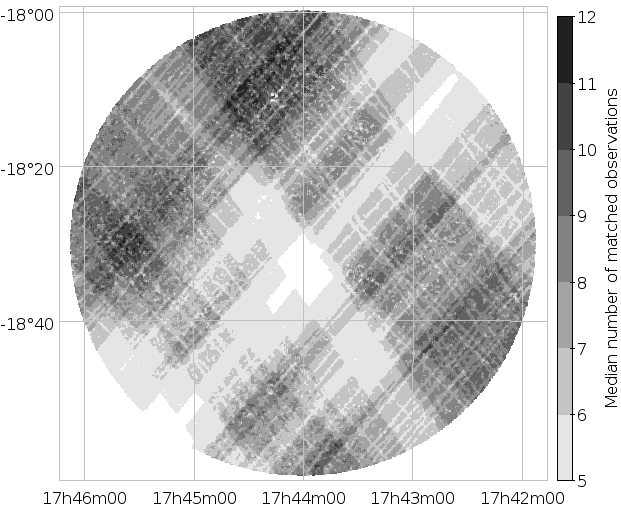}}\\
  \resizebox{\hsize}{!}{\includegraphics{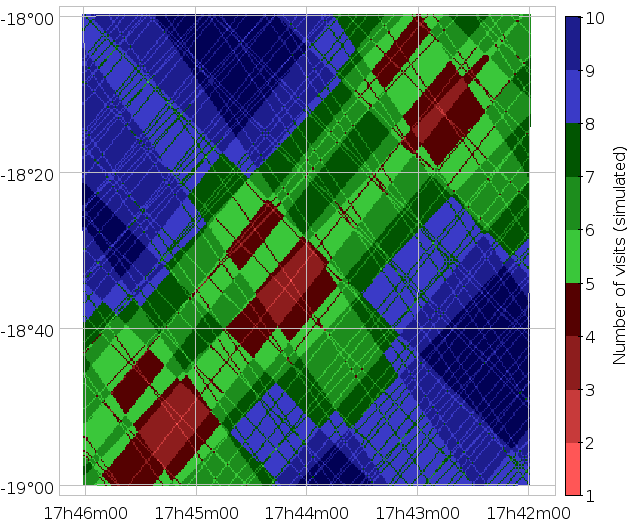}}
  \caption{Illustration of how the combination of scan law coverage and data filtering leads to
  gaps in the {\gdr} source distribution. The top panel shows the source density in the area of
  $0.5$ degree radius around $(\alpha,\delta)=(266^\circ,-18.5^\circ)$. The middle panel shows the
  median number of observations (i.e.\ focal plane transits) that were matched to each source. The
  bottom panel shows the predicted number of visits by {\gaia} according to the nominal scanning
  law.}
  \label{fig:scangaps}
\end{figure}

This is illustrated in more detail for the Milky Way bulge region in \figref{fig:bulge}. The pattern
of dark stripes, with a clear lack of sources, is again related to the {\gaia} scanning law. The
bulge lies in the ecliptic region and thus suffers from poor scan law coverage in {\gdr}. In
combination with the filtering on the astrometric solution quality prior to {\gdr} publication this
can even lead to areas where sources are entirely missing. This is illustrated in
\figref{fig:scangaps}, which shows the circle on the sky around
$(\alpha,\delta)=(266^\circ,-18.5^\circ)$ with a $0.5$ degree radius. The top panel shows the
distribution of the $268\,435$ sources in this area. The distribution shows the striping pattern and
also contains very thin strips where fewer sources than the average are found. Most prominent,
however, are three large gaps where no sources occur. The middle panel shows the median number of
observations matched to each source in this region and there the pattern is even richer. We note
that the minimum number of observations is five (as demanded by the filtering applied, see
\secref{sec:gdrvalidation}), suggesting that the gaps are related to the number of times a
particular coordinate on the sky was visited by {\gaia}. This is confirmed in the bottom panel which
shows a simulation of the expected number of visits corresponding to the scanning law as executed
between September 2014 and September 2015. This time period does not cover the ecliptic pole
scanning phase, but during that phase this region on the sky was not observed. The gaps in the
source distribution correspond to the areas in the simulation where fewer than five visits by
{\gaia} occur, which thus explains the gaps as being due to the filtering applied for {\gdr}. In
addition the simulation shows the same very thin strips where {\gaia} has collected fewer
observations than the maximum of 12 occurring in this area on the sky. Whenever few observations are
collected there is a good chance that the source gets filtered out if focal plane transits from
particular visits by {\gaia} are discarded for other reasons and thus the total number of
observations drops below five. Although the simulated scan law coverage very much resembles the
pattern in the number of matched observations, there are differences in detail because the actually
executed scanning law differs somewhat from the nominal scanning law used in the simulation.

The striping pattern seen over the bulge region in \figref{fig:bulge} can thus be explained as a
consequence of the {\gaia} scan law coverage over the first 14 months of the mission combined with
the filtering applied to the astrometric results before including them in {\gdr}. Although the
striping and gaps are most prominently visible in the bulge region this pattern also occurs in other
parts of the sky in the ecliptic region, notably along the Milky Way plane in the anti-centre
direction. In these areas the step changes in the number of observations collected by {\gaia}
combined with the filtering has in some unlucky cases led to one half of an open cluster partly
missing from the catalogue.

Further remarks on the catalogue completeness are the following:
\begin{itemize}
  \item Many bright stars at $G\lesssim7$ are missing from {\gdr} as the corresponding measurements
    cannot yet be treated routinely by the DPAC. The images are heavily saturated and the instrument
    configuration (TDI gate setting used) is difficult to calibrate due to the sparsity of bright
    sources on the sky.
  \item High proper motion stars ($\mu>3.5\text{ arcsec yr}^{-1}$) are missing from the catalogue
    due to a technical issue in the construction of the IGSL \citep[cf.][]{DPACP-14}.
  \item As mentioned in \secref{sec:gdrvalidation} extremely blue and red sources are missing from
    {\gdr} which, for example, affects the completeness of the white dwarf population in {\gdr}
    and that of sources in extincted regions \citep[cf.][]{DPACP-16}.
  \item In dense areas on the sky (with source densities above a few hundred thousand per square
    degree) the crowding of sources will lead to the truncation of the observation windows for some
    stars when they overlap with the window of another star. These truncated windows have not been
    used in the data processing for {\gdr}. This means that in dense areas the average number of
    transits used per source will be smaller (especially for fainter sources), which in combination
    with the filtering on the number of observations and the astrometric or photometric solution
    quality means these sources may have been removed from {\gdr}.
  \item The survey completeness is also affected by the way the data is treated onboard {\gaia},
    meaning both the detection of sources and the assignment of observation windows. The details are
    provided in \cite{DPACP-1}. We note here that in very dense areas (above $\sim400\,000$ stars
    per square degree) the effective magnitude limit of the {\gaia} survey may be brighter by up to
    several magnitudes, with data for faint sources being collected for a reduced number of focal
    plane transits.
  \item An examination of double stars from the Washington Visual Double Star Catalog
    \citep{2001AJ....122.3466M} contained in {\gdr} shows that below about $4$ arcsec there is a
    notable decrease in the completeness of the detection of the secondaries, which is related to
    the above mentioned limitations in crowded regions \citep{DPACP-16}. The implication of this
    finding and the previous two items is that the effective angular resolution on the sky of
    {\gdr}, in particular in dense areas, is not yet at the levels expected for the $1.5$~m {\gaia}
    telescope mirrors (which should lead to an angular resolution comparable to that of the Hubble
    Space Telescope).
\end{itemize}

The limitations to the {\gdr} source list described above lead to a catalogue which is not complete
in any sense and for which the faint magnitude limit is ill-defined and dependent on celestial
position. No attempt was made to derive a detailed completeness function. Hence when using the
catalogue for scientific analyses, care needs to be taken with the interpretation of source
distributions both on the sky and in apparent magnitude.

\begin{figure*}
  \sidecaption
  \includegraphics[width=12cm]{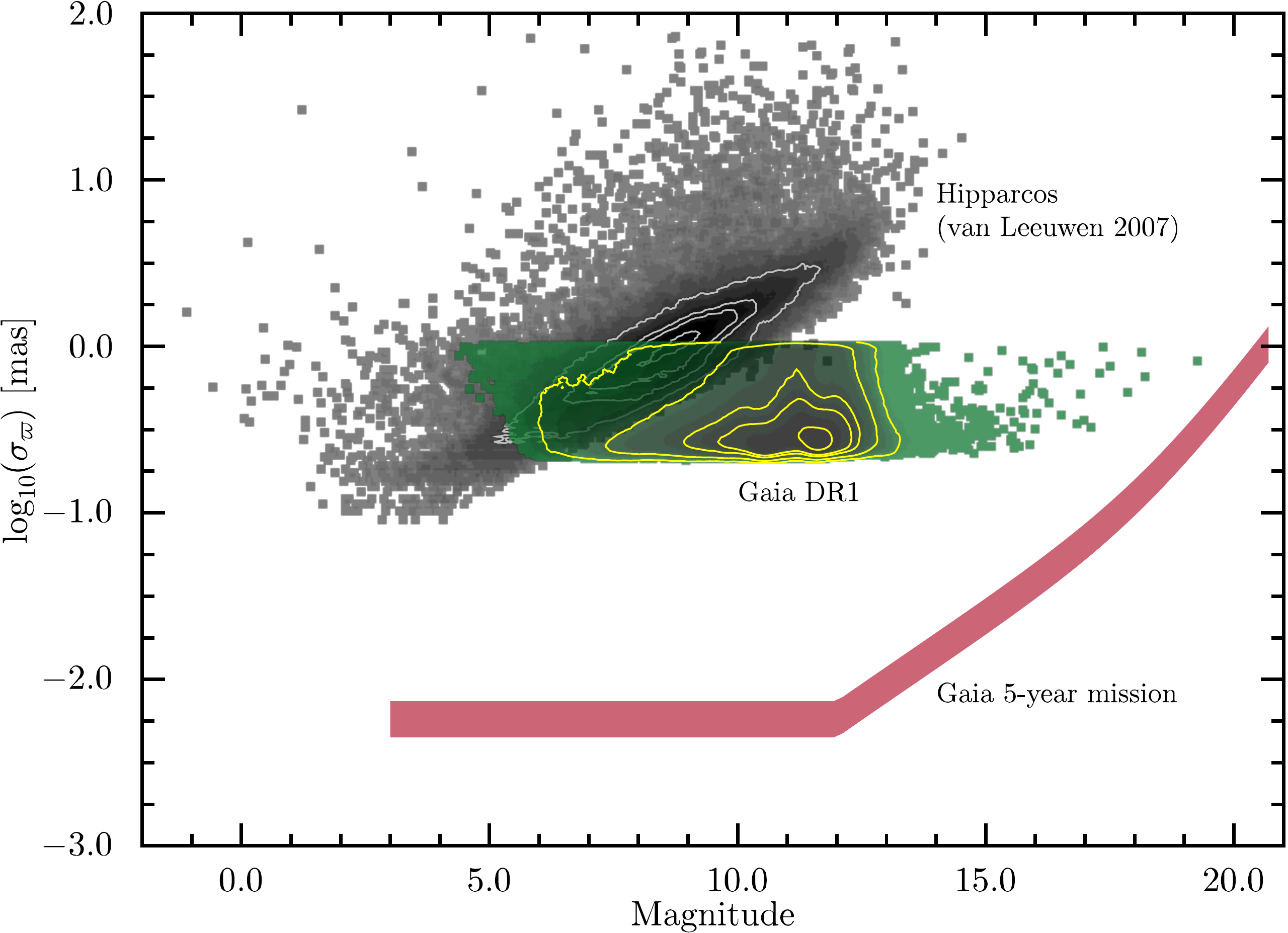}
  \caption{Parallax standard uncertainties as a function of magnitude for {\hip} \citep{book:newhip}
  and the primary astrometric data set in {\gdr}, compared to the predicted 5-year {\gaia} mission
  parallax standard uncertainties. The band for the 5-year mission predictions indicates the
  expected variation as a function of celestial position. The colour coding for the {\hip} and
  {\gdr} parallax uncertainty distributions indicates increasing numbers of sources from light to
  dark colours (logarithmic scale). The contours enclose 10, 50, $68.3$, and 90 per cent of the data
  in the case of {\hip}, while for {\gdr} they enclose 10, 50, $68.3$, $95.4$, and $99.7$ per cent
  of the data.}
  \label{fig:hipgdr1gaia}
\end{figure*}

\subsection{Known problems in the {\gdr} photometry}

Although the $G$-band fluxes and magnitudes provided with {\gdr} have standard uncertainties as good
as a few per cent in magnitude at the survey limit and down to the milli-magnitude level at the
bright end, there are nevertheless limitations inherent to this first {\gaia} data release. The
$G$-band fluxes were derived as part of the image parameter determination in the initial data
treatment \citep[see \secref{sec:shortcuts} above and][]{DPACP-7} and thus suffer from the lack of
an accurate PSF model. In addition at the bright end ($G<12$) the calibrations of the photometry are
complicated by the use of TDI gates, while over the range $G=12$--$17$ the effects of different
observation window sizes make the calibration more complex. The result is that for the brightest,
$G<12$, stars the photometric accuracy is estimated to currently be limited to a calibration floor
of $\sim 3$~mmag for the individual CCD transits, \citep{DPACP-12, DPACP-11}. The quoted standard
uncertainties on the mean $G$-band magnitudes at the bright end can vary by an order of magnitude
(caused by poorly calibrated transitions from one TDI gate setting to another). Over the range
$G=12$--$17$ the distribution of photometric standard errors as a function of magnitude shows two
bumps at $G\sim13$ and $G\sim16$ which are related to the transition from one observation window
type to another \citep{DPACP-12}. An examination of the scatter in repeated photometric measurements
for well-observed sources indicates that the quoted standard uncertainties on the $G$-band
photometry are largely realistic as indicators of the photometric precision \citep[see][for
details]{DPACP-11}, however unaccounted for systematic errors cannot be excluded. Potential
systematic errors in the photometry are discussed in \cite{DPACP-11} and \cite{DPACP-16}. There is a
small fraction of sources for which the mean value of $G$ is clearly wrong. These are sources with
magnitudes well beyond the {\gaia} survey limit of $G=20.7$ and also at brighter magnitudes such
errors occur as evidenced by the presence of a small number of {\tyctwo} sources with magnitudes up
to $G\sim19$ (cf.\ \figref{fig:hipgdr1gaia}), although it should be noted that a number of these
sources may well be variables with large brightness excursions, leading to faint magnitudes at the
{\gdr} observation epoch.

\subsection{Known problems in the {\gdr} astrometry}
\label{sec:astrometry-limitations}

The data processing shortcuts and simplifications discussed in \secref{sec:shortcuts} have
introduced a number of known weaknesses in the astrometric solution for {\gdr}, which are described
and explained extensively in \cite{DPACP-14}. Here we highlight the weaknesses most directly
relevant to the scientific exploitation of the {\gdr} data.

\paragraph{Source modelling} All sources were treated as single stars without taking their radial
velocity into account. Hence any astrometric effects due to the orbital motion in binaries or due to
perspective acceleration were ignored. In addition for resolved binaries the positions used to
derive the mean proper motion over the time period between the {\hip}/{\tyc} (around J1991.25) and
the {\gdr} (J2015.0) epochs may be inconsistent \citep[cf.][]{DPACP-14}. The {\gdr} catalogue does
provide the so-called excess source noise, which is meant to represent the astrometric modelling
errors for a specific source, and thus could in principle be used to identify candidate astrometric
binaries or otherwise problematic sources. However in {\gdr} all sources have significant excess
source noise because currently unmodelled attitude and calibration errors are partly `absorbed' in
this quantity \citep[see][for more details]{DPACP-14}. The level at which the excess source noise is
indicative of a source being different from a single star should thus be calibrated against a sample
of known non-single star sources in {\gdr} before it can be used in scientific analyses.

\paragraph{Periodic basic angle variations} As described in \cite{DPACP-1}, a number of issues
affecting the performance of the {\gaia} instruments came to light during the commissioning period.
The most relevant issue for the astrometric quality of {\gdr} is the periodic variation of the basic
angle between the two telescopes of {\gaia}. This angle enters into all the measurements of angular
separations between sources on the sky and its value should either be stable or its variations known
at the level of $\sim 1$~\muas. The actual basic angle variations, measured both through the on
board metrology system and from the daily astrometric solution carried out as part of the DPAC
First-Look analysis \citep[see][]{DPACP-7}, have a component which varies periodically with the
satellite spin period and with a significant amplitude of roughly 1~mas. The harmonic component that
varies as the cosine of the spacecraft heliotropic spin phase cannot be distinguished from a
zero-point offset in the parallaxes, making the calibrations of the basic angle variations an
essential component of the success of {\gaia} \citep[for more detail see][]{2016A&A...586A..26M}.
For {\gdr} the corrections for the basic angle variations were done by adopting the variations as
measured by the onboard metrology system. At the accuracy level of {\gdr} this is sufficient.
However \cite{DPACP-14} do conclude that a global parallax zero point offset of $\pm0.1$~mas may be
present, which is confirmed by the zero-point offset of about $-0.04$~mas found during the
validation of {\gdr} \citep{DPACP-16}. For future data releases the basic angle variations will
largely be determined as calibration parameters within the astrometric solution
\citep[cf.][]{DPACP-14} with the aim to fully account for the variations.

\paragraph{Strongly correlated astrometric parameters} Figure 7 in \cite{DPACP-14} presents a
statistical overview of the standard uncertainties and the correlations between the astrometric
parameters of each source in the primary astrometric data set. In {\gdr} the correlation levels are
high, reaching median values near $-1$ or $+1$ over large regions of the sky. It is thus very
important to make use of the full covariance matrix when taking the standard uncertainties on
(subsets and linear combinations of) the astrometric parameters into account in any scientific
analysis of the data. The correlations will decrease in future data releases as the number of
observations per source and the scan direction diversity increase.

\paragraph{Spatially correlated systematics} Several of the weaknesses in the astrometric solution
identified in \cite{DPACP-14} will lead to systematic errors that are colour dependent and spatially
correlated over areas on the sky that may extend up to tens of degrees. One important contributor to
these correlations is the incomplete modelling of the spacecraft attitude, which is extensively
described in appendix D.3 of \cite{DPACP-14}. Special astrometric validation solutions indeed
point to the presence of spatially correlated and colour-dependent systematics of $\pm0.2$~mas. The
global validation of the astrometric results confirms the presence of spatial variations of the
parallax zero-point \citep[see][]{DPACP-16}. Over large spatial scales the parallax zero-point
variations reach an amplitude of $\sim0.3$~mas, while over a few smaller areas ($\sim2$ degree
radius) much larger parallaxes biases may occur of up to $\pm1$~mas. {\em The recommendation is to
consider the quoted uncertainties on the parallaxes as $\varpi\pm\sigma_\varpi \text{ (random)
}\pm0.3$~mas (systematic). Furthermore averaging parallaxes over small regions of the sky will not
reduce the uncertainty on the mean below the $0.3$ mas level.} Similar studies into proper motion
biases are not possible due to the limited accuracy of ground-based proper motion catalogues.

Finally, we illustrate graphically the preliminary nature of the {\gdr} astrometry in
\figref{fig:hipgdr1gaia}. It shows the distribution of parallax standard uncertainties as a function
of magnitude for {\hip} \citep{book:newhip} and {\gdr}, and the expected parallax standard
uncertainties achievable after a 5-year {\gaia} mission \citep[as provided in][]{DPACP-1}. Note how
in contrast to {\hip} the {\gdr} parallax standard uncertainties do not decrease with increasing
source brightness but stay at the same level. This is partly related to the gating strategy for
bright ($G<12$) sources which prevents significant gains in signal to noise ratio, but the
uncertainty levels are much more than a factor of 2 away from the 5-year mission uncertainty floor
at the bright end (where the factor of 2 is the gain in signal to noise going from 14 to 60 months
of observations). This indicates that the parallax uncertainties are dominated by calibration errors at
this stage, the calibration floor being $\sim0.2$~mas. The second important point in this figure is
that the expected 5-year parallax standard uncertainties are much better than what can be achieved
for {\gdr}, with the current parallax standard uncertainty levels being comparable to the standard
uncertainty levels that can ultimately be achieved at the {\gaia} survey limit.

%
%

\section{{\gdr} access facilities}
\label{sec:access}

Access to the data contained in {\gdr} is provided through various channels. The main access point
is the ESA {\gaia} Archive, which can be accessed through \url{http://archives.esac.esa.int/gaia/}.
The archive provides access to the data through simple query forms but also allows the submission of
sophisticated data base queries in the Astronomical Data Query Language \citep{2008ivoa.spec.1030O}.
The electronic tables comprising {\gdr} contain descriptions of each data field which can be
inspected online. The {\gaia} archive is Virtual Observatory (\url{http://www.ivoa.net/}) compatible
and also allows for access through the Table Access Protocol \citep{2010ivoa.spec.0327D}. More
extensive documentation, providing more detail on the data processing than is possible to include in
peer-reviewed papers, is available from the archive in various electronic formats. Further tools
provided are a visualisation application, graphics with statistical overviews of the data, an online
help system, and the means to upload user generated tables which can be combined with {\gaia} data
and shared with other users of the {\gaia} archive. More details on the data access facilities are
provided in \cite{DPACP-19}.

As part of the archive services pre-computed cross-match tables linking {\gdr} to other large
surveys are provided to facilitate the analysis of combined data sets. The details on how these
cross-match tables were computed are provided in \cite{DPACP-17}.

Finally, the {\gdr} data is also made available through a number of partner and affiliated data
centres located in Europe, the United States, South Africa, and Japan. These data centres do not
necessarily hold all the data contained in the {\gaia} archive and may layer their own access and
analysis facilities on top of the {\gaia} data.

%
%

\section{Conclusions}
\label{sec:conclusions}

Less than three years after the launch of {\gaia} we present the first {\gaia} data release, where
the use of positional information from the {\hip} and {\tyctwo} catalogues allowed the derivation of
positions, parallaxes, and proper motions for about 2 million sources from the first 14 months of
observations. This represents a data release that was not foreseen in the original {\gaia} mission
planning and presents the astronomical community with advanced access to a large set of parallaxes
and proper motions for sources to magnitude $11.5$, at precisions substantially better than
previously available. The release contains the positions and the mean $G$-band magnitudes for an
additional $1141$ million sources to the {\gaia} survey limit at $G\approx20.7$, as well as the light
curves for a sample of about three thousand variable stars.

The typical uncertainty for the position and parallaxes for sources in the primary astrometric data
set is about $0.3$~mas, and about $1$~mas~yr$^{-1}$ for the proper motions. We stress again that a
systematic component of $\sim0.3$~mas should be added to the parallax uncertainties and that
averaging parallaxes over small regions on the sky will not lead to a gain in precision. For the
subset of {\hip} stars in the primary astrometric data set the proper motions are much more precise,
at about $0.06$~mas~yr$^{-1}$ (albeit with a systematic uncertainty at the same level). The
positions of the sources in the secondary astrometric data set are typically known to $\sim10$~mas.
The positions and proper motions are given in a reference frame that is aligned with the
International Celestial Reference Frame (ICRF) to better than $0.1$~mas at epoch J2015.0, and
non-rotating with respect to ICRF to within $0.03$~mas~yr$^{-1}$.

The photometric data comprises the mean {\gaia} $G$-band magnitudes for all the sources contained in
{\gdr}, with uncertainties ranging from a few milli-magnitudes at the bright end to $\sim0.03$ mag
at the survey limit (although systematic errors cannot be excluded), as well as light curves for
{\cepnum} Cepheids and {\rrlnum} RR Lyrae variables observed at high cadence around the south
ecliptic pole.

We have illustrated the scientific quality of the {\gdr} and have also pointed out the substantial
shortcomings and the preliminary nature of this first {\gaia} data release. When using the data
presented here the warnings given in \secref{sec:gdrlimitations} should be considered carefully.
However, we are confident of the overall quality of the data, which represents a major advance in
terms of available precise positions, parallaxes, proper motions, and homogeneous all-sky
photometry. In addition, the scientific exploitation of the data at this early stage will surely
improve the quality of future {\gaia} data releases.

We note in closing that all of the shortcomings listed in this and the accompanying {\gdr} papers
will be addressed in future {\gaia} data releases with very substantial improvements already
expected for {\gaia}~DR2.

%
%

\begin{acknowledgements}
This work has made use of results from the European Space Agency (ESA) space mission {\it Gaia}, the
data from which were processed by the {\it Gaia} Data Processing and Analysis Consortium (DPAC).
Funding for the DPAC has been provided by national institutions, in particular the institutions
participating in the {\it Gaia} Multilateral Agreement. The {\it Gaia} mission website is
\url{http://www.cosmos.esa.int/gaia}. The authors are current or past members of the ESA {\it Gaia}
mission team and of the {\it Gaia} DPAC.
This work has received financial supported from
the Algerian Centre de Recherche en Astronomie, Astrophysique et G\'{e}ophysique of Bouzareah Observatory;
the Austrian FWF Hertha Firnberg Programme through grants T359, P20046, and P23737;
the BELgian federal Science Policy Office (BELSPO) through various PROgramme de D\'eveloppement d'Exp\'eriences scientifiques (PRODEX) grants;
the Brazil-France exchange programmes FAPESP-COFECUB and CAPES-COFECUB;
the Chinese National Science Foundation through grant NSFC 11573054;
the Czech-Republic Ministry of Education, Youth, and Sports through grant LG 15010;
the Danish Ministry of Science;
the Estonian Ministry of Education and Research through grant IUT40-1;
the European Commission’s Sixth Framework Programme through the European Leadership in Space Astrometry (ELSA) Marie Curie Research Training Network (MRTN-CT-2006-033481), through Marie Curie project PIOF-GA-2009-255267 (SAS-RRL), and through a Marie Curie Transfer-of-Knowledge (ToK) fellowship (MTKD-CT-2004-014188); the European Commission's Seventh Framework Programme through grant FP7-606740 (FP7-SPACE-2013-1) for the {\it Gaia} European Network for Improved data User Services (GENIUS) and through grant 264895 for the {\it Gaia} Research for European Astronomy Training (GREAT-ITN) network;
the European Research Council (ERC) through grant 320360 and through the European Union’s Horizon 2020 research and innovation programme through grant agreement 670519 (Mixing and Angular Momentum tranSport of massIvE stars -- MAMSIE);
the European Science Foundation (ESF), in the framework of the {\it Gaia} Research for European Astronomy Training Research Network Programme (GREAT-ESF);
the European Space Agency in the framework of the {\it Gaia} project;
the European Space Agency Plan for European Cooperating States (PECS) programme through grants for Slovenia; the Czech Space Office through ESA PECS contract 98058;
the Academy of Finland; the Magnus Ehrnrooth Foundation;
the French Centre National de la Recherche Scientifique (CNRS) through action `D\'efi MASTODONS';
the French Centre National d'Etudes Spatiales (CNES);
the French L'Agence Nationale de la Recherche (ANR) `investissements d'avenir' Initiatives D’EXcellence (IDEX) programme PSL$\ast$ through grant ANR-10-IDEX-0001-02;
the R\'egion Aquitaine;
the Universit\'e de Bordeaux;
the French Utinam Institute of the Universit\'e de Franche-Comt\'e, supported by the R\'egion de Franche-Comt\'e and the Institut des Sciences de l'Univers (INSU);
the German Aerospace Agency (Deutsches Zentrum f\"{u}r Luft- und Raumfahrt e.V., DLR) through grants 50QG0501, 50QG0601, 50QG0602, 50QG0701, 50QG0901, 50QG1001, 50QG1101, 50QG140, 50QG1401, 50QG1402, and 50QG1404;
the Hungarian Academy of Sciences through Lend\"ulet Programme LP2014-17;
the Hungarian National Research, Development, and Innovation Office through grants NKFIH K-115709 and PD-116175;
the Israel Ministry of Science and Technology through grant 3-9082;
the Agenzia Spaziale Italiana (ASI) through grants I/037/08/0, I/058/10/0, 2014-025-R.0, and 2014-025-R.1.2015 to INAF and contracts I/008/10/0 and 2013/030/I.0 to ALTEC S.p.A.;
the Italian Istituto Nazionale di Astrofisica (INAF);
the Netherlands Organisation for Scientific Research (NWO) through grant NWO-M-614.061.414 and through a VICI grant to A.~Helmi;
the Netherlands Research School for Astronomy (NOVA);
the Polish National Science Centre through HARMONIA grant 2015/18/M/ST9/00544;
the Portugese Funda\c{c}\~ao para a Ci\^{e}ncia e a Tecnologia (FCT) through grants PTDC/CTE-SPA/118692/2010, PDCTE/CTE-AST/81711/2003, and SFRH/BPD/74697/2010; the Strategic Programmes PEst-OE/AMB/UI4006/2011 for SIM, UID/FIS/00099/2013 for CENTRA, and UID/EEA/00066/2013 for UNINOVA;
the Slovenian Research Agency;
the Spanish Ministry of Economy MINECO-FEDER through grants AyA2014-55216, AyA2011-24052, ESP2013-48318-C2-R, and ESP2014-55996-C2-R and MDM-2014-0369 of ICCUB (Unidad de Excelencia `Mar\'{\i}a de Maeztu);
the Swedish National Space Board (SNSB/Rymdstyrelsen);
the Swiss State Secretariat for Education, Research, and Innovation through the ESA PRODEX programme, the Mesures d’Accompagnement, and the Activit\'es Nationales Compl\'ementaires;
the Swiss National Science Foundation, including an Early Postdoc.Mobility fellowship;
the United Kingdom Rutherford Appleton Laboratory;
the United Kingdom Science and Technology Facilities Council (STFC) through grants PP/C506756/1 and ST/I00047X/1; and
the United Kingdom Space Agency (UKSA) through grants ST/K000578/1 and ST/N000978/1.
We acknowledge the valuable advice provided by Vincenzo~Innocente (CERN) during two pre-launch reviews of DPAC.

This research has made use of the Set of Identifications, Measurements, and Bibliography for
Astronomical Data \citep{2000A&AS..143....9W} and of the `Aladin sky atlas'
\citep{2000A&AS..143...33B,2014ASPC..485..277B}, which are developed and operated at Centre de
Donn\'ees astronomiques de Strasbourg (CDS), France. Some of the figures in this paper were made
with TOPCAT (\url{http://www.starlink.ac.uk/topcat/}) or through the use of the STIL library
(\url{http://www.starlink.ac.uk/stil}). This research made use of the AAVSO Photometric All-Sky
Survey (APASS, \url{https://www.aavso.org/apass}), funded by the Robert Martin Ayers Sciences Fund.
This publication made use of data products from the Two Micron All Sky Survey, which is a joint
project of the University of Massachusetts and the Infrared Processing and Analysis
Center/California Institute of Technology, funded by the National Aeronautics and Space
Administration and the National Science Foundation. We thank the anonymous referee for suggestions
that helped improve this paper.
\end{acknowledgements}

%
%

\bibliographystyle{aa} 
\bibliography{aa201629512} 

%
%

\begin{appendix}

  \section{List of acronyms}
  \label{app:acronyms}
  Below, we give a list of acronyms used in this paper.\hfill\\
  \begin{tabular}{ll}
    \hline\hline
    \noalign{\smallskip}
    \textbf{Acronym} & \textbf{Description} \\
    \noalign{\smallskip}
    \hline
    \noalign{\smallskip}
    2MASS&Two-Micron All Sky Survey \\
    AAVSO&American Association of Variable Star Observers \\
    APASS&AAVSO Photometric All-Sky Survey \\
    BP&Blue Photometer \\
    CCD&Charge-Coupled Device \\
    CTI&Charge Transfer Inefficiency \\
    DPAC&Data Processing and Analysis Consortium \\
    ICRF&International Celestial Reference Frame \\
    IGSL&Initial {\gaia} Source List \\
    OBMT&OnBoard Mission Timeline \\
    PSF&Point Spread Function \\
    RP&Red Photometer \\
    RVS&Radial Velocity Spectrometer \\
    TCB&Barycentric Coordinate Time \\
    TDI&Time-Delayed Integration (CCD) \\
    TGAS&{\tyc}-{\gaia} Astrometric Solution \\
    \noalign{\smallskip}
    \hline
  \end{tabular}

  \section{Example {\gaia} archive queries}
  \label{app:queries}

  Tables \ref{tab:queries}--\ref{tab:queriesvperp} list the queries in Astronomical Data Query
  Language form that can be submitted to the Gaia archive in order to retrieve the data necessary to
  reproduce \figsref{fig:hrdcomp}, \ref{fig:mgdistcolourslice}, \ref{fig:hrdlarge},
  \ref{fig:pleiades2}, and \ref{fig:hrdvperp}. The selection on the standard uncertainty in $G$
  ignores the contribution of the $G$-band magnitude zero point error. Including this small
  ($\sim0.003$~mag) error term does not alter the query results, except for the selection for
  \figref{fig:hrdlarge}.

  \begin{table*}
    \caption{Minimal queries that can be submitted to the {\gaia} archive in the Astronomical Data
    Query Language to retrieve the data necessary to reproduce the HR-diagrams and the magnitude
    distribution in Figs.~\ref{fig:hrdcomp} and \ref{fig:mgdistcolourslice}.}
    \label{tab:queries}
    \centering
    \begin{tabular}{p{\textwidth}}
      Query to reproduce panels \textbf{a} and \textbf{b} of \figref{fig:hrdcomp}. This results in a
      table of $43\,546$ rows listing the {\gaia} source identifier, the {\hip} number, the values
      of $M_G$ based on the {\hip} and {\gdr} parallaxes respectively, and the value of $(B-V)$ from
      the {\hip} catalogue \citep{book:newhip}.
      \begin{verbatim}
      select gaia.source_id, gaia.hip,
        gaia.phot_g_mean_mag+5*log10(gaia.parallax)-10 as g_mag_abs_gaia,
        gaia.phot_g_mean_mag+5*log10(hip.plx)-10 as g_mag_abs_hip,
        hip.b_v
      from gaiadr1.tgas_source as gaia
      inner join public.hipparcos_newreduction as hip
        on gaia.hip = hip.hip
      where gaia.parallax/gaia.parallax_error >= 5 and
        hip.plx/hip.e_plx >= 5 and
        hip.e_b_v > 0.0 and hip.e_b_v <= 0.05 and
        2.5/log(10)*gaia.phot_g_mean_flux_error/gaia.phot_g_mean_flux <= 0.05
      \end{verbatim} \\

      Query to reproduce panel \textbf{c} of \figref{fig:hrdcomp}. This results in a table of
      $74\,771$ rows listing the {\gaia} source identifier, the values of $M_G$ based on the {\gdr}
      parallax, and the value of $(B-V)$ from the {\hip} catalogue \citep{book:newhip}.
      \begin{verbatim}
      select gaia.source_id, gaia.hip,
        gaia.phot_g_mean_mag+5*log10(gaia.parallax)-10 as g_mag_abs,
        hip.b_v
      from gaiadr1.tgas_source as gaia
      inner join public.hipparcos_newreduction as hip
        on gaia.hip = hip.hip
      where gaia.parallax/gaia.parallax_error >= 5 and
        hip.e_b_v > 0.0 and hip.e_b_v <= 0.05 and
        2.5/log(10)*gaia.phot_g_mean_flux_error/gaia.phot_g_mean_flux <= 0.05
      \end{verbatim} \\

      Query to reproduce \figref{fig:mgdistcolourslice}. This results in a table of $3174$
      rows listing the {\gaia} source identifier, and the values of $M_G$ based on the {\gdr} and
      {\hip} \citep{book:newhip} parallaxes, respectively.
      \begin{verbatim}
      select gaia.source_id, gaia.hip, 
        gaia.phot_g_mean_mag+5*log10(gaia.parallax)-10 as g_mag_abs_gaia,
        gaia.phot_g_mean_mag+5*log10(hip.plx)-10 as g_mag_abs_hip
      from gaiadr1.tgas_source as gaia
      inner join public.hipparcos_newreduction as hip
        on gaia.hip = hip.hip
      where gaia.parallax/gaia.parallax_error >= 5 and
        hip.plx/hip.e_plx >= 5 and
        hip.e_b_v > 0.0 and hip.e_b_v <= 0.05 and
        hip.b_v >= 1.0 and hip.b_v <= 1.1 and
        2.5/log(10)*gaia.phot_g_mean_flux_error/gaia.phot_g_mean_flux <= 0.05
      \end{verbatim} \\
    \end{tabular}
  \end{table*}


  \begin{table*}
    \caption{Minimal queries that can be submitted to the {\gaia} archive in the Astronomical Data
    Query Language to retrieve the data necessary to reproduce the HR diagram in
    \figref{fig:hrdlarge} as well as the Pleiades parallax histogram in \figref{fig:pleiades2}.}
    \label{tab:queriesB}
    \centering
    \begin{tabular}{p{\textwidth}}
      Query to reproduce \figref{fig:hrdlarge}. This results in a table of $1\,004\,207$ rows
      ($1\,004\,204$ when including the zero-point uncertainty on $G$ in the selection criteria)
      listing the {\gaia} source identifier and the values of $M_G$ and $(G-K_\mathrm{s})$.
      \begin{verbatim}
      select gaia.source_id, 
        gaia.phot_g_mean_mag+5*log10(gaia.parallax)-10 as g_mag_abs,
        gaia.phot_g_mean_mag-tmass.ks_m as g_min_ks
      from gaiadr1.tgas_source as gaia
      inner join gaiadr1.tmass_best_neighbour as xmatch
        on gaia.source_id = xmatch.source_id
      inner join gaiadr1.tmass_original_valid as tmass
        on tmass.tmass_oid = xmatch.tmass_oid 
      where gaia.parallax/gaia.parallax_error >= 5 and ph_qual = 'AAA' and
        sqrt(power(2.5/log(10)*gaia.phot_g_mean_flux_error/gaia.phot_g_mean_flux,2)) <= 0.05 and
        sqrt(power(2.5/log(10)*gaia.phot_g_mean_flux_error/gaia.phot_g_mean_flux,2)
             + power(tmass.ks_msigcom,2)) <= 0.05
      \end{verbatim} \\

      Query to carry out a simplistic selection of Pleiades cluster members and reproduce
      \figref{fig:pleiades2}. This results in a table of 164 rows listing the {\gaia} source
      identifier and the {\gaia} parallax.
      \begin{verbatim}
      select gaia.source_id, 
        gaia.parallax
      from gaiadr1.tgas_source as gaia
      where contains(point('ICRS',gaia.ra,gaia.dec),circle('ICRS',56.75,24.12,5)) = 1
        and sqrt(power(gaia.pmra-20.5,2)+power(gaia.pmdec+45.5,2)) < 6.0
      \end{verbatim}
    \end{tabular}
  \end{table*}


  \begin{table*}
    \caption{Minimal queries that can be submitted to the {\gaia} archive in the Astronomical Data
    Query Language to retrieve the data necessary to reproduce the HR diagram in
    \figref{fig:hrdvperp}. In this case the query is split into three parts.}
    \label{tab:queriesvperp}
    \centering
    \begin{tabular}{p{\textwidth}}
      Query to retrieve stars with {\hip} colour indices. This results in a table of $30\,009$ rows
      listing the {\gaia} source identifier, the value of $M_G$ based on the {\gdr} parallax, the
      {\hip} \citep{book:newhip} value of $(B-V)$, and the value of $v_\perp$.
      \begin{verbatim}
      select gaia.source_id,
        gaia.phot_g_mean_mag+5*log10(gaia.parallax)-10 as g_mag_abs,
        hip.b_v as b_min_v,
        sqrt(power(gaia.pmra,2)+power(gaia.pmdec,2))/gaia.parallax*4.74047 as vperp
      from gaiadr1.tgas_source as gaia
      inner join public.hipparcos_newreduction as hip
        on gaia.hip = hip.hip
      where gaia.parallax/gaia.parallax_error >= 5 and
        hip.e_b_v > 0.0 and hip.e_b_v <= 0.05 and
        2.5/log(10)*gaia.phot_g_mean_flux_error/gaia.phot_g_mean_flux <= 0.05 and
        (gaia.parallax >= 10.0 or
        sqrt(power(gaia.pmra,2)+power(gaia.pmdec,2)) >= 200 or
        gaia.phot_g_mean_mag <= 7.5)
      \end{verbatim} \\

      Query to retrieve stars with {\tyctwo} colour indices. This results in a table of $8983$ rows
      listing the {\gaia} source identifier, the value of $M_G$ based on the {\gdr} parallax, the
      {\tyctwo} value of $(B-V)$ calculated as $0.85(B-V)_\mathrm{T}$, and the value of $v_\perp$.
      \begin{verbatim}
      select gaia.source_id,
        gaia.phot_g_mean_mag+5*log10(gaia.parallax)-10 as g_mag_abs,
        0.85*(tycho2.bt_mag-tycho2.vt_mag) as b_min_v,
        sqrt(power(gaia.pmra,2)+power(gaia.pmdec,2))/gaia.parallax*4.74047 as vperp
      from gaiadr1.tgas_source as gaia
      inner join public.tycho2 as tycho2
        on gaia.tycho2_id = tycho2.id
      where gaia.parallax/gaia.parallax_error >= 5 and
        sqrt(power(tycho2.e_bt_mag,2) + power(tycho2.e_vt_mag,2)) <= 0.05 and
        2.5/log(10)*gaia.phot_g_mean_flux_error/gaia.phot_g_mean_flux <= 0.05 and
        (gaia.parallax >= 10.0 or
        sqrt(power(gaia.pmra,2)+power(gaia.pmdec,2)) >= 200 or
        gaia.phot_g_mean_mag <= 7.5)
      \end{verbatim} \\

      Query to retrieve stars with APASS colour indices. This results in a table of $2144$ rows
      listing the {\gaia} source identifier, the value of $M_G$ based on the
      {\gdr} parallax, the APASS value of $(B-V)$, and the value of $v_\perp$.
      \begin{verbatim}
      select gaia.source_id,
        gaia.phot_g_mean_mag+5*log10(gaia.parallax)-10 as g_mag_abs,
        (urat.b_mag-urat.v_mag) as b_min_v,
        sqrt(power(gaia.pmra,2)+power(gaia.pmdec,2))/gaia.parallax*4.74047 as vperp
      from gaiadr1.tgas_source as gaia
      inner join public.tycho2 as tycho2
        on gaia.tycho2_id = tycho2.id
      inner join gaiadr1.urat1_best_neighbour as uratxmatch
        on gaia.source_id = uratxmatch.source_id
      inner join gaiadr1.urat1_original_valid as urat
        on uratxmatch.urat1_oid = urat.urat1_oid
      where gaia.parallax/gaia.parallax_error >= 5 and
        sqrt(power(tycho2.e_bt_mag,2) + power(tycho2.e_vt_mag,2)) > 0.05 and
        sqrt(power(urat.b_mag_error,2) + power(urat.v_mag_error,2)) <= 0.05 and
        2.5/log(10)*gaia.phot_g_mean_flux_error/gaia.phot_g_mean_flux <= 0.05 and
        (gaia.parallax >= 10.0 or
        sqrt(power(gaia.pmra,2)+power(gaia.pmdec,2)) >= 200 or
        gaia.phot_g_mean_mag <= 7.5)
      \end{verbatim} \\
    \end{tabular}
  \end{table*}
\end{appendix}

\end{document}

%% file: authors_full.tex
\author{
{\it Gaia} Collaboration
\and A.G.A.    ~Brown                         \inst{\ref{inst:0001}}
\and A.        ~Vallenari                     \inst{\ref{inst:0002}}
\and T.        ~Prusti                        \inst{\ref{inst:0003}}
\and J.H.J.    ~de Bruijne                    \inst{\ref{inst:0003}}
\and F.        ~Mignard                       \inst{\ref{inst:0005}}
\and R.        ~Drimmel                       \inst{\ref{inst:0006}}
\and C.        ~Babusiaux                     \inst{\ref{inst:0007}}
\and C.A.L.    ~Bailer-Jones                  \inst{\ref{inst:0008}}
\and U.        ~Bastian                       \inst{\ref{inst:0009}}
\and M.        ~Biermann                      \inst{\ref{inst:0009}}
\and D.W.      ~Evans                         \inst{\ref{inst:0011}}
\and L.        ~Eyer                          \inst{\ref{inst:0012}}
\and F.        ~Jansen                        \inst{\ref{inst:0013}}
\and C.        ~Jordi                         \inst{\ref{inst:0014}}
\and D.        ~Katz                          \inst{\ref{inst:0007}}
\and S.A.      ~Klioner                       \inst{\ref{inst:0016}}
\and U.        ~Lammers                       \inst{\ref{inst:0017}}
\and L.        ~Lindegren                     \inst{\ref{inst:0018}}
\and X.        ~Luri                          \inst{\ref{inst:0014}}
\and W.        ~O'Mullane                     \inst{\ref{inst:0017}}
\and C.        ~Panem                         \inst{\ref{inst:0021}}
\and D.        ~Pourbaix                      \inst{\ref{inst:0022},\ref{inst:0023}}
\and S.        ~Randich                       \inst{\ref{inst:0024}}
\and P.        ~Sartoretti                    \inst{\ref{inst:0007}}
\and H.I.      ~Siddiqui                      \inst{\ref{inst:0026}}
\and C.        ~Soubiran                      \inst{\ref{inst:0027}}
\and V.        ~Valette                       \inst{\ref{inst:0021}}
\and F.        ~van Leeuwen                   \inst{\ref{inst:0011}}
\and N.A.      ~Walton                        \inst{\ref{inst:0011}}
\and C.        ~Aerts                         \inst{\ref{inst:0031},\ref{inst:0032}}
\and F.        ~Arenou                        \inst{\ref{inst:0007}}
\and M.        ~Cropper                       \inst{\ref{inst:0034}}
\and E.        ~H{\o}g                        \inst{\ref{inst:0035}}
\and M.G.      ~Lattanzi                      \inst{\ref{inst:0006}}
\and E.K.      ~Grebel                        \inst{\ref{inst:0009}}
\and A.D.      ~Holland                       \inst{\ref{inst:0038}}
\and C.        ~Huc                           \inst{\ref{inst:0021}}
\and X.        ~Passot                        \inst{\ref{inst:0021}}
\and M.        ~Perryman                      \inst{\ref{inst:0003}}
\and L.        ~Bramante                      \inst{\ref{inst:0042}}
\and C.        ~Cacciari                      \inst{\ref{inst:0043}}
\and J.        ~Casta\~{n}eda                 \inst{\ref{inst:0014}}
\and L.        ~Chaoul                        \inst{\ref{inst:0021}}
\and N.        ~Cheek                         \inst{\ref{inst:0046}}
\and F.        ~De Angeli                     \inst{\ref{inst:0011}}
\and C.        ~Fabricius                     \inst{\ref{inst:0014}}
\and R.        ~Guerra                        \inst{\ref{inst:0017}}
\and J.        ~Hern\'{a}ndez                 \inst{\ref{inst:0017}}
\and A.        ~Jean-Antoine-Piccolo          \inst{\ref{inst:0021}}
\and E.        ~Masana                        \inst{\ref{inst:0014}}
\and R.        ~Messineo                      \inst{\ref{inst:0042}}
\and N.        ~Mowlavi                       \inst{\ref{inst:0012}}
\and K.        ~Nienartowicz                  \inst{\ref{inst:0055}}
\and D.        ~Ord\'{o}\~{n}ez-Blanco        \inst{\ref{inst:0055}}
\and P.        ~Panuzzo                       \inst{\ref{inst:0007}}
\and J.        ~Portell                       \inst{\ref{inst:0014}}
\and P.J.      ~Richards                      \inst{\ref{inst:0059}}
\and M.        ~Riello                        \inst{\ref{inst:0011}}
\and G.M.      ~Seabroke                      \inst{\ref{inst:0034}}
\and P.        ~Tanga                         \inst{\ref{inst:0005}}
\and F.        ~Th\'{e}venin                  \inst{\ref{inst:0005}}
\and J.        ~Torra                         \inst{\ref{inst:0014}}
\and S.G.      ~Els                           \inst{\ref{inst:0065},\ref{inst:0009}}
\and G.        ~Gracia-Abril                  \inst{\ref{inst:0065},\ref{inst:0014}}
\and G.        ~Comoretto                     \inst{\ref{inst:0026}}
\and M.        ~Garcia-Reinaldos              \inst{\ref{inst:0017}}
\and T.        ~Lock                          \inst{\ref{inst:0017}}
\and E.        ~Mercier                       \inst{\ref{inst:0065},\ref{inst:0009}}
\and M.        ~Altmann                       \inst{\ref{inst:0009},\ref{inst:0075}}
\and R.        ~Andrae                        \inst{\ref{inst:0008}}
\and T.L.      ~Astraatmadja                  \inst{\ref{inst:0008}}
\and I.        ~Bellas-Velidis                \inst{\ref{inst:0078}}
\and K.        ~Benson                        \inst{\ref{inst:0034}}
\and J.        ~Berthier                      \inst{\ref{inst:0080}}
\and R.        ~Blomme                        \inst{\ref{inst:0081}}
\and G.        ~Busso                         \inst{\ref{inst:0011}}
\and B.        ~Carry                         \inst{\ref{inst:0005},\ref{inst:0080}}
\and A.        ~Cellino                       \inst{\ref{inst:0006}}
\and G.        ~Clementini                    \inst{\ref{inst:0043}}
\and S.        ~Cowell                        \inst{\ref{inst:0011}}
\and O.        ~Creevey                       \inst{\ref{inst:0005},\ref{inst:0089}}
\and J.        ~Cuypers                       \inst{\ref{inst:0081}}
\and M.        ~Davidson                      \inst{\ref{inst:0091}}
\and J.        ~De Ridder                     \inst{\ref{inst:0031}}
\and A.        ~de Torres                     \inst{\ref{inst:0093}}
\and L.        ~Delchambre                    \inst{\ref{inst:0094}}
\and A.        ~Dell'Oro                      \inst{\ref{inst:0024}}
\and C.        ~Ducourant                     \inst{\ref{inst:0027}}
\and Y.        ~Fr\'{e}mat                    \inst{\ref{inst:0081}}
\and M.        ~Garc\'{i}a-Torres             \inst{\ref{inst:0098}}
\and E.        ~Gosset                        \inst{\ref{inst:0094},\ref{inst:0023}}
\and J.-L.     ~Halbwachs                     \inst{\ref{inst:0101}}
\and N.C.      ~Hambly                        \inst{\ref{inst:0091}}
\and D.L.      ~Harrison                      \inst{\ref{inst:0011},\ref{inst:0104}}
\and M.        ~Hauser                        \inst{\ref{inst:0009}}
\and D.        ~Hestroffer                    \inst{\ref{inst:0080}}
\and S.T.      ~Hodgkin                       \inst{\ref{inst:0011}}
\and H.E.      ~Huckle                        \inst{\ref{inst:0034}}
\and A.        ~Hutton                        \inst{\ref{inst:0109}}
\and G.        ~Jasniewicz                    \inst{\ref{inst:0110}}
\and S.        ~Jordan                        \inst{\ref{inst:0009}}
\and M.        ~Kontizas                      \inst{\ref{inst:0112}}
\and A.J.      ~Korn                          \inst{\ref{inst:0113}}
\and A.C.      ~Lanzafame                     \inst{\ref{inst:0114},\ref{inst:0115}}
\and M.        ~Manteiga                      \inst{\ref{inst:0116}}
\and A.        ~Moitinho                      \inst{\ref{inst:0117}}
\and K.        ~Muinonen                      \inst{\ref{inst:0118},\ref{inst:0119}}
\and J.        ~Osinde                        \inst{\ref{inst:0120}}
\and E.        ~Pancino                       \inst{\ref{inst:0024},\ref{inst:0122}}
\and T.        ~Pauwels                       \inst{\ref{inst:0081}}
\and J.-M.     ~Petit                         \inst{\ref{inst:0124}}
\and A.        ~Recio-Blanco                  \inst{\ref{inst:0005}}
\and A.C.      ~Robin                         \inst{\ref{inst:0124}}
\and L.M.      ~Sarro                         \inst{\ref{inst:0127}}
\and C.        ~Siopis                        \inst{\ref{inst:0022}}
\and M.        ~Smith                         \inst{\ref{inst:0034}}
\and K.W.      ~Smith                         \inst{\ref{inst:0008}}
\and A.        ~Sozzetti                      \inst{\ref{inst:0006}}
\and W.        ~Thuillot                      \inst{\ref{inst:0080}}
\and W.        ~van Reeven                    \inst{\ref{inst:0109}}
\and Y.        ~Viala                         \inst{\ref{inst:0007}}
\and U.        ~Abbas                         \inst{\ref{inst:0006}}
\and A.        ~Abreu Aramburu                \inst{\ref{inst:0136}}
\and S.        ~Accart                        \inst{\ref{inst:0137}}
\and J.J.      ~Aguado                        \inst{\ref{inst:0127}}
\and P.M.      ~Allan                         \inst{\ref{inst:0059}}
\and W.        ~Allasia                       \inst{\ref{inst:0140}}
\and G.        ~Altavilla                     \inst{\ref{inst:0043}}
\and M.A.      ~\'{A}lvarez                   \inst{\ref{inst:0116}}
\and J.        ~Alves                         \inst{\ref{inst:0143}}
\and R.I.      ~Anderson                      \inst{\ref{inst:0144},\ref{inst:0012}}
\and A.H.      ~Andrei                        \inst{\ref{inst:0146},\ref{inst:0147},\ref{inst:0075}}
\and E.        ~Anglada Varela                \inst{\ref{inst:0120},\ref{inst:0046}}
\and E.        ~Antiche                       \inst{\ref{inst:0014}}
\and T.        ~Antoja                        \inst{\ref{inst:0003}}
\and S.        ~Ant\'{o}n                     \inst{\ref{inst:0153},\ref{inst:0154}}
\and B.        ~Arcay                         \inst{\ref{inst:0116}}
\and N.        ~Bach                          \inst{\ref{inst:0109}}
\and S.G.      ~Baker                         \inst{\ref{inst:0034}}
\and L.        ~Balaguer-N\'{u}\~{n}ez        \inst{\ref{inst:0014}}
\and C.        ~Barache                       \inst{\ref{inst:0075}}
\and C.        ~Barata                        \inst{\ref{inst:0117}}
\and A.        ~Barbier                       \inst{\ref{inst:0137}}
\and F.        ~Barblan                       \inst{\ref{inst:0012}}
\and D.        ~Barrado y Navascu\'{e}s       \inst{\ref{inst:0163}}
\and M.        ~Barros                        \inst{\ref{inst:0117}}
\and M.A.      ~Barstow                       \inst{\ref{inst:0165}}
\and U.        ~Becciani                      \inst{\ref{inst:0115}}
\and M.        ~Bellazzini                    \inst{\ref{inst:0043}}
\and A.        ~Bello Garc\'{i}a              \inst{\ref{inst:0168}}
\and V.        ~Belokurov                     \inst{\ref{inst:0011}}
\and P.        ~Bendjoya                      \inst{\ref{inst:0005}}
\and A.        ~Berihuete                     \inst{\ref{inst:0171}}
\and L.        ~Bianchi                       \inst{\ref{inst:0140}}
\and O.        ~Bienaym\'{e}                  \inst{\ref{inst:0101}}
\and F.        ~Billebaud                     \inst{\ref{inst:0027}}
\and N.        ~Blagorodnova                  \inst{\ref{inst:0011}}
\and S.        ~Blanco-Cuaresma               \inst{\ref{inst:0012},\ref{inst:0027}}
\and T.        ~Boch                          \inst{\ref{inst:0101}}
\and A.        ~Bombrun                       \inst{\ref{inst:0093}}
\and R.        ~Borrachero                    \inst{\ref{inst:0014}}
\and S.        ~Bouquillon                    \inst{\ref{inst:0075}}
\and G.        ~Bourda                        \inst{\ref{inst:0027}}
\and H.        ~Bouy                          \inst{\ref{inst:0163}}
\and A.        ~Bragaglia                     \inst{\ref{inst:0043}}
\and M.A.      ~Breddels                      \inst{\ref{inst:0185}}
\and N.        ~Brouillet                     \inst{\ref{inst:0027}}
\and T.        ~Br\"{ u}semeister             \inst{\ref{inst:0009}}
\and B.        ~Bucciarelli                   \inst{\ref{inst:0006}}
\and P.        ~Burgess                       \inst{\ref{inst:0011}}
\and R.        ~Burgon                        \inst{\ref{inst:0038}}
\and A.        ~Burlacu                       \inst{\ref{inst:0021}}
\and D.        ~Busonero                      \inst{\ref{inst:0006}}
\and R.        ~Buzzi                         \inst{\ref{inst:0006}}
\and E.        ~Caffau                        \inst{\ref{inst:0007}}
\and J.        ~Cambras                       \inst{\ref{inst:0195}}
\and H.        ~Campbell                      \inst{\ref{inst:0011}}
\and R.        ~Cancelliere                   \inst{\ref{inst:0197}}
\and T.        ~Cantat-Gaudin                 \inst{\ref{inst:0002}}
\and T.        ~Carlucci                      \inst{\ref{inst:0075}}
\and J.M.      ~Carrasco                      \inst{\ref{inst:0014}}
\and M.        ~Castellani                    \inst{\ref{inst:0201}}
\and P.        ~Charlot                       \inst{\ref{inst:0027}}
\and J.        ~Charnas                       \inst{\ref{inst:0055}}
\and A.        ~Chiavassa                     \inst{\ref{inst:0005}}
\and M.        ~Clotet                        \inst{\ref{inst:0014}}
\and G.        ~Cocozza                       \inst{\ref{inst:0043}}
\and R.S.      ~Collins                       \inst{\ref{inst:0091}}
\and G.        ~Costigan                      \inst{\ref{inst:0001}}
\and F.        ~Crifo                         \inst{\ref{inst:0007}}
\and N.J.G.    ~Cross                         \inst{\ref{inst:0091}}
\and M.        ~Crosta                        \inst{\ref{inst:0006}}
\and C.        ~Crowley                       \inst{\ref{inst:0093}}
\and C.        ~Dafonte                       \inst{\ref{inst:0116}}
\and Y.        ~Damerdji                      \inst{\ref{inst:0094},\ref{inst:0215}}
\and A.        ~Dapergolas                    \inst{\ref{inst:0078}}
\and P.        ~David                         \inst{\ref{inst:0080}}
\and M.        ~David                         \inst{\ref{inst:0218}}
\and P.        ~De Cat                        \inst{\ref{inst:0081}}
\and F.        ~de Felice                     \inst{\ref{inst:0220}}
\and P.        ~de Laverny                    \inst{\ref{inst:0005}}
\and F.        ~De Luise                      \inst{\ref{inst:0222}}
\and R.        ~De March                      \inst{\ref{inst:0042}}
\and D.        ~de Martino                    \inst{\ref{inst:0224}}
\and R.        ~de Souza                      \inst{\ref{inst:0225}}
\and J.        ~Debosscher                    \inst{\ref{inst:0031}}
\and E.        ~del Pozo                      \inst{\ref{inst:0109}}
\and M.        ~Delbo                         \inst{\ref{inst:0005}}
\and A.        ~Delgado                       \inst{\ref{inst:0011}}
\and H.E.      ~Delgado                       \inst{\ref{inst:0127}}
\and P.        ~Di Matteo                     \inst{\ref{inst:0007}}
\and S.        ~Diakite                       \inst{\ref{inst:0124}}
\and E.        ~Distefano                     \inst{\ref{inst:0115}}
\and C.        ~Dolding                       \inst{\ref{inst:0034}}
\and S.        ~Dos Anjos                     \inst{\ref{inst:0225}}
\and P.        ~Drazinos                      \inst{\ref{inst:0112}}
\and J.        ~Duran                         \inst{\ref{inst:0120}}
\and Y.        ~Dzigan                        \inst{\ref{inst:0238},\ref{inst:0239}}
\and B.        ~Edvardsson                    \inst{\ref{inst:0113}}
\and H.        ~Enke                          \inst{\ref{inst:0241}}
\and N.W.      ~Evans                         \inst{\ref{inst:0011}}
\and G.        ~Eynard Bontemps               \inst{\ref{inst:0137}}
\and C.        ~Fabre                         \inst{\ref{inst:0244}}
\and M.        ~Fabrizio                      \inst{\ref{inst:0122},\ref{inst:0222}}
\and S.        ~Faigler                       \inst{\ref{inst:0247}}
\and A.J.      ~Falc\~{a}o                    \inst{\ref{inst:0248}}
\and M.        ~Farr\`{a}s Casas              \inst{\ref{inst:0014}}
\and L.        ~Federici                      \inst{\ref{inst:0043}}
\and G.        ~Fedorets                      \inst{\ref{inst:0118}}
\and J.        ~Fern\'{a}ndez-Hern\'{a}ndez   \inst{\ref{inst:0046}}
\and P.        ~Fernique                      \inst{\ref{inst:0101}}
\and A.        ~Fienga                        \inst{\ref{inst:0254}}
\and F.        ~Figueras                      \inst{\ref{inst:0014}}
\and F.        ~Filippi                       \inst{\ref{inst:0042}}
\and K.        ~Findeisen                     \inst{\ref{inst:0007}}
\and A.        ~Fonti                         \inst{\ref{inst:0042}}
\and M.        ~Fouesneau                     \inst{\ref{inst:0008}}
\and E.        ~Fraile                        \inst{\ref{inst:0260}}
\and M.        ~Fraser                        \inst{\ref{inst:0011}}
\and J.        ~Fuchs                         \inst{\ref{inst:0262}}
\and M.        ~Gai                           \inst{\ref{inst:0006}}
\and S.        ~Galleti                       \inst{\ref{inst:0043}}
\and L.        ~Galluccio                     \inst{\ref{inst:0005}}
\and D.        ~Garabato                      \inst{\ref{inst:0116}}
\and F.        ~Garc\'{i}a-Sedano             \inst{\ref{inst:0127}}
\and A.        ~Garofalo                      \inst{\ref{inst:0043}}
\and N.        ~Garralda                      \inst{\ref{inst:0014}}
\and P.        ~Gavras                        \inst{\ref{inst:0007},\ref{inst:0078},\ref{inst:0112}}
\and J.        ~Gerssen                       \inst{\ref{inst:0241}}
\and R.        ~Geyer                         \inst{\ref{inst:0016}}
\and G.        ~Gilmore                       \inst{\ref{inst:0011}}
\and S.        ~Girona                        \inst{\ref{inst:0276}}
\and G.        ~Giuffrida                     \inst{\ref{inst:0122}}
\and M.        ~Gomes                         \inst{\ref{inst:0117}}
\and A.        ~Gonz\'{a}lez-Marcos           \inst{\ref{inst:0279}}
\and J.        ~Gonz\'{a}lez-N\'{u}\~{n}ez    \inst{\ref{inst:0046},\ref{inst:0281}}
\and J.J.      ~Gonz\'{a}lez-Vidal            \inst{\ref{inst:0014}}
\and M.        ~Granvik                       \inst{\ref{inst:0118}}
\and A.        ~Guerrier                      \inst{\ref{inst:0137}}
\and P.        ~Guillout                      \inst{\ref{inst:0101}}
\and J.        ~Guiraud                       \inst{\ref{inst:0021}}
\and A.        ~G\'{u}rpide                   \inst{\ref{inst:0014}}
\and R.        ~Guti\'{e}rrez-S\'{a}nchez     \inst{\ref{inst:0026}}
\and L.P.      ~Guy                           \inst{\ref{inst:0055}}
\and R.        ~Haigron                       \inst{\ref{inst:0007}}
\and D.        ~Hatzidimitriou                \inst{\ref{inst:0112}}
\and M.        ~Haywood                       \inst{\ref{inst:0007}}
\and U.        ~Heiter                        \inst{\ref{inst:0113}}
\and A.        ~Helmi                         \inst{\ref{inst:0185}}
\and D.        ~Hobbs                         \inst{\ref{inst:0018}}
\and W.        ~Hofmann                       \inst{\ref{inst:0009}}
\and B.        ~Holl                          \inst{\ref{inst:0012}}
\and G.        ~Holland                       \inst{\ref{inst:0011}}
\and J.A.S.    ~Hunt                          \inst{\ref{inst:0034}}
\and A.        ~Hypki                         \inst{\ref{inst:0001}}
\and V.        ~Icardi                        \inst{\ref{inst:0042}}
\and M.        ~Irwin                         \inst{\ref{inst:0011}}
\and G.        ~Jevardat de Fombelle          \inst{\ref{inst:0055}}
\and P.        ~Jofr\'{e}                     \inst{\ref{inst:0011},\ref{inst:0027}}
\and P.G.      ~Jonker                        \inst{\ref{inst:0306},\ref{inst:0032}}
\and A.        ~Jorissen                      \inst{\ref{inst:0022}}
\and F.        ~Julbe                         \inst{\ref{inst:0014}}
\and A.        ~Karampelas                    \inst{\ref{inst:0112},\ref{inst:0078}}
\and A.        ~Kochoska                      \inst{\ref{inst:0312}}
\and R.        ~Kohley                        \inst{\ref{inst:0017}}
\and K.        ~Kolenberg                     \inst{\ref{inst:0314},\ref{inst:0031},\ref{inst:0316}}
\and E.        ~Kontizas                      \inst{\ref{inst:0078}}
\and S.E.      ~Koposov                       \inst{\ref{inst:0011}}
\and G.        ~Kordopatis                    \inst{\ref{inst:0241},\ref{inst:0005}}
\and P.        ~Koubsky                       \inst{\ref{inst:0262}}
\and A.        ~Krone-Martins                 \inst{\ref{inst:0117}}
\and M.        ~Kudryashova                   \inst{\ref{inst:0080}}
\and I.        ~Kull                          \inst{\ref{inst:0247}}
\and R.K.      ~Bachchan                      \inst{\ref{inst:0018}}
\and F.        ~Lacoste-Seris                 \inst{\ref{inst:0137}}
\and A.F.      ~Lanza                         \inst{\ref{inst:0115}}
\and J.-B.     ~Lavigne                       \inst{\ref{inst:0137}}
\and C.        ~Le Poncin-Lafitte             \inst{\ref{inst:0075}}
\and Y.        ~Lebreton                      \inst{\ref{inst:0007},\ref{inst:0331}}
\and T.        ~Lebzelter                     \inst{\ref{inst:0143}}
\and S.        ~Leccia                        \inst{\ref{inst:0224}}
\and N.        ~Leclerc                       \inst{\ref{inst:0007}}
\and I.        ~Lecoeur-Taibi                 \inst{\ref{inst:0055}}
\and V.        ~Lemaitre                      \inst{\ref{inst:0137}}
\and H.        ~Lenhardt                      \inst{\ref{inst:0009}}
\and F.        ~Leroux                        \inst{\ref{inst:0137}}
\and S.        ~Liao                          \inst{\ref{inst:0006},\ref{inst:0340}}
\and E.        ~Licata                        \inst{\ref{inst:0140}}
\and H.E.P.    ~Lindstr{\o}m                  \inst{\ref{inst:0035},\ref{inst:0343}}
\and T.A.      ~Lister                        \inst{\ref{inst:0344}}
\and E.        ~Livanou                       \inst{\ref{inst:0112}}
\and A.        ~Lobel                         \inst{\ref{inst:0081}}
\and W.        ~L\"{ o}ffler                  \inst{\ref{inst:0009}}
\and M.        ~L\'{o}pez                     \inst{\ref{inst:0163}}
\and D.        ~Lorenz                        \inst{\ref{inst:0143}}
\and I.        ~MacDonald                     \inst{\ref{inst:0091}}
\and T.        ~Magalh\~{a}es Fernandes       \inst{\ref{inst:0248}}
\and S.        ~Managau                       \inst{\ref{inst:0137}}
\and R.G.      ~Mann                          \inst{\ref{inst:0091}}
\and G.        ~Mantelet                      \inst{\ref{inst:0009}}
\and O.        ~Marchal                       \inst{\ref{inst:0007}}
\and J.M.      ~Marchant                      \inst{\ref{inst:0356}}
\and M.        ~Marconi                       \inst{\ref{inst:0224}}
\and S.        ~Marinoni                      \inst{\ref{inst:0201},\ref{inst:0122}}
\and P.M.      ~Marrese                       \inst{\ref{inst:0201},\ref{inst:0122}}
\and G.        ~Marschalk\'{o}                \inst{\ref{inst:0362},\ref{inst:0363}}
\and D.J.      ~Marshall                      \inst{\ref{inst:0364}}
\and J.M.      ~Mart\'{i}n-Fleitas            \inst{\ref{inst:0109}}
\and M.        ~Martino                       \inst{\ref{inst:0042}}
\and N.        ~Mary                          \inst{\ref{inst:0137}}
\and G.        ~Matijevi\v{c}                 \inst{\ref{inst:0241}}
\and T.        ~Mazeh                         \inst{\ref{inst:0247}}
\and P.J.      ~McMillan                      \inst{\ref{inst:0018}}
\and S.        ~Messina                       \inst{\ref{inst:0115}}
\and D.        ~Michalik                      \inst{\ref{inst:0018}}
\and N.R.      ~Millar                        \inst{\ref{inst:0011}}
\and B.M.H.    ~Miranda                       \inst{\ref{inst:0117}}
\and D.        ~Molina                        \inst{\ref{inst:0014}}
\and R.        ~Molinaro                      \inst{\ref{inst:0224}}
\and M.        ~Molinaro                      \inst{\ref{inst:0377}}
\and L.        ~Moln\'{a}r                    \inst{\ref{inst:0362}}
\and M.        ~Moniez                        \inst{\ref{inst:0379}}
\and P.        ~Montegriffo                   \inst{\ref{inst:0043}}
\and R.        ~Mor                           \inst{\ref{inst:0014}}
\and A.        ~Mora                          \inst{\ref{inst:0109}}
\and R.        ~Morbidelli                    \inst{\ref{inst:0006}}
\and T.        ~Morel                         \inst{\ref{inst:0094}}
\and S.        ~Morgenthaler                  \inst{\ref{inst:0385}}
\and D.        ~Morris                        \inst{\ref{inst:0091}}
\and A.F.      ~Mulone                        \inst{\ref{inst:0042}}
\and T.        ~Muraveva                      \inst{\ref{inst:0043}}
\and I.        ~Musella                       \inst{\ref{inst:0224}}
\and J.        ~Narbonne                      \inst{\ref{inst:0137}}
\and G.        ~Nelemans                      \inst{\ref{inst:0032},\ref{inst:0031}}
\and L.        ~Nicastro                      \inst{\ref{inst:0393}}
\and L.        ~Noval                         \inst{\ref{inst:0137}}
\and C.        ~Ord\'{e}novic                 \inst{\ref{inst:0005}}
\and J.        ~Ordieres-Mer\'{e}             \inst{\ref{inst:0396}}
\and P.        ~Osborne                       \inst{\ref{inst:0011}}
\and C.        ~Pagani                        \inst{\ref{inst:0165}}
\and I.        ~Pagano                        \inst{\ref{inst:0115}}
\and F.        ~Pailler                       \inst{\ref{inst:0021}}
\and H.        ~Palacin                       \inst{\ref{inst:0137}}
\and L.        ~Palaversa                     \inst{\ref{inst:0012}}
\and P.        ~Parsons                       \inst{\ref{inst:0026}}
\and M.        ~Pecoraro                      \inst{\ref{inst:0140}}
\and R.        ~Pedrosa                       \inst{\ref{inst:0405}}
\and H.        ~Pentik\"{ a}inen              \inst{\ref{inst:0118}}
\and B.        ~Pichon                        \inst{\ref{inst:0005}}
\and A.M.      ~Piersimoni                    \inst{\ref{inst:0222}}
\and F.-X.     ~Pineau                        \inst{\ref{inst:0101}}
\and E.        ~Plachy                        \inst{\ref{inst:0362}}
\and G.        ~Plum                          \inst{\ref{inst:0007}}
\and E.        ~Poujoulet                     \inst{\ref{inst:0412}}
\and A.        ~Pr\v{s}a                      \inst{\ref{inst:0413}}
\and L.        ~Pulone                        \inst{\ref{inst:0201}}
\and S.        ~Ragaini                       \inst{\ref{inst:0043}}
\and S.        ~Rago                          \inst{\ref{inst:0006}}
\and N.        ~Rambaux                       \inst{\ref{inst:0080}}
\and M.        ~Ramos-Lerate                  \inst{\ref{inst:0418}}
\and P.        ~Ranalli                       \inst{\ref{inst:0018}}
\and G.        ~Rauw                          \inst{\ref{inst:0094}}
\and A.        ~Read                          \inst{\ref{inst:0165}}
\and S.        ~Regibo                        \inst{\ref{inst:0031}}
\and C.        ~Reyl\'{e}                     \inst{\ref{inst:0124}}
\and R.A.      ~Ribeiro                       \inst{\ref{inst:0248}}
\and L.        ~Rimoldini                     \inst{\ref{inst:0055}}
\and V.        ~Ripepi                        \inst{\ref{inst:0224}}
\and A.        ~Riva                          \inst{\ref{inst:0006}}
\and G.        ~Rixon                         \inst{\ref{inst:0011}}
\and M.        ~Roelens                       \inst{\ref{inst:0012}}
\and M.        ~Romero-G\'{o}mez              \inst{\ref{inst:0014}}
\and N.        ~Rowell                        \inst{\ref{inst:0091}}
\and F.        ~Royer                         \inst{\ref{inst:0007}}
\and L.        ~Ruiz-Dern                     \inst{\ref{inst:0007}}
\and G.        ~Sadowski                      \inst{\ref{inst:0022}}
\and T.        ~Sagrist\`{a} Sell\'{e}s       \inst{\ref{inst:0009}}
\and J.        ~Sahlmann                      \inst{\ref{inst:0017}}
\and J.        ~Salgado                       \inst{\ref{inst:0120}}
\and E.        ~Salguero                      \inst{\ref{inst:0120}}
\and M.        ~Sarasso                       \inst{\ref{inst:0006}}
\and H.        ~Savietto                      \inst{\ref{inst:0440}}
\and M.        ~Schultheis                    \inst{\ref{inst:0005}}
\and E.        ~Sciacca                       \inst{\ref{inst:0115}}
\and M.        ~Segol                         \inst{\ref{inst:0443}}
\and J.C.      ~Segovia                       \inst{\ref{inst:0046}}
\and D.        ~Segransan                     \inst{\ref{inst:0012}}
\and I-C.      ~Shih                          \inst{\ref{inst:0007}}
\and R.        ~Smareglia                     \inst{\ref{inst:0377}}
\and R.L.      ~Smart                         \inst{\ref{inst:0006}}
\and E.        ~Solano                        \inst{\ref{inst:0163},\ref{inst:0450}}
\and F.        ~Solitro                       \inst{\ref{inst:0042}}
\and R.        ~Sordo                         \inst{\ref{inst:0002}}
\and S.        ~Soria Nieto                   \inst{\ref{inst:0014}}
\and J.        ~Souchay                       \inst{\ref{inst:0075}}
\and A.        ~Spagna                        \inst{\ref{inst:0006}}
\and F.        ~Spoto                         \inst{\ref{inst:0005}}
\and U.        ~Stampa                        \inst{\ref{inst:0009}}
\and I.A.      ~Steele                        \inst{\ref{inst:0356}}
\and H.        ~Steidelm\"{ u}ller            \inst{\ref{inst:0016}}
\and C.A.      ~Stephenson                    \inst{\ref{inst:0026}}
\and H.        ~Stoev                         \inst{\ref{inst:0461}}
\and F.F.      ~Suess                         \inst{\ref{inst:0011}}
\and M.        ~S\"{ u}veges                  \inst{\ref{inst:0055}}
\and J.        ~Surdej                        \inst{\ref{inst:0094}}
\and L.        ~Szabados                      \inst{\ref{inst:0362}}
\and E.        ~Szegedi-Elek                  \inst{\ref{inst:0362}}
\and D.        ~Tapiador                      \inst{\ref{inst:0467},\ref{inst:0468}}
\and F.        ~Taris                         \inst{\ref{inst:0075}}
\and G.        ~Tauran                        \inst{\ref{inst:0137}}
\and M.B.      ~Taylor                        \inst{\ref{inst:0471}}
\and R.        ~Teixeira                      \inst{\ref{inst:0225}}
\and D.        ~Terrett                       \inst{\ref{inst:0059}}
\and B.        ~Tingley                       \inst{\ref{inst:0474}}
\and S.C.      ~Trager                        \inst{\ref{inst:0185}}
\and C.        ~Turon                         \inst{\ref{inst:0007}}
\and A.        ~Ulla                          \inst{\ref{inst:0477}}
\and E.        ~Utrilla                       \inst{\ref{inst:0109}}
\and G.        ~Valentini                     \inst{\ref{inst:0222}}
\and A.        ~van Elteren                   \inst{\ref{inst:0001}}
\and E.        ~Van Hemelryck                 \inst{\ref{inst:0081}}
\and M.        ~van Leeuwen                   \inst{\ref{inst:0011}}
\and M.        ~Varadi                        \inst{\ref{inst:0012},\ref{inst:0362}}
\and A.        ~Vecchiato                     \inst{\ref{inst:0006}}
\and J.        ~Veljanoski                    \inst{\ref{inst:0185}}
\and T.        ~Via                           \inst{\ref{inst:0195}}
\and D.        ~Vicente                       \inst{\ref{inst:0276}}
\and S.        ~Vogt                          \inst{\ref{inst:0489}}
\and H.        ~Voss                          \inst{\ref{inst:0014}}
\and V.        ~Votruba                       \inst{\ref{inst:0262}}
\and S.        ~Voutsinas                     \inst{\ref{inst:0091}}
\and G.        ~Walmsley                      \inst{\ref{inst:0021}}
\and M.        ~Weiler                        \inst{\ref{inst:0014}}
\and K.        ~Weingrill                     \inst{\ref{inst:0241}}
\and T.        ~Wevers                        \inst{\ref{inst:0032}}
\and \L{}.     ~Wyrzykowski                   \inst{\ref{inst:0011},\ref{inst:0498}}
\and A.        ~Yoldas                        \inst{\ref{inst:0011}}
\and M.        ~\v{Z}erjal                    \inst{\ref{inst:0312}}
\and S.        ~Zucker                        \inst{\ref{inst:0238}}
\and C.        ~Zurbach                       \inst{\ref{inst:0110}}
\and T.        ~Zwitter                       \inst{\ref{inst:0312}}
\and A.        ~Alecu                         \inst{\ref{inst:0011}}
\and M.        ~Allen                         \inst{\ref{inst:0003}}
\and C.        ~Allende Prieto                \inst{\ref{inst:0034},\ref{inst:0507},\ref{inst:0508}}
\and A.        ~Amorim                        \inst{\ref{inst:0117}}
\and G.        ~Anglada-Escud\'{e}            \inst{\ref{inst:0014}}
\and V.        ~Arsenijevic                   \inst{\ref{inst:0117}}
\and S.        ~Azaz                          \inst{\ref{inst:0003}}
\and P.        ~Balm                          \inst{\ref{inst:0026}}
\and M.        ~Beck                          \inst{\ref{inst:0055}}
\and H.-H.     ~Bernstein$^\dagger$           \inst{\ref{inst:0009}}
\and L.        ~Bigot                         \inst{\ref{inst:0005}}
\and A.        ~Bijaoui                       \inst{\ref{inst:0005}}
\and C.        ~Blasco                        \inst{\ref{inst:0518}}
\and M.        ~Bonfigli                      \inst{\ref{inst:0222}}
\and G.        ~Bono                          \inst{\ref{inst:0201}}
\and S.        ~Boudreault                    \inst{\ref{inst:0034},\ref{inst:0522}}
\and A.        ~Bressan                       \inst{\ref{inst:0523}}
\and S.        ~Brown                         \inst{\ref{inst:0011}}
\and P.-M.     ~Brunet                        \inst{\ref{inst:0021}}
\and P.        ~Bunclark$^\dagger$            \inst{\ref{inst:0011}}
\and R.        ~Buonanno                      \inst{\ref{inst:0201}}
\and A.G.      ~Butkevich                     \inst{\ref{inst:0016}}
\and C.        ~Carret                        \inst{\ref{inst:0405}}
\and C.        ~Carrion                       \inst{\ref{inst:0127}}
\and L.        ~Chemin                        \inst{\ref{inst:0027},\ref{inst:0532}}
\and F.        ~Ch\'{e}reau                   \inst{\ref{inst:0007}}
\and L.        ~Corcione                      \inst{\ref{inst:0006}}
\and E.        ~Darmigny                      \inst{\ref{inst:0021}}
\and K.S.      ~de Boer                       \inst{\ref{inst:0536}}
\and P.        ~de Teodoro                    \inst{\ref{inst:0046}}
\and P.T.      ~de Zeeuw                      \inst{\ref{inst:0001},\ref{inst:0539}}
\and C.        ~Delle Luche                   \inst{\ref{inst:0007},\ref{inst:0137}}
\and C.D.      ~Domingues                     \inst{\ref{inst:0542}}
\and P.        ~Dubath                        \inst{\ref{inst:0055}}
\and F.        ~Fodor                         \inst{\ref{inst:0021}}
\and B.        ~Fr\'{e}zouls                  \inst{\ref{inst:0021}}
\and A.        ~Fries                         \inst{\ref{inst:0014}}
\and D.        ~Fustes                        \inst{\ref{inst:0116}}
\and D.        ~Fyfe                          \inst{\ref{inst:0165}}
\and E.        ~Gallardo                      \inst{\ref{inst:0014}}
\and J.        ~Gallegos                      \inst{\ref{inst:0046}}
\and D.        ~Gardiol                       \inst{\ref{inst:0006}}
\and M.        ~Gebran                        \inst{\ref{inst:0014},\ref{inst:0553}}
\and A.        ~Gomboc                        \inst{\ref{inst:0312},\ref{inst:0555}}
\and A.        ~G\'{o}mez                     \inst{\ref{inst:0007}}
\and E.        ~Grux                          \inst{\ref{inst:0124}}
\and A.        ~Gueguen                       \inst{\ref{inst:0007},\ref{inst:0559}}
\and A.        ~Heyrovsky                     \inst{\ref{inst:0091}}
\and J.        ~Hoar                          \inst{\ref{inst:0017}}
\and G.        ~Iannicola                     \inst{\ref{inst:0201}}
\and Y.        ~Isasi Parache                 \inst{\ref{inst:0014}}
\and A.-M.     ~Janotto                       \inst{\ref{inst:0021}}
\and E.        ~Joliet                        \inst{\ref{inst:0093},\ref{inst:0566}}
\and A.        ~Jonckheere                    \inst{\ref{inst:0081}}
\and R.        ~Keil                          \inst{\ref{inst:0568},\ref{inst:0569}}
\and D.-W.     ~Kim                           \inst{\ref{inst:0008}}
\and P.        ~Klagyivik                     \inst{\ref{inst:0362}}
\and J.        ~Klar                          \inst{\ref{inst:0241}}
\and J.        ~Knude                         \inst{\ref{inst:0035}}
\and O.        ~Kochukhov                     \inst{\ref{inst:0113}}
\and I.        ~Kolka                         \inst{\ref{inst:0575}}
\and J.        ~Kos                           \inst{\ref{inst:0312},\ref{inst:0577}}
\and A.        ~Kutka                         \inst{\ref{inst:0262},\ref{inst:0579}}
\and V.        ~Lainey                        \inst{\ref{inst:0080}}
\and D.        ~LeBouquin                     \inst{\ref{inst:0137}}
\and C.        ~Liu                           \inst{\ref{inst:0008},\ref{inst:0583}}
\and D.        ~Loreggia                      \inst{\ref{inst:0006}}
\and V.V.      ~Makarov                       \inst{\ref{inst:0585}}
\and M.G.      ~Marseille                     \inst{\ref{inst:0137}}
\and C.        ~Martayan                      \inst{\ref{inst:0081},\ref{inst:0588}}
\and O.        ~Martinez-Rubi                 \inst{\ref{inst:0014}}
\and B.        ~Massart                       \inst{\ref{inst:0005},\ref{inst:0137},\ref{inst:0592}}
\and F.        ~Meynadier                     \inst{\ref{inst:0007},\ref{inst:0075}}
\and S.        ~Mignot                        \inst{\ref{inst:0007}}
\and U.        ~Munari                        \inst{\ref{inst:0002}}
\and A.-T.     ~Nguyen                        \inst{\ref{inst:0021}}
\and T.        ~Nordlander                    \inst{\ref{inst:0113}}
\and P.        ~Ocvirk                        \inst{\ref{inst:0241},\ref{inst:0101}}
\and K.S.      ~O'Flaherty                    \inst{\ref{inst:0601}}
\and A.        ~Olias Sanz                    \inst{\ref{inst:0602}}
\and P.        ~Ortiz                         \inst{\ref{inst:0165}}
\and J.        ~Osorio                        \inst{\ref{inst:0153}}
\and D.        ~Oszkiewicz                    \inst{\ref{inst:0118},\ref{inst:0606}}
\and A.        ~Ouzounis                      \inst{\ref{inst:0091}}
\and M.        ~Palmer                        \inst{\ref{inst:0014}}
\and P.        ~Park                          \inst{\ref{inst:0012}}
\and E.        ~Pasquato                      \inst{\ref{inst:0022}}
\and C.        ~Peltzer                       \inst{\ref{inst:0011}}
\and J.        ~Peralta                       \inst{\ref{inst:0014}}
\and F.        ~P\'{e}turaud                  \inst{\ref{inst:0007}}
\and T.        ~Pieniluoma                    \inst{\ref{inst:0118}}
\and E.        ~Pigozzi                       \inst{\ref{inst:0042}}
\and J.        ~Poels$^\dagger$               \inst{\ref{inst:0094}}
\and G.        ~Prat                          \inst{\ref{inst:0617}}
\and T.        ~Prod'homme                    \inst{\ref{inst:0001},\ref{inst:0619}}
\and F.        ~Raison                        \inst{\ref{inst:0620},\ref{inst:0559}}
\and J.M.      ~Rebordao                      \inst{\ref{inst:0542}}
\and D.        ~Risquez                       \inst{\ref{inst:0001}}
\and B.        ~Rocca-Volmerange              \inst{\ref{inst:0624}}
\and S.        ~Rosen                         \inst{\ref{inst:0034},\ref{inst:0165}}
\and M.I.      ~Ruiz-Fuertes                  \inst{\ref{inst:0055}}
\and F.        ~Russo                         \inst{\ref{inst:0006}}
\and S.        ~Sembay                        \inst{\ref{inst:0165}}
\and I.        ~Serraller Vizcaino            \inst{\ref{inst:0630}}
\and A.        ~Short                         \inst{\ref{inst:0003}}
\and A.        ~Siebert                       \inst{\ref{inst:0101},\ref{inst:0241}}
\and H.        ~Silva                         \inst{\ref{inst:0248}}
\and D.        ~Sinachopoulos                 \inst{\ref{inst:0078}}
\and E.        ~Slezak                        \inst{\ref{inst:0005}}
\and M.        ~Soffel                        \inst{\ref{inst:0016}}
\and D.        ~Sosnowska                     \inst{\ref{inst:0012}}
\and V.        ~Strai\v{z}ys                  \inst{\ref{inst:0639}}
\and M.        ~ter Linden                    \inst{\ref{inst:0093},\ref{inst:0641}}
\and D.        ~Terrell                       \inst{\ref{inst:0642}}
\and S.        ~Theil                         \inst{\ref{inst:0643}}
\and C.        ~Tiede                         \inst{\ref{inst:0008},\ref{inst:0645}}
\and L.        ~Troisi                        \inst{\ref{inst:0122},\ref{inst:0647}}
\and P.        ~Tsalmantza                    \inst{\ref{inst:0008}}
\and D.        ~Tur                           \inst{\ref{inst:0195}}
\and M.        ~Vaccari                       \inst{\ref{inst:0650},\ref{inst:0651}}
\and F.        ~Vachier                       \inst{\ref{inst:0080}}
\and P.        ~Valles                        \inst{\ref{inst:0014}}
\and W.        ~Van Hamme                     \inst{\ref{inst:0654}}
\and L.        ~Veltz                         \inst{\ref{inst:0241},\ref{inst:0089}}
\and J.        ~Virtanen                      \inst{\ref{inst:0118},\ref{inst:0119}}
\and J.-M.     ~Wallut                        \inst{\ref{inst:0021}}
\and R.        ~Wichmann                      \inst{\ref{inst:0660}}
\and M.I.      ~Wilkinson                     \inst{\ref{inst:0011},\ref{inst:0165}}
\and H.        ~Ziaeepour                     \inst{\ref{inst:0124}}
\and S.        ~Zschocke                      \inst{\ref{inst:0016}}
}

\institute{
     Leiden Observatory, Leiden University, Niels Bohrweg 2, 2333 CA Leiden, The Netherlands\relax                                                                                                           \label{inst:0001}
\and INAF - Osservatorio astronomico di Padova, Vicolo Osservatorio 5, 35122 Padova, Italy\relax                                                                                                             \label{inst:0002}
\and Scientific Support Office, Directorate of Science, European Space Research and Technology Centre (ESA/ESTEC), Keplerlaan 1, 2201AZ, Noordwijk, The Netherlands\relax                                    \label{inst:0003}
\and Laboratoire Lagrange, Universit\'{e} Nice Sophia-Antipolis, Observatoire de la C\^{o}te d'Azur, CNRS, CS 34229, F-06304 Nice Cedex, France\relax                                                        \label{inst:0005}
\and INAF - Osservatorio Astrofisico di Torino, via Osservatorio 20, 10025 Pino Torinese (TO), Italy\relax                                                                                                   \label{inst:0006}
\and GEPI, Observatoire de Paris, PSL Research University, CNRS, Univ. Paris Diderot, Sorbonne Paris Cit{\'e}, 5 Place Jules Janssen, 92190 Meudon, France\relax                                             \label{inst:0007}
\and Max Planck Institute for Astronomy, K\"{ o}nigstuhl 17, 69117 Heidelberg, Germany\relax                                                                                                                 \label{inst:0008}
\and Astronomisches Rechen-Institut, Zentrum f\"{ u}r Astronomie der Universit\"{ a}t Heidelberg, M\"{ o}nchhofstr. 12-14, D-69120 Heidelberg, Germany\relax                                                 \label{inst:0009}
\and Institute of Astronomy, University of Cambridge, Madingley Road, Cambridge CB3 0HA, United Kingdom\relax                                                                                                \label{inst:0011}
\and Department of Astronomy, University of Geneva, Chemin des Maillettes 51, CH-1290 Versoix, Switzerland\relax                                                                                             \label{inst:0012}
\and Mission Operations Division, Operations Department, Directorate of Science, European Space Research and Technology Centre (ESA/ESTEC), Keplerlaan 1, 2201 AZ, Noordwijk, The Netherlands\relax          \label{inst:0013}
\and Institut de Ci\`{e}ncies del Cosmos, Universitat  de  Barcelona  (IEEC-UB), Mart\'{i}  Franqu\`{e}s  1, E-08028 Barcelona, Spain\relax                                                                  \label{inst:0014}
\and Lohrmann Observatory, Technische Universit\"{ a}t Dresden, Mommsenstra{\ss}e 13, 01062 Dresden, Germany\relax                                                                                           \label{inst:0016}
\and European Space Astronomy Centre (ESA/ESAC), Camino bajo del Castillo, s/n, Urbanizacion Villafranca del Castillo, Villanueva de la Ca\~{n}ada, E-28692 Madrid, Spain\relax                              \label{inst:0017}
\and Lund Observatory, Department of Astronomy and Theoretical Physics, Lund University, Box 43, SE-22100 Lund, Sweden\relax                                                                                 \label{inst:0018}
\and CNES Centre Spatial de Toulouse, 18 avenue Edouard Belin, 31401 Toulouse Cedex 9, France\relax                                                                                                          \label{inst:0021}
\and Institut d'Astronomie et d'Astrophysique, Universit\'{e} Libre de Bruxelles CP 226, Boulevard du Triomphe, 1050 Brussels, Belgium\relax                                                                 \label{inst:0022}
\and F.R.S.-FNRS, Rue d'Egmont 5, 1000 Brussels, Belgium\relax                                                                                                                                               \label{inst:0023}
\and INAF - Osservatorio Astrofisico di Arcetri, Largo Enrico Fermi 5, I-50125 Firenze, Italy\relax                                                                                                          \label{inst:0024}
\and Telespazio Vega UK Ltd for ESA/ESAC, Camino bajo del Castillo, s/n, Urbanizacion Villafranca del Castillo, Villanueva de la Ca\~{n}ada, E-28692 Madrid, Spain\relax                                     \label{inst:0026}
\and Laboratoire d'astrophysique de Bordeaux, Universit\'{e} de Bordeaux, CNRS, B18N, all{\'e}e Geoffroy Saint-Hilaire, 33615 Pessac, France\relax                                                           \label{inst:0027}
\and Instituut voor Sterrenkunde, KU Leuven, Celestijnenlaan 200D, 3001 Leuven, Belgium\relax                                                                                                                \label{inst:0031}
\and Department of Astrophysics/IMAPP, Radboud University Nijmegen, P.O.Box 9010, 6500 GL Nijmegen, The Netherlands\relax                                                                                    \label{inst:0032}
\and Mullard Space Science Laboratory, University College London, Holmbury St Mary, Dorking, Surrey RH5 6NT, United Kingdom\relax                                                                            \label{inst:0034}
\and Niels Bohr Institute, University of Copenhagen, Juliane Maries Vej 30, 2100 Copenhagen {\O}, Denmark\relax                                                                                              \label{inst:0035}
\and Centre for Electronic Imaging, Department of Physical Sciences, The Open University, Walton Hall MK7 6AA Milton Keynes, United Kingdom\relax                                                            \label{inst:0038}
\newpage
\and ALTEC S.p.a, Corso Marche, 79,10146 Torino, Italy\relax                                                                                                                                                 \label{inst:0042}
\and INAF - Osservatorio Astronomico di Bologna, via Ranzani 1, 40127 Bologna,  Italy\relax                                                                                                                  \label{inst:0043}
\and Serco Gesti\'{o}n de Negocios for ESA/ESAC, Camino bajo del Castillo, s/n, Urbanizacion Villafranca del Castillo, Villanueva de la Ca\~{n}ada, E-28692 Madrid, Spain\relax                              \label{inst:0046}
\and Department of Astronomy, University of Geneva, Chemin d'Ecogia 16, CH-1290 Versoix, Switzerland\relax                                                                                                   \label{inst:0055}
\and STFC, Rutherford Appleton Laboratory, Harwell, Didcot, OX11 0QX, United Kingdom\relax                                                                                                                   \label{inst:0059}
\and Gaia DPAC Project Office, ESAC, Camino bajo del Castillo, s/n, Urbanizacion Villafranca del Castillo, Villanueva de la Ca\~{n}ada, E-28692 Madrid, Spain\relax                                          \label{inst:0065}
\and SYRTE, Observatoire de Paris, PSL Research University, CNRS, Sorbonne Universit{\'e}s, UPMC Univ. Paris 06, LNE, 61 avenue de l'Observatoire, 75014 Paris, France\relax                                 \label{inst:0075}
\and National Observatory of Athens, I. Metaxa and Vas. Pavlou, Palaia Penteli, 15236 Athens, Greece\relax                                                                                                   \label{inst:0078}
\and IMCCE, Observatoire de Paris, PSL Research University, CNRS, Sorbonne Universit{\'e}s, UPMC Univ. Paris 06, Univ. Lille, 77 av. Denfert-Rochereau, 75014 Paris, France\relax                            \label{inst:0080}
\and Royal Observatory of Belgium, Ringlaan 3, 1180 Brussels, Belgium\relax                                                                                                                                  \label{inst:0081}
\and Institut d'Astrophysique Spatiale, Universit\'{e} Paris XI, UMR 8617, CNRS, B\^{a}timent 121, 91405, Orsay Cedex, France\relax                                                                          \label{inst:0089}
\and Institute for Astronomy, Royal Observatory, University of Edinburgh, Blackford Hill, Edinburgh EH9 3HJ, United Kingdom\relax                                                                            \label{inst:0091}
\and HE Space Operations BV for ESA/ESAC, Camino bajo del Castillo, s/n, Urbanizacion Villafranca del Castillo, Villanueva de la Ca\~{n}ada, E-28692 Madrid, Spain\relax                                     \label{inst:0093}
\and Institut d'Astrophysique et de G\'{e}ophysique, Universit\'{e} de Li\`{e}ge, 19c, All\'{e}e du 6 Ao\^{u}t, B-4000 Li\`{e}ge, Belgium\relax                                                              \label{inst:0094}
\and \'{A}rea de Lenguajes y Sistemas Inform\'{a}ticos, Universidad Pablo de Olavide, Ctra. de Utrera, km 1. 41013, Sevilla, Spain\relax                                                                     \label{inst:0098}
\and Observatoire Astronomique de Strasbourg, Universit\'{e} de Strasbourg, CNRS, UMR 7550, 11 rue de l'Universit\'{e}, 67000 Strasbourg, France\relax                                                       \label{inst:0101}
\and Kavli Institute for Cosmology, University of Cambridge, Madingley Road, Cambridge CB3 0HA, United Kingdom\relax                                                                                          \label{inst:0104}
\and Aurora Technology for ESA/ESAC, Camino bajo del Castillo, s/n, Urbanizacion Villafranca del Castillo, Villanueva de la Ca\~{n}ada, E-28692 Madrid, Spain\relax                                          \label{inst:0109}
\and Laboratoire Univers et Particules de Montpellier, Universit\'{e} Montpellier, Place Eug\`{e}ne Bataillon, CC72, 34095 Montpellier Cedex 05, France\relax                                                \label{inst:0110}
\and Department of Astrophysics, Astronomy and Mechanics, National and Kapodistrian University of Athens, Panepistimiopolis, Zografos, 15783 Athens, Greece\relax                                            \label{inst:0112}
\and Department of Physics and Astronomy, Division of Astronomy and Space Physics, Uppsala University, Box 516, 75120 Uppsala, Sweden\relax                                                                  \label{inst:0113}
\and Universit\`{a} di Catania, Dipartimento di Fisica e Astronomia, Sezione Astrofisica, Via S. Sofia 78, I-95123 Catania, Italy\relax                                                                      \label{inst:0114}
\and INAF - Osservatorio Astrofisico di Catania, via S. Sofia 78, 95123 Catania, Italy\relax                                                                                                                 \label{inst:0115}
\and Universidade da Coru\~{n}a, Facultade de Inform\'{a}tica, Campus de Elvi\~{n}a S/N, 15071, A Coru\~{n}a, Spain\relax                                                                                    \label{inst:0116}
\and CENTRA, Universidade de Lisboa, FCUL, Campo Grande, Edif. C8, 1749-016 Lisboa, Portugal\relax                                                                                                           \label{inst:0117}
\and University of Helsinki, Department of Physics, P.O. Box 64, FI-00014 University of Helsinki, Finland\relax                                                                                              \label{inst:0118}
\and Finnish Geospatial Research Institute FGI, Geodeetinrinne 2, FI-02430 Masala, Finland\relax                                                                                                             \label{inst:0119}
\and Isdefe for ESA/ESAC, Camino bajo del Castillo, s/n, Urbanizacion Villafranca del Castillo, Villanueva de la Ca\~{n}ada, E-28692 Madrid, Spain\relax                                                     \label{inst:0120}
\and ASI Science Data Center, via del Politecnico SNC, 00133 Roma, Italy\relax                                                                                                                               \label{inst:0122}
\newpage
\and Institut UTINAM UMR6213, CNRS, OSU THETA Franche-Comt\'{e} Bourgogne, Universit\'{e} Bourgogne Franche-Comt\'{e}, F-25000 Besan\c{c}on, France\relax                                                    \label{inst:0124}
\and Dpto. de Inteligencia Artificial, UNED, c/ Juan del Rosal 16, 28040 Madrid, Spain\relax                                                                                                                 \label{inst:0127}
\and Elecnor Deimos Space for ESA/ESAC, Camino bajo del Castillo, s/n, Urbanizacion Villafranca del Castillo, Villanueva de la Ca\~{n}ada, E-28692 Madrid, Spain\relax                                       \label{inst:0136}
\and Thales Services for CNES Centre Spatial de Toulouse, 18 avenue Edouard Belin, 31401 Toulouse Cedex 9, France\relax                                                                                      \label{inst:0137}
\and EURIX S.r.l., via Carcano 26, 10153, Torino, Italy\relax                                                                                                                                                \label{inst:0140}
\and University of Vienna, Department of Astrophysics, T\"{ u}rkenschanzstra{\ss}e 17, A1180 Vienna, Austria\relax                                                                                           \label{inst:0143}
\and Department of Physics and Astronomy, The Johns Hopkins University, 3400 N Charles St, Baltimore, MD 21218, USA\relax                                                                                    \label{inst:0144}
\and ON/MCTI-BR, Rua Gal. Jos\'{e} Cristino 77, Rio de Janeiro, CEP 20921-400, RJ,  Brazil\relax                                                                                                             \label{inst:0146}
\and OV/UFRJ-BR, Ladeira Pedro Ant\^{o}nio 43, Rio de Janeiro, CEP 20080-090, RJ, Brazil\relax                                                                                                               \label{inst:0147}
\and Faculdade Ciencias, Universidade do Porto, Departamento Matematica Aplicada, Rua do Campo Alegre, 687 4169-007 Porto, Portugal\relax                                                                    \label{inst:0153}
\and Instituto de Astrof\'{\i}sica e Ci\^encias do Espa\,co, Universidade de Lisboa Faculdade de Ci\^encias, Campo Grande, PT1749-016 Lisboa, Portugal\relax                                                 \label{inst:0154}
\and Departamento de Astrof\'{i}sica, Centro de Astrobiolog\'{i}a (CSIC-INTA), ESA-ESAC. Camino Bajo del Castillo s/n. 28692 Villanueva de la Ca\~{n}ada, Madrid, Spain\relax                                \label{inst:0163}
\and Department of Physics and Astronomy, University of Leicester, University Road, Leicester LE1 7RH, United Kingdom\relax                                                                                  \label{inst:0165}
\and University of Oviedo, Campus Universitario, 33203 Gij\'{o}n, Spain\relax                                                                                                                                \label{inst:0168}
\and University of C\'{a}diz, Avd. De la universidad, Jerez de la Frontera, C\'{a}diz, Spain\relax                                                                                                           \label{inst:0171}
\and Kapteyn Astronomical Institute, University of Groningen, Landleven 12, 9747 AD Groningen, The Netherlands\relax                                                                                         \label{inst:0185}
\and Consorci de Serveis Universitaris de Catalunya, C/ Gran Capit\`{a}, 2-4 3rd floor, 08034 Barcelona, Spain\relax                                                                                         \label{inst:0195}
\and University of Turin, Department of Computer Sciences, Corso Svizzera 185, 10149 Torino, Italy\relax                                                                                                     \label{inst:0197}
\and INAF - Osservatorio Astronomico di Roma, Via di Frascati 33, 00078 Monte Porzio Catone (Roma), Italy\relax                                                                                              \label{inst:0201}
\and CRAAG - Centre de Recherche en Astronomie, Astrophysique et G\'{e}ophysique, Route de l'Observatoire Bp 63 Bouzareah 16340 Algiers, Algeria\relax                                                       \label{inst:0215}
\and Universiteit Antwerpen, Onderzoeksgroep Toegepaste Wiskunde, Middelheimlaan 1, 2020 Antwerpen, Belgium\relax                                                                                            \label{inst:0218}
\and Department of Physics and Astronomy, University of Padova, Via Marzolo 8, I-35131 Padova, Italy\relax                                                                                                   \label{inst:0220}
\and INAF - Osservatorio Astronomico di Teramo, Via Mentore Maggini, 64100 Teramo, Italy\relax                                                                                                               \label{inst:0222}
\and INAF - Osservatorio Astronomico di Capodimonte, Via Moiariello 16, 80131, Napoli, Italy\relax                                                                                                           \label{inst:0224}
\and Instituto de Astronomia, Geof\`{i}sica e Ci\^{e}ncias Atmosf\'{e}ricas, Universidade de S\~{a}o Paulo, Rua do Mat\~{a}o, 1226, Cidade Universitaria, 05508-900 S\~{a}o Paulo, SP, Brazil\relax          \label{inst:0225}
\and Department of Geosciences, Tel Aviv University, Tel Aviv 6997801, Israel\relax                                                                                                                          \label{inst:0238}
\and Astronomical Institute Anton Pannekoek, University of Amsterdam, PO Box 94249, 1090 GE, Amsterdam, The Netherlands\relax                                                                                \label{inst:0239}
\and Leibniz Institute for Astrophysics Potsdam (AIP), An der Sternwarte 16, 14482 Potsdam, Germany\relax                                                                                                    \label{inst:0241}
\and ATOS for CNES Centre Spatial de Toulouse, 18 avenue Edouard Belin, 31401 Toulouse Cedex 9, France\relax                                                                                                 \label{inst:0244}
\and School of Physics and Astronomy, Tel Aviv University, Tel Aviv 6997801, Israel\relax                                                                                                                    \label{inst:0247}
\and UNINOVA - CTS, Campus FCT-UNL, Monte da Caparica, 2829-516 Caparica, Portugal\relax                                                                                                                     \label{inst:0248}
\and Laboratoire G\'{e}oazur, Universit\'{e} Nice Sophia-Antipolis, UMR 7329, CNRS, Observatoire de la C\^{o}te d'Azur, 250 rue A. Einstein, F-06560 Valbonne, France\relax                                  \label{inst:0254}
\newpage
\and RHEA for ESA/ESAC, Camino bajo del Castillo, s/n, Urbanizacion Villafranca del Castillo, Villanueva de la Ca\~{n}ada, E-28692 Madrid, Spain\relax                                                       \label{inst:0260}
\and Astronomical Institute, Academy of Sciences of the Czech Republic, Fri\v{c}ova 298, 25165 Ond\v{r}ejov, Czech Republic\relax                                                                            \label{inst:0262}
\and Barcelona Supercomputing Center - Centro Nacional de Supercomputaci\'{o}n, c/ Jordi Girona 29, Ed. Nexus II, 08034 Barcelona, Spain\relax                                                               \label{inst:0276}
\and Department of Mechanical Engineering, University of La Rioja, c/ San Jos\'{e} de Calasanz, 31, 26004 Logro\~{n}o, La Rioja, Spain\relax                                                                 \label{inst:0279}
\and ETSE Telecomunicaci\'{o}n, Universidade de Vigo, Campus Lagoas-Marcosende, 36310 Vigo, Galicia, Spain\relax                                                                                             \label{inst:0281}
\and SRON, Netherlands Institute for Space Research, Sorbonnelaan 2, 3584CA, Utrecht, The Netherlands\relax                                                                                                  \label{inst:0306}
\and Faculty of Mathematics and Physics, University of Ljubljana, Jadranska ulica 19, 1000 Ljubljana, Slovenia\relax                                                                                         \label{inst:0312}
\and Physics Department, University of Antwerp, Groenenborgerlaan 171, 2020 Antwerp, Belgium\relax                                                                                                           \label{inst:0314}
\and Harvard-Smithsonian Center for Astrophysics, 60 Garden Street, Cambridge MA 02138, USA\relax                                                                                                            \label{inst:0316}
\and Institut de Physique de Rennes, Universit{\'e} de Rennes 1, F-35042 Rennes, France\relax                                                                                                                \label{inst:0331}
\and Shanghai Astronomical Observatory, Chinese Academy of Sciences, 80 Nandan Rd, 200030 Shanghai, China\relax                                                                                              \label{inst:0340}
\and CSC Danmark A/S, Retortvej 8, 2500 Valby, Denmark\relax                                                                                                                                                 \label{inst:0343}
\and Las Cumbres Observatory Global Telescope Network, Inc., 6740 Cortona Drive, Suite 102, Goleta, CA  93117, USA\relax                                                                                     \label{inst:0344}
\and Astrophysics Research Institute, Liverpool John Moores University, L3 5RF, United Kingdom\relax                                                                                                         \label{inst:0356}
\and Konkoly Observatory, Research Centre for Astronomy and Earth Sciences, Hungarian Academy of Sciences, Konkoly Thege Mikl\'{o}s \'{u}t 15-17, 1121 Budapest, Hungary\relax                               \label{inst:0362}
\and Baja Observatory of University of Szeged, Szegedi \'{u}t III/70, 6500 Baja, Hungary\relax                                                                                                               \label{inst:0363}
\and Laboratoire AIM, IRFU/Service d'Astrophysique - CEA/DSM - CNRS - Universit\'{e} Paris Diderot, B\^{a}t 709, CEA-Saclay, F-91191 Gif-sur-Yvette Cedex, France\relax                                      \label{inst:0364}
\and INAF - Osservatorio Astronomico di Trieste, Via G.B. Tiepolo 11, 34143, Trieste, Italy\relax                                                                                                            \label{inst:0377}
\and Laboratoire de l'Acc\'{e}l\'{e}rateur Lin\'{e}aire, Universit\'{e} Paris-Sud, CNRS/IN2P3, Universit\'{e} Paris-Saclay, 91898 Orsay Cedex, France\relax                                                  \label{inst:0379}
\and \'{E}cole polytechnique f\'{e}d\'{e}rale de Lausanne, SB MATHAA STAP, MA B1 473 (B\^{a}timent MA), Station 8, CH-1015 Lausanne, Switzerland\relax                                                       \label{inst:0385}
\and INAF/IASF-Bologna, Via P. Gobetti 101, 40129 Bologna, Italy\relax                                                                                                                                       \label{inst:0393}
\and Technical University of Madrid, Jos\'{e} Guti\'{e}rrez Abascal 2, 28006 Madrid, Spain\relax                                                                                                             \label{inst:0396}
\and EQUERT International for CNES Centre Spatial de Toulouse, 18 avenue Edouard Belin, 31401 Toulouse Cedex 9, France\relax                                                                                 \label{inst:0405}
\and AKKA for CNES Centre Spatial de Toulouse, 18 avenue Edouard Belin, 31401 Toulouse Cedex 9, France\relax                                                                                                 \label{inst:0412}
\and Villanova University, Dept. of Astrophysics and Planetary Science, 800 E Lancaster Ave, Villanova PA 19085, USA\relax                                                                                   \label{inst:0413}
\and Vitrociset Belgium for ESA/ESAC, Camino bajo del Castillo, s/n, Urbanizacion Villafranca del Castillo, Villanueva de la Ca\~{n}ada, E-28692 Madrid, Spain\relax                                         \label{inst:0418}
\and Fork Research, Rua do Cruzado Osberno, Lt. 1, 9 esq., Lisboa, Portugal\relax                                                                                                                            \label{inst:0440}
\and APAVE SUDEUROPE SAS for CNES Centre Spatial de Toulouse, 18 avenue Edouard Belin, 31401 Toulouse Cedex 9, France\relax                                                                                  \label{inst:0443}
\and Spanish Virtual Observatory\relax                                                                                                                                                                       \label{inst:0450}
\and Fundaci\'{o}n Galileo Galilei - INAF, Rambla Jos\'{e} Ana Fern\'{a}ndez P\'{e}rez 7, E-38712 Bre\~{n}a Baja, Santa Cruz de Tenerife, Spain\relax                                                        \label{inst:0461}
\and INSA for ESA/ESAC, Camino bajo del Castillo, s/n, Urbanizacion Villafranca del Castillo, Villanueva de la Ca\~{n}ada, E-28692 Madrid, Spain\relax                                                       \label{inst:0467}
\and Dpto. Arquitectura de Computadores y Autom\'{a}tica, Facultad de Inform\'{a}tica, Universidad Complutense de Madrid, C/ Prof. Jos\'{e} Garc\'{i}a Santesmases s/n, 28040 Madrid, Spain\relax            \label{inst:0468}
\newpage
\and H H Wills Physics Laboratory, University of Bristol, Tyndall Avenue, Bristol BS8 1TL, United Kingdom\relax                                                                                              \label{inst:0471}
\and Stellar Astrophysics Centre, Aarhus University, Department of Physics and Astronomy, 120 Ny Munkegade, Building 1520, DK-8000 Aarhus C, Denmark\relax                                                   \label{inst:0474}
\and Applied Physics Department, University of Vigo, E-36310 Vigo, Spain\relax                                                                                                                               \label{inst:0477}
\and HE Space Operations BV for ESA/ESTEC, Keplerlaan 1, 2201AZ, Noordwijk, The Netherlands\relax                                                                                                            \label{inst:0489}
\and Warsaw University Observatory, Al. Ujazdowskie 4, 00-478 Warszawa, Poland\relax                                                                                                                         \label{inst:0498}
\and Instituto de Astrof\'{\i}sica de Canarias, E-38205 La Laguna, Tenerife, Spain\relax                                                                                                                     \label{inst:0507}
\and Universidad de La Laguna, Departamento de Astrof\'{\i}sica, E-38206 La Laguna, Tenerife, Spain\relax                                                                                                    \label{inst:0508}
\and RHEA for ESA/ESTEC, Keplerlaan 1, 2201AZ, Noordwijk, The Netherlands\relax                                                                                                                              \label{inst:0518}
\and Max Planck Institute for Solar System Research, Justus-von-Liebig-Weg 3, 37077 G\"{ o}ttingen, Germany\relax                                                                                            \label{inst:0522}
\and SISSA (Scuola Internazionale Superiore di Studi Avanzati), via Bonomea 265, 34136 Trieste, Italy\relax                                                                                                  \label{inst:0523}
\and Instituto Nacional de Pesquisas Espaciais/Minist\'{e}rio da Ciencia Tecnologia, Avenida dos Astronautas 1758, S\~{a}o Jos\'{e} Dos Campos, SP 12227-010, Brazil\relax                                   \label{inst:0532}
\and Argelander Institut f\"{ u}r Astronomie der Universit\"{ a}t Bonn, Auf dem H\"{ u}gel 71, 53121 Bonn, Germany\relax                                                                                     \label{inst:0536}
\and European Southern Observatory (ESO), Karl-Schwarzschild-Stra{\ss}e 2, 85748 Garching bei M\"{ u}nchen, Germany\relax                                                                                    \label{inst:0539}
\and Laboratory of Optics, Lasers and Systems, Faculty of Sciences, University of Lisbon, Campus do Lumiar, Estrada do Pa\c{c}o do Lumiar, 22, 1649-038 Lisboa, Portugal\relax                               \label{inst:0542}
\and Department of Physics and Astronomy, Notre Dame University, Louaize, PO Box 72, Zouk Mika\"{ e}l, Lebanon\relax                                                                                         \label{inst:0553}
\and University of Nova Gorica, Vipavska 13, 5000 Nova Gorica, Slovenia\relax                                                                                                                                \label{inst:0555}
\and Max Planck Institute for Extraterrestrial Physics, OPINAS, Gie{\ss}enbachstra{\ss}e, 85741 Garching, Germany\relax                                                                                      \label{inst:0559}
\and NASA/IPAC Infrared Science Archive, California Institute of Technology, Mail Code 100-22, 770 South Wilson Avenue, Pasadena, CA, 91125, USA\relax                                                       \label{inst:0566}
\and Center of Applied Space Technology and Microgravity (ZARM), c/o Universit\"{ a}t Bremen, Am Fallturm 1, 28359 Bremen, Germany\relax                                                                     \label{inst:0568}
\and RHEA System for ESA/ESOC, Robert Bosch Stra{\ss}e 5, 64293 Darmstadt, Germany\relax                                                                                                                     \label{inst:0569}
\and Tartu Observatory, 61602 T\~{o}ravere, Estonia\relax                                                                                                                                                    \label{inst:0575}
\and Sydney Institute for Astronomy, School of Physics A28, The University of Sydney, NSW 2006, Australia\relax                                                                                              \label{inst:0577}
\and Slovak Organisation for Space Activities, Zamocka 18, 85101 Bratislava, Slovak Republic\relax                                                                                                           \label{inst:0579}
\and National Astronomical Observatories, CAS, 100012 Beijing, China\relax                                                                                                                                   \label{inst:0583}
\and US Naval Observatory, Astrometry Department, 3450 Massachusetts Ave. NW, Washington DC 20392-5420 D.C., USA\relax                                                                                       \label{inst:0585}
\and European Southern Observatory (ESO), Alonso de C\'{o}rdova 3107, Vitacura, Casilla 19001, Santiago de Chile, Chile\relax                                                                                \label{inst:0588}
\and Airbus Defence and Space SAS, 31 Rue des Cosmonautes, 31402 Toulouse Cedex 4, France\relax                                                                                                              \label{inst:0592}
\and EJR-Quartz BV for ESA/ESTEC, Keplerlaan 1, 2201AZ, Noordwijk, The Netherlands\relax                                                                                                                     \label{inst:0601}
\and The Server Labs for ESA/ESAC, Camino bajo del Castillo, s/n, Urbanizacion Villafranca del Castillo, Villanueva de la Ca\~{n}ada, E-28692 Madrid, Spain\relax                                            \label{inst:0602}
\and Astronomical Observatory Institute, Faculty of Physics, A. Mickiewicz University, ul. S\l{}oneczna 36, 60-286 Pozna\'{n}, Poland\relax                                                                  \label{inst:0606}
\and CS Syst\`{e}mes d'Information for CNES Centre Spatial de Toulouse, 18 avenue Edouard Belin, 31401 Toulouse Cedex 9, France\relax                                                                        \label{inst:0617}
\and Directorate of Science, European Space Research and Technology Centre (ESA/ESTEC), Keplerlaan 1, 2201AZ, Noordwijk, The Netherlands\relax                                                               \label{inst:0619}
\newpage
\and Praesepe BV for ESA/ESAC, Camino bajo del Castillo, s/n, Urbanizacion Villafranca del Castillo, Villanueva de la Ca\~{n}ada, E-28692 Madrid, Spain\relax                                                \label{inst:0620}
\and Sorbonne Universit\'{e}s UPMC et CNRS, UMR7095, Institut d'Astrophysique de Paris, F75014, Paris, France\relax                                                                                          \label{inst:0624}
\and GMV for ESA/ESAC, Camino bajo del Castillo, s/n, Urbanizacion Villafranca del Castillo, Villanueva de la Ca\~{n}ada, E-28692 Madrid, Spain\relax                                                        \label{inst:0630}
\and Institute of Theoretical Physics and Astronomy, Vilnius University, Sauletekio al. 3, Vilnius, LT-10222, Lithuania\relax                                                                                \label{inst:0639}
\and S[\&]T Corporation, PO Box 608, 2600 AP, Delft, The Netherlands\relax                                                                                                                                   \label{inst:0641}
\and Department of Space Studies, Southwest Research Institute (SwRI), 1050 Walnut Street, Suite 300, Boulder, Colorado 80302, USA\relax                                                                     \label{inst:0642}
\and Deutsches Zentrum f\"{ u}r Luft- und Raumfahrt, Institute of Space Systems, Am Fallturm 1, D-28359 Bremen, Germany\relax                                                                                \label{inst:0643}
\and University of Applied Sciences Munich, Karlstr. 6, 80333 Munich, Germany\relax                                                                                                                          \label{inst:0645}
\and Dipartimento di Fisica, Universit\`{a} di Roma Tor Vergata, via della Ricerca Scientifica 1, 00133 Rome, Italy\relax                                                                                    \label{inst:0647}
\and Department of Physics and Astronomy, University of the Western Cape, Robert Sobukwe Road, 7535 Bellville, Cape Town, South Africa\relax                                                                 \label{inst:0650}
\and INAF - Istituto di Radioastronomia, via Gobetti 101, 40129 Bologna, Italy\relax                                                                                                                         \label{inst:0651}
\and Department of Physics, Florida International University, 11200 SW 8th Street, Miami, FL 33199, USA\relax                                                                                                \label{inst:0654}
\and Hamburger Sternwarte, Gojenbergsweg 112, D-21029 Hamburg, Germany\relax                                                                                                                                 \label{inst:0660}
}

%% file: aa201629512.bbl
\begin{thebibliography}{45}
\expandafter\ifx\csname natexlab\endcsname\relax\def\natexlab#1{#1}\fi

\bibitem[{{Arenou} {et~al.}(2016){Arenou}, {Luri}, {Babusiaux}, {Fabricius},
  {Helmi}, \& {Robin}}]{DPACP-16}
{Arenou}, F., {Luri}, X., {Babusiaux}, C., {et~al.} 2016, \aap, this volume

\bibitem[{{Boch} \& {Fernique}(2014)}]{2014ASPC..485..277B}
{Boch}, T. \& {Fernique}, P. 2014, in Astronomical Society of the Pacific
  Conference Series, Vol. 485, Astronomical Data Analysis Software and Systems
  XXIII, ed. N.~{Manset} \& P.~{Forshay}, 277

\bibitem[{{Bonnarel} {et~al.}(2000){Bonnarel}, {Fernique}, {Bienaym{\'e}},
  {Egret}, {Genova}, {Louys}, {Ochsenbein}, {Wenger}, \&
  {Bartlett}}]{2000A&AS..143...33B}
{Bonnarel}, F., {Fernique}, P., {Bienaym{\'e}}, O., {et~al.} 2000, \aaps, 143,
  33

\bibitem[{{Campbell} {et~al.}(2015){Campbell}, {Marsh}, {Fraser}, {Hodgkin},
  {de Miguel}, {G{\"a}nsicke}, {Steeghs}, {Hourihane}, {Breedt}, {Littlefair},
  {Koposov}, {Wyrzykowski}, {Altavilla}, {Blagorodnova}, {Clementini},
  {Damljanovic}, {Delgado}, {Dennefeld}, {Drake},
  {Fern{\'a}ndez-Hern{\'a}ndez}, {Gilmore}, {Gualandi}, {Hamanowicz},
  {Handzlik}, {Hardy}, {Harrison}, {I{\l}kiewicz}, {Jonker}, {Kochanek},
  {Ko{\l}aczkowski}, {Kostrzewa-Rutkowska}, {Kotak}, {van Leeuwen}, {Leto},
  {Ochner}, {Pawlak}, {Palaversa}, {Rixon}, {Rybicki}, {Shappee}, {Smartt},
  {Torres}, {Tomasella}, {Turatto}, {Ulaczyk}, {van Velzen}, {Vince}, {Walton},
  {Wielg{\'o}rski}, {Wevers}, {Whitelock}, {Yoldas}, {De Angeli}, {Burgess},
  {Busso}, {Busuttil}, {Butterley}, {Chambers}, {Copperwheat}, {Danilet},
  {Dhillon}, {Evans}, {Eyer}, {Froebrich}, {Gomboc}, {Holland}, {Holoien},
  {Jarvis}, {Kaiser}, {Kann}, {Koester}, {Kolb}, {Komossa}, {Magnier},
  {Mahabal}, {Polshaw}, {Prieto}, {Prusti}, {Riello}, {Scholz}, {Simonian},
  {Stanek}, {Szabados}, {Waters}, \& {Wilson}}]{2015MNRAS.452.1060C}
{Campbell}, H.~C., {Marsh}, T.~R., {Fraser}, M., {et~al.} 2015, \mnras, 452,
  1060

\bibitem[{{Carrasco} {et~al.}(2016){Carrasco}, {Evans}, {Montegriffo}, {Jordi},
  {van Leeuwen}, \& {Riello}}]{DPACP-9}
{Carrasco}, J.~M., {Evans}, D.~W., {Montegriffo}, P., {et~al.} 2016, \aap, this
  volume

\bibitem[{{Clementini} {et~al.}(2016){Clementini}, {Ripepi}, {Leccia},
  {Mowlavi}, {Lecoeur-Taibi}, \& {Marconi}}]{DPACP-13}
{Clementini}, G., {Ripepi}, V., {Leccia}, S., {et~al.} 2016, \aap, this volume

\bibitem[{{Cropper} \& {Katz}(2011)}]{2011EAS....45..181C}
{Cropper}, M. \& {Katz}, D. 2011, in EAS Publications Series, Vol.~45, EAS
  Publications Series, 181--188

\bibitem[{{Crowley} {et~al.}(2016{\natexlab{a}}){Crowley}, {Abreu}, {Kohley},
  {Prod'homme}, \& {Beaufort}}]{2016arXiv160801476C}
{Crowley}, C., {Abreu}, A., {Kohley}, R., {Prod'homme}, T., \& {Beaufort}, T.
  2016{\natexlab{a}}, ArXiv e-prints [\eprint[arXiv]{1608.01476}]

\bibitem[{{Crowley} {et~al.}(2016{\natexlab{b}}){Crowley}, {Kohley}, {Hambly},
  {Davidson}, {Abreu}, \& {van Leeuwen}}]{DPACP-20}
{Crowley}, C., {Kohley}, R., {Hambly}, N.~C., {et~al.} 2016{\natexlab{b}},
  \aap, this volume

\bibitem[{{Dowler} {et~al.}(2010){Dowler}, {Rixon}, \&
  {Tody}}]{2010ivoa.spec.0327D}
{Dowler}, P., {Rixon}, G., \& {Tody}, D. 2010, {Table Access Protocol Version
  1.0}, IVOA Recommendation 27 March 2010

\bibitem[{{Epanechnikov}(1969)}]{doi:10.1137/1114019}
{Epanechnikov}, V.~A. 1969, Theory of Probability \& Its Applications, 14, 153

\bibitem[{{ESA}(1997)}]{1997ESASP1200.....E}
{ESA}, ed. 1997, ESA Special Publication, Vol. 1200, {The HIPPARCOS and TYCHO
  catalogues. Astrometric and photometric star catalogues derived from the ESA
  HIPPARCOS Space Astrometry Mission}

\bibitem[{{Evans} {et~al.}(2016){Evans}, {Riello}, {De Angeli}, {Busso}, {van
  Leeuwen}, \& {Jordi}}]{DPACP-11}
{Evans}, D., {Riello}, M., {De Angeli}, F., {et~al.} 2016, \aap, this volume

\bibitem[{{Eyer} {et~al.}(2016){Eyer}, {Mowlavi}, {Evans}, {Nienartowicz},
  {Ordo\'{o}n\~{n}ez}, \& {Holl}}]{DPACP-15}
{Eyer}, L., {Mowlavi}, N., {Evans}, D.~W., {et~al.} 2016, \aap, this volume

\bibitem[{{Fabricius} {et~al.}(2016){Fabricius}, {Bastian}, {Portell},
  {Casta\~{n}eda}, {Davidson}, {Hambly}, {Clotet}, \& {Biermann}}]{DPACP-7}
{Fabricius}, C., {Bastian}, U., {Portell}, J., {et~al.} 2016, \aap, this volume

\bibitem[{{Gould}(2004)}]{2004astro.ph..3506G}
{Gould}, A. 2004, ArXiv Astrophysics e-prints [\eprint{astro-ph/0403506}]

\bibitem[{{Henden} \& {Munari}(2014)}]{2014CoSka..43..518H}
{Henden}, A. \& {Munari}, U. 2014, Contributions of the Astronomical
  Observatory Skalnate Pleso, 43, 518

\bibitem[{{H{\o}g} {et~al.}(2000){H{\o}g}, {Fabricius}, {Makarov}, {Urban},
  {Corbin}, {Wycoff}, {Bastian}, {Schwekendiek}, \&
  {Wicenec}}]{2000A&A...355L..27H}
{H{\o}g}, E., {Fabricius}, C., {Makarov}, V.~V., {et~al.} 2000, \aap, 355, L27

\bibitem[{{{\it Gaia}~Collaboration} {et~al.}(2016{\natexlab{a}}){{\it
  Gaia}~Collaboration}, {Clementini}, {Eyer}, {Ripepi}, {Marconi}, {Muraveva},
  \& {Garofalo}}]{DPACP-24}
{{\it Gaia}~Collaboration}, {Clementini}, G., {Eyer}, E., {et~al.}
  2016{\natexlab{a}}, \aap, this volume

\bibitem[{{{\it Gaia}~Collaboration} {et~al.}(2016{\natexlab{b}}){{\it
  Gaia}~Collaboration}, {Prusti}, {de Bruijne}, {Brown}, {Vallenari},
  {Babusiaux}, \& {Bailer-Jones}}]{DPACP-1}
{{\it Gaia}~Collaboration}, {Prusti}, T., {de Bruijne}, J.~H.~J., {et~al.}
  2016{\natexlab{b}}, \aap, this volume

\bibitem[{{{\it Gaia}~Collaboration} {et~al.}(2016{\natexlab{c}}){{\it
  Gaia}~Collaboration}, {van Leeuwen}, {Bastian}, {Jordi}, {Vallenari},
  {Prusti}, \& {de Bruijne}}]{DPACP-23}
{{\it Gaia}~Collaboration}, {van Leeuwen}, F., {Bastian}, U., {et~al.}
  2016{\natexlab{c}}, \aap, this volume

\bibitem[{{Jordi} {et~al.}(2010){Jordi}, {Gebran}, {Carrasco}, {de Bruijne},
  {Voss}, {Fabricius}, {Knude}, {Vallenari}, {Kohley}, \&
  {Mora}}]{2010A&A...523A..48J}
{Jordi}, C., {Gebran}, M., {Carrasco}, J.~M., {et~al.} 2010, \aap, 523, A48

\bibitem[{{Lindegren} {et~al.}(2016){Lindegren}, {Lammers}, {Bastian},
  {Hern\'andez}, \& {Klioner}}]{DPACP-14}
{Lindegren}, L., {Lammers}, U., {Bastian}, U., {Hern\'andez}, J., \& {Klioner},
  S. 2016, \aap, this volume

\bibitem[{{Lindegren} {et~al.}(2012){Lindegren}, {Lammers}, {Hobbs},
  {O'Mullane}, {Bastian}, \& {Hern{\'a}ndez}}]{2012A&A...538A..78L}
{Lindegren}, L., {Lammers}, U., {Hobbs}, D., {et~al.} 2012, \aap, 538, A78

\bibitem[{{M{\"a}dler} {et~al.}(2016){M{\"a}dler}, {Jofr{\'e}}, {Gilmore},
  {Worley}, {Soubiran}, {Blanco-Cuaresma}, {Hawkins}, \&
  {Casey}}]{2016arXiv160603015M}
{M{\"a}dler}, T., {Jofr{\'e}}, P., {Gilmore}, G., {et~al.} 2016, ArXiv e-prints
  [\eprint[arXiv]{1606.03015}]

\bibitem[{{Marrese} {et~al.}(2016){Marrese}, {Marinoni}, {Giuffrida}, \&
  {Fabrizio}}]{DPACP-17}
{Marrese}, P.~M., {Marinoni}, S., {Giuffrida}, G., \& {Fabrizio}, M. 2016,
  \aap, this volume

\bibitem[{{Mason} {et~al.}(2001){Mason}, {Wycoff}, {Hartkopf}, {Douglass}, \&
  {Worley}}]{2001AJ....122.3466M}
{Mason}, B.~D., {Wycoff}, G.~L., {Hartkopf}, W.~I., {Douglass}, G.~G., \&
  {Worley}, C.~E. 2001, \aj, 122, 3466

\bibitem[{{Melis} {et~al.}(2014){Melis}, {Reid}, {Mioduszewski}, {Stauffer}, \&
  {Bower}}]{2014Sci...345.1029M}
{Melis}, C., {Reid}, M.~J., {Mioduszewski}, A.~J., {Stauffer}, J.~R., \&
  {Bower}, G.~C. 2014, Science, 345, 1029

\bibitem[{{Michalik} \& {Lindegren}(2016)}]{2016A&A...586A..26M}
{Michalik}, D. \& {Lindegren}, L. 2016, \aap, 586, A26

\bibitem[{{Michalik} {et~al.}(2015){Michalik}, {Lindegren}, \&
  {Hobbs}}]{2015A&A...574A.115M}
{Michalik}, D., {Lindegren}, L., \& {Hobbs}, D. 2015, \aap, 574, A115

\bibitem[{{Mignard} {et~al.}(2016){Mignard}, {Klioner}, {Lindegren}, {Bastian},
  {Bombrun}, \& {Hern\'andez}}]{DPACP-26}
{Mignard}, F., {Klioner}, S., {Lindegren}, L., {et~al.} 2016, \aap, this volume

\bibitem[{{Osuna} {et~al.}(2008){Osuna}, {Ortiz}, {Lusted}, {Dowler}, {Szalay},
  {Shirasaki}, {Nieto-Santisteban}, {Ohishi}, {O'Mullane}, {VOQL-TEG Group}, \&
  {VOQL Working Group.}}]{2008ivoa.spec.1030O}
{Osuna}, P., {Ortiz}, I., {Lusted}, J., {et~al.} 2008, {IVOA Astronomical Data
  Query Language Version 2.00}, IVOA Recommendation 30 October 2008

\bibitem[{{Perryman} {et~al.}(2001){Perryman}, {de Boer}, {Gilmore}, {H{\o}g},
  {Lattanzi}, {Lindegren}, {Luri}, {Mignard}, {Pace}, \& {de
  Zeeuw}}]{2001A&A...369..339P}
{Perryman}, M.~A.~C., {de Boer}, K.~S., {Gilmore}, G., {et~al.} 2001, \aap,
  369, 339

\bibitem[{{Recio-Blanco} {et~al.}(2016){Recio-Blanco}, {de Laverny}, {Allende
  Prieto}, {Fustes}, {Manteiga}, {Arcay}, {Bijaoui}, {Dafonte}, {Ordenovic}, \&
  {Ordo{\~n}ez Blanco}}]{2016A&A...585A..93R}
{Recio-Blanco}, A., {de Laverny}, P., {Allende Prieto}, C., {et~al.} 2016,
  \aap, 585, A93

\bibitem[{{Riello} {et~al.}(2016){Riello}, {De Angeli}, {Evans}, {Busso},
  {Montegriffo}, \& {Holland}}]{DPACP-10}
{Riello}, M., {De Angeli}, F., {Evans}, D.~W., {et~al.} 2016, \aap, this volume

\bibitem[{{Salgado} {et~al.}(2016){Salgado}, {Gonz\'{a}lez-N\'{u}\~{n}ez},
  {Guti\'{e}rrez-S\'{a}nchez}, {Segovia}, \& {Dur\'{a}n}}]{DPACP-19}
{Salgado}, J., {Gonz\'{a}lez-N\'{u}\~{n}ez}, J., {Guti\'{e}rrez-S\'{a}nchez},
  R., {Segovia}, J.~C., \& {Dur\'{a}n}, J. 2016, \aap, this volume

\bibitem[{{Skrutskie} {et~al.}(2006){Skrutskie}, {Cutri}, {Stiening},
  {Weinberg}, {Schneider}, {Carpenter}, {Beichman}, {Capps}, {Chester},
  {Elias}, {Huchra}, {Liebert}, {Lonsdale}, {Monet}, {Price}, {Seitzer},
  {Jarrett}, {Kirkpatrick}, {Gizis}, {Howard}, {Evans}, {Fowler}, {Fullmer},
  {Hurt}, {Light}, {Kopan}, {Marsh}, {McCallon}, {Tam}, {Van Dyk}, \&
  {Wheelock}}]{2006AJ....131.1163S}
{Skrutskie}, M.~F., {Cutri}, R.~M., {Stiening}, R., {et~al.} 2006, \aj, 131,
  1163

\bibitem[{{Smart} \& {Nicastro}(2014)}]{2014AA...570A..87S}
{Smart}, R.~L. \& {Nicastro}, L. 2014, \aap, 570, 8

\bibitem[{{van Leeuwen}(1999)}]{1999A&A...341L..71V}
{van Leeuwen}, F. 1999, \aap, 341, L71

\bibitem[{van Leeuwen(2007)}]{book:newhip}
van Leeuwen, F. 2007, {H}ipparcos, the {N}ew {R}eduction of the {R}aw {D}ata,
  {A}strophysics and {S}pace {S}cience {L}ibrary. {V}ol. 350 edn. (Springer)

\bibitem[{{van Leeuwen}(2009)}]{2009A&A...497..209V}
{van Leeuwen}, F. 2009, \aap, 497, 209

\bibitem[{{van Leeuwen} {et~al.}(2016){van Leeuwen}, {Evans}, {De Angeli},
  {Jordi}, {Busso}, \& {Cacciari}}]{DPACP-12}
{van Leeuwen}, F., {Evans}, D.~W., {De Angeli}, F., {et~al.} 2016, \aap, this
  volume

\bibitem[{{Wenger} {et~al.}(2000){Wenger}, {Ochsenbein}, {Egret}, {Dubois},
  {Bonnarel}, {Borde}, {Genova}, {Jasniewicz}, {Lalo{\"e}}, {Lesteven}, \&
  {Monier}}]{2000A&AS..143....9W}
{Wenger}, M., {Ochsenbein}, F., {Egret}, D., {et~al.} 2000, \aaps, 143, 9

\bibitem[{{Whitmore} {et~al.}(2016){Whitmore}, {Allam}, {Budav{\'a}ri},
  {Casertano}, {Downes}, {Donaldson}, {Fall}, {Lubow}, {Quick}, {Strolger},
  {Wallace}, \& {White}}]{2016AJ....151..134W}
{Whitmore}, B.~C., {Allam}, S.~S., {Budav{\'a}ri}, T., {et~al.} 2016, \aj, 151,
  134

\bibitem[{{York} {et~al.}(2000){York}, {Adelman}, {Anderson}, {Anderson},
  {Annis}, {Bahcall}, {Bakken}, {Barkhouser}, {Bastian}, {Berman}, {Boroski},
  {Bracker}, {Briegel}, {Briggs}, {Brinkmann}, {Brunner}, {Burles}, {Carey},
  {Carr}, {Castander}, {Chen}, {Colestock}, {Connolly}, {Crocker}, {Csabai},
  {Czarapata}, {Davis}, {Doi}, {Dombeck}, {Eisenstein}, {Ellman}, {Elms},
  {Evans}, {Fan}, {Federwitz}, {Fiscelli}, {Friedman}, {Frieman}, {Fukugita},
  {Gillespie}, {Gunn}, {Gurbani}, {de Haas}, {Haldeman}, {Harris}, {Hayes},
  {Heckman}, {Hennessy}, {Hindsley}, {Holm}, {Holmgren}, {Huang}, {Hull},
  {Husby}, {Ichikawa}, {Ichikawa}, {Ivezi{\'c}}, {Kent}, {Kim}, {Kinney},
  {Klaene}, {Kleinman}, {Kleinman}, {Knapp}, {Korienek}, {Kron}, {Kunszt},
  {Lamb}, {Lee}, {Leger}, {Limmongkol}, {Lindenmeyer}, {Long}, {Loomis},
  {Loveday}, {Lucinio}, {Lupton}, {MacKinnon}, {Mannery}, {Mantsch}, {Margon},
  {McGehee}, {McKay}, {Meiksin}, {Merelli}, {Monet}, {Munn}, {Narayanan},
  {Nash}, {Neilsen}, {Neswold}, {Newberg}, {Nichol}, {Nicinski}, {Nonino},
  {Okada}, {Okamura}, {Ostriker}, {Owen}, {Pauls}, {Peoples}, {Peterson},
  {Petravick}, {Pier}, {Pope}, {Pordes}, {Prosapio}, {Rechenmacher}, {Quinn},
  {Richards}, {Richmond}, {Rivetta}, {Rockosi}, {Ruthmansdorfer}, {Sandford},
  {Schlegel}, {Schneider}, {Sekiguchi}, {Sergey}, {Shimasaku}, {Siegmund},
  {Smee}, {Smith}, {Snedden}, {Stone}, {Stoughton}, {Strauss}, {Stubbs},
  {SubbaRao}, {Szalay}, {Szapudi}, {Szokoly}, {Thakar}, {Tremonti}, {Tucker},
  {Uomoto}, {Vanden Berk}, {Vogeley}, {Waddell}, {Wang}, {Watanabe},
  {Weinberg}, {Yanny}, {Yasuda}, \& {SDSS Collaboration}}]{2000AJ....120.1579Y}
{York}, D.~G., {Adelman}, J., {Anderson}, Jr., J.~E., {et~al.} 2000, \aj, 120,
  1579

\end{thebibliography}
